\documentclass[final,3p,times]{elsarticle}
\usepackage{amssymb}
\usepackage{amsmath}
\usepackage{amsthm}
\usepackage{mathrsfs}
\usepackage[colorlinks,citecolor=blue]{hyperref}
\usepackage{appendix}
\usepackage{booktabs}
\usepackage{multirow}
\usepackage{enumitem}
\usepackage{cases}
\usepackage{underscore} 
\usepackage[numbers]{natbib}
\usepackage{lineno}
\usepackage{color}
\usepackage{algorithm}
\usepackage{algorithmic}
\let\scrS\scriptscriptstyle

\usepackage{geometry}
\usepackage{graphicx}
\usepackage{subfigure}
\usepackage{caption}
\usepackage{rotating}
\usepackage{setspace}
\usepackage{cases}
\journal{{  }}
\date{}

\newtheorem{mytheom}{Theorem}[subsection]
\newtheorem{mylemma}{Lemma}[subsection]
\newtheorem{myprop}{Proposition}[subsection]

\newtheorem{myhypo}{Assumption}[section]

\begin{document}
\begin{frontmatter}
\title{First-order multivariate integer-valued autoregressive model with multivariate mixture distributions}

\author[1]{Weiyang Yu}
\ead{ywyschol@outlook.com}

\author[1]{Haitao Zheng \corref{cor1}}
\ead{htzheng@swjtu.edu.cn}

\cortext[cor1]{Corresponding author}
\address[1]{School of Mathematics, Southwest Jiaotong University, Chengdu, 611756, P.R. China}

\begin{abstract}
The univariate integer-valued time series has been extensively studied, but literature on multivariate integer-valued time series models is quite limited and the complex correlation structure among the multivariate integer-valued time series is barely discussed. In this study, we proposed a first-order multivariate integer-valued autoregressive model to characterize the correlation among multivariate integer-valued time series with higher flexibility. Under the general conditions, we established the stationarity and ergodicity of the proposed model. With the proposed method, we discussed the models with multivariate Poisson-lognormal distribution and multivariate geometric-logitnormal distribution and the corresponding properties. The estimation method based on EM algorithm was developed for the model parameters and extensive simulation studies were performed to evaluate the effectiveness of proposed estimation method. Finally, a real crime data was analyzed to demonstrate the advantage of the proposed model with comparison to the other models.
\end{abstract}

\begin{keyword}
EM algorithm, multivariate Poisson-lognormal distribution, multivariate geometric-logitnormal distribution, thinning operator
\end{keyword}

\end{frontmatter}

\section{Introduction}

There has been extensive research on modeling integer-valued time series data over the past few decades. Counting time series occurs frequently in the fields such as epidemiology, biology, engineering, and finance. Nevertheless, modeling multivariate integer-valued time series exhibits certain challenges, as it necessitates consideration of dependence among multivariate time series. Although some models have focused on modeling univariate and bivariate integer-valued time series data, the majority of existing methods can not model the complex correlation structure of multivariate counting time series well.

Most of integer-valued time series models are built on the basis of the thinning operator "$\circ$" which was initially proposed by \citet{steutel1979}. Following this, \citet{mckenzie1985} and \citet{alosh1987} proposed the first-order integer-valued autoregressive model (INAR(1)) which can be expressed as
$$
X_t=\alpha \circ X_{t-1}+R_t,
$$
where $\alpha\in[0,1]$ and $R_t$ is an independent integer-valued random variable with finite mean $\mu$ and variance $\sigma^{2}$. Here, $\alpha \circ X_{t-1}$ represents $\sum_{i=1}^{X_{t-1}}Y_i$, where $Y_i$ is independent and identically distributed Bernoulli random variable with parameter $\alpha$. Such models have been widely reviewed, seeing \citet{jung2006binomial} and \citet{weiss2018introduction}.

To improve the performance of univariate integer-valued autoregressive model with fixed parameter $\alpha$, \citet{zheng2007} introduced the first-order random coefficient integer-valued autoregressive model. This model generalized the fixed parameter $\alpha$ in the first-order integer-valued autoregressive model to a random coefficient $\alpha_{t}$, where $\alpha_{t} \in [0,1]$. \citet{kim2008} introduced the signed binomial thinning operator, \citet{ristic2009} introduced the negative binomial thinning operator, and \citet{zhang2010} introduced the signed generalized power series thinning operator. \citet{popovic2015} introduced a two-dimensional first-order integer-valued autoregressive model with random truncation coefficients, considering the correlation between the two variables. \citet{yu2020} extended the results of \citet{zheng2007} and established a two-dimensional random coefficient integer-valued autoregressive model.

In practice, a large number of integer-valued time series involve periodic feature such as tourism demand, fire activity, and social science time series. To address this issue, \citet{monteiro2010} proposed a class of univariate INAR models for periodic changes. Based on this, \citet{monteiro2015} further extended the model to the bivariate case and investigated related periodic changes. More recently, \citet{santos2021} generalized the results of \citet{monteiro2015} to the multivariate case and obtained more comprehensive results.

For multivariate integer-valued time series, there exist correlations among multivariate random vectors and dependence between multivariate count time series. To deal with this issue, \citet{franke1993} extended the one-dimensional binomial thinning operator "$\circ$" to the binomial thinning operator matrix. They defined the rule of operation for thinning operator matrices, replacing element-wise operations in general matrix operations with thinning operator operations, and studied the properties of the operator matrix. 
\citet{latour1997} generalized the thinning operator matrix, allowing the operations of the operator matrix to go beyond the binomial distribution. Furthermore, they established a first-order multivariate integer-valued autoregressive (MINAR(1)) model based on thinning operator matrices. However, this paper didn't pay attention to the correlation among innovation processes.
To address this limitation, \citet{quoreshi2006} proposed a bivariate integer-valued moving average (BINMA) model. This model allows for both positive and negative correlations among the random vector.
In addition, to handle the correlation and overdispersion in multivariate integer-valued processes, \citet{karlis2007finite} proposed finite mixture models. Furthermore, \citet{heinen2007} introduced the multivariate autoregressive conditional double Poisson model for multivariate integer-valued time series.
For further improvement in modeling correlation, \citet{pedeli2011} considered the correlation between bivariate integer-valued random vector and developed a first-order bivariate integer-valued autoregressive model(BINAR(1)). They studied the properties of this model based on the bivariate Poisson distribution and bivariate negative binomial distributions. \citet{pedeli2013} extended this model to the multivariate case by considering the correlation among elements of the multivariate integer-valued random vectors. 
Recently, \citet{liu2016} proposed a new stationary bivariate INAR(1) process with zero-truncated Poisson marginal distribution, and \citet{popovic2016} introduced a bivariate INAR(1) model with geometric marginal distribution. To describe the correlation between bivariate integer-valued sequences, \citet{cui2018new} introduced a special bivariate Poisson distribution and established a new integer-valued generalized autoregressive conditional heteroscedasticity (INGARCH) model which can flexibly handle the correlation between bivariate integer-valued sequences. \citet{piancastelli2023flexible} introduced another bivariate Poisson distribution and developed a new integer-valued generalized autoregressive conditional heteroscedasticity model which addressed some issues in the work of \citet{cui2018new}.
To further improve the flexibility of modeling the multivariate integer-valued time series, \citet{karlis2013flexible} employed the Coupla method to establish a bivariate INAR(1) models that can capture the correlation in multivariate count time series. \citet{cui2020flexible} proposed a new bivariate Poisson distribution based on the multiplicative factor approach and developed a new integer-valued generalized autoregressive conditional heteroscedasticity model that can describe positive, negative, and zero correlations among multivariate time series.

It can be seen that the existing methods always have some limitations on capturing the correlation structure among multivariate integer-valued time series. In this study, we propose a model that can flexibly describe the correlation among multivariate integer-valued time series. Two kinds of multivariate integer-valued mixture distributions are discussed to model the innovation processes. The conditions for stationarity and ergodicity of the proposed models are studied under general conditions, as well as related moment properties. The corresponding estimation algorithm for the unknown parameters of the models is developed by using the EM algorithm.

The main content of this paper includes the following sections. In Section 2, the general form of the MINAR(1) model is introduced. The models with multivariate Poisson-lognormal distribution and multivariate geometric-logitno rmal distribution are studied and some corresponding theoretical properties are given. In Section 3, parameter estimation method and algorithm are developed. In Section 4, extensive simulation studies are performed to verify the effectiveness of the proposed methods. In Section 5, the proposed method is applied to a real dataset to show the advantage of new models by comparing with the other methods.

\section{Multivariate integer-valued autoregressive process}
\subsection{General form}
The usual first-order integer-valued autoregressive process $\left\{ X_t : t=0,\pm1,\pm2,\dots \right\}$ can be generally defined as follows
$$
X_t = \alpha \circ X_{t-1} + R_{t},
$$
where $\alpha \in [0,1]$ and $R_{t}$ is a set of uncorrelated random variables with finite mean $\mu$ and variance $\sigma^{2}$ (\citet{alosh1987}). The symbol "$\circ$" represents the binomial thinning operator proposed by \citet{steutel1979}, defined as
$$
\alpha \circ X = \sum\limits_{i=1}^{X} Z_{i},
$$
where $X$ is a non-negative integer-valued random variable and $\{Z_{i}\}$ are independent and identically distributed Bernoulli random variables with $P(Z_{i}=1)=1-P(Z_{i}=0)=\alpha$. The incorporation of such binomial thinning operators is the foundation for the analysis of integer-valued time series.

Following the method proposed by \citet{pedeli2013}, we extend the univariate INAR(1) process by establishing the operational rules of the thinning operator matrix and the multivariate integer-valued autoregressive models. Let $\boldsymbol{X}_t = [X_{1,t}, X_{2,t}, \cdots, X_{N,t}]^{\prime }$ and $\boldsymbol{R}_t = [R_{1,t}, R_{2,t}, \cdots, R_{N,t}]^{\prime }$ be an $N$-dimensional non-negative integer-valued random vector, and $\boldsymbol{A}$ be an $N \times N$ diagonal matrix defined as follows:
$$
\boldsymbol{A}=\left[ \begin{matrix}
	\alpha _1&		0&		\cdots&		0\\
	0&		\alpha _2&		&		0\\
	\vdots&		&		\ddots&		\vdots\\
	0&		0&		\cdots&		\alpha _{\scrS N}\\
\end{matrix} \right] ,
$$
where the elements $\alpha_{i}$, $i = 1,2,\dots,N$, are constant parameters. We can define the first-order multivariate integer-valued autoregressive process as follows:
\begin{equation}
\label{geneq1}
\boldsymbol{X}_t =\boldsymbol{A} \circ \boldsymbol{X}_{t-1} + \boldsymbol{R}_{t}, \ t \in \mathbb{T},
\end{equation}
where $\left\{ \boldsymbol{R }_t \right\} _{t\in \mathbb{T}}$ is an independent and identically distributed non-negative integer-valued innovation process with $N$-dimensional mean vector $\boldsymbol{\mu }_{\scrS   R}$ and $N\times N$-dimensional variance matrix $\boldsymbol{\Sigma }_{\scrS   R}$. The operations of binomial thinning operator matrix $\boldsymbol{A} \circ$ are independent of each other and independent of $\left\{ \boldsymbol{R }_t \right\} _{t\in \mathbb{T}}$. The mean vector and variance matrix of $\left\{ \boldsymbol{X }_t \right\} _{t\in \mathbb{T}}$ are denoted by $\boldsymbol{\mu }_{\scrS X }$ and $\boldsymbol{\Sigma }_{\scrS X}$ respectively.

For convenience, we assume that $\boldsymbol{\mu }_{\scrS R }=\left\{ \mu _{\scrS   R,  i} \right\} _{i=1}^{N}$, $ \boldsymbol{\Sigma }_{\scrS R}=\left\{ \sigma _{\scrS R,ij} \right\} _{i,j=1}^{N\times N} $, $\boldsymbol{\mu }_{\scrS X}=\left\{ \mu _{\scrS   X,i} \right\} _{i=1}^{N}$, $\boldsymbol{\Sigma }_{\scrS X}  =\left\{ \sigma _{\scrS   X,ij} \right\} _{i,j=1}^{N\times N} $. Assuming the first and second moments of the innovation process exist and \begin{myhypo}
\label{genhypo1}
$0 < \alpha_{s} < 1$, $s \in \{1,2,\dots,N\}$. 
\end{myhypo}

Clearly in equation \eqref{geneq1}, we replace the general univariate binomial thinning operator with binomial thinning operator matrix. In fact, $\boldsymbol{A}\circ$ has the operational rules of the binomial thinning operator in addition to the general matrix operations, i.e. we replace numerical operations in general matrix operations with the thinning operator operations defined above. When the maximum eigenvalue of matrix $\boldsymbol{A}$ is less than 1, i.e., $\max(\alpha_j) < 1$, $j=1,2,\dots,N$, the non-negative integer-valued random process $\{\boldsymbol{X}_t\}_{t\in \mathbb{T}}$ is the unique strictly stationary process (\citet{franke1993}).

\begin{mytheom}
\label{gentheo3}
Let $\{\boldsymbol{X}_t\}_{t\in \mathbb{T}}$ be a first-order multivariate integer-valued autoregressive (MINAR(1)) process  defined by equation \eqref{geneq1}. Then, this process is strictly stationary and ergodic.
\end{mytheom}
See proof in the appendix.

In model \eqref{geneq1}, we can obtain the following moment properties of  $\{\boldsymbol{X}_t\}_{t\in \mathbb{T}}$.

\begin{myprop}
\label{genprop1}
Let $\{\boldsymbol{X}_t\}_{t\in \mathbb{T}}$ be a process defined by \eqref{geneq1}. Then
\begin{enumerate}[label=(\roman*)]
\item $\mu_{\scrS X, i} = \frac{\mu_{\scrS   R,  i }}{1-\alpha_i}$;
\item $\sigma_{\scrS X, ii} = \frac{\alpha_i\mu_{\scrS   R,i}+\sigma_{\scrS   R,ii}}{1-\alpha_{i}^{2}}$;
\item $\sigma_{\scrS X, ij} = \frac{\sigma_{\scrS   R,ij}}{1-\alpha_i\alpha_j}$;
\item $Cov\left( X_{i,t},_{i+h,t} \right) = \gamma_{\scrS   X_i}(h) = \alpha_{i}^{h}\sigma_{\scrS   X,i}$,  $h=0,1,2,\dots$;
\item $Cov\left( X_{i,t+h},X_{j,t} \right) = \frac{\alpha_{i}^{h}}{1-\alpha_i\alpha_j}\sigma_{\scrS   R,ij}$,  $h=0,1,2,\dots,i\ne j$. 
\end{enumerate}
\end{myprop}

In the following content, we will establish two multivariate integer-valued autoregressive models with different distributions of the innovation processes. The first one is the multivariate Poisson-lognormal MINAR(1) model, where the innovation process follows a multivariate Poisson-lognormal distribution. The second one is the multivariate geometric-logitnormal MINAR(1) model, where the innovation process follows a multivariate geometric-logitnormal distribution. These two mixture distributions can explain the over-dispersion phenomenon in the data and model the potential correlation between multivariate integer-valued time series. Moreover, for convenience, we assume that the mean parameters of the two different distributions is denoted as N-dimensional vector $\boldsymbol{\mu}$, and the covariance matrix of those distributions is denoted as $N \times N$ matrix $\boldsymbol{\Sigma}$, where $\boldsymbol{\mu}=\{\mu_{s}  \}_{s=1}^{N}$ and $\boldsymbol{\Sigma}=\{\sigma_{\scrS ij}\}_{i,j=1}^{N\times N}$. Thus, we represent the parameters in the model as $\boldsymbol{\theta}=\{\boldsymbol{\alpha}, \boldsymbol{\mu}, \mathrm{vech}(\boldsymbol{\Sigma})\}=\{\alpha_{1}, \alpha_{2}, \dots, \alpha_{\scrS N}, \mu_{1}, \mu_{2}, \dots, \mu_{\scrS N},$ $\sigma_{11}, \sigma_{12}, \dots, \sigma_{\scrS NN}\}$, where "vech" means vectorization operation without duplicate elements (\citet{hamilton2020time}), and the parameter space as $\Theta$.

\subsection{The Poisson-lognormal MINAR(1) process}
In univariate non-negative integer-valued distributions, the Poisson distribution is commonly used to describe the variation of integer data. Naturally, when dealing with multivariate integer-valued data, one would want to generalize the univariate Poisson distribution to the multivariate case, so that the generalized Poisson distribution can capture the variations in multivariate count data. However, in multivariate discrete distributions, the most popular classical definition of the multivariate Poisson distribution (\citet{johnson1997}) is too restrictive, as it only allows different pairs of random variables to have the same covariance. \citet{karlis2005} proposed a more general multivariate Poisson distribution that allows different pairs of random variables to have different covariance. This distribution can be seen as a discrete counterpart of the multivariate normal distribution. Unfortunately, the extended multivariate Poisson distribution obtained through this approach does not support negative correlation between two count random variables so that it lacks the ability of modeling multivariate integer-valued time series flexibly.

To extend the multivariate integer-valued distribution to support the potential positive or negative correlation among elements of integer-valued random vector, one can accomplish this by mixing $n$ independent Poisson distributions.
Suppose there are $N$ independent Poisson distributions with parameters $\theta_i$. \citet{steyn1976} assumed that $\boldsymbol{\theta}=\left(\theta_1,\theta_2,\dots,\theta_{\scrS N}\right)^{\prime}$ follows a multivariate normal distribution $N\left(\boldsymbol{\mu}, \boldsymbol{\Sigma}\right)$, and constructed an $N$-dimensional Poisson mixture distribution.
Although this method allows the existence of negative correlation between two count random variables, it is logically problematic to assume that the strictly positive parameter vector $\boldsymbol{\theta}$ follows a normal distribution.
To address this logical issue and retain the rich covariance structure of the normal distribution, \citet{aitchison1989} used the multivariate lognormal distribution to construct a multivariate Poisson-lognormal mixture distribution for describing multivariate integer-valued random vector. In this section, we will construct a new MINAR(1) process using the multivariate Poisson-lognormal mixture distribution to introduce flexible correlation relationships into the MINAR(1) process.

Let $\boldsymbol{R}=\left(R_1,R_2,\dots,R_{\scrS N}\right)^{\prime }$ be the random vector obtained from the following equations
\begin{equation}
\label{pleq1}
\begin{aligned}
&\boldsymbol{R} \sim \text{Pois}\left(\text{exp}\left(\boldsymbol{\lambda}\right)\right) \\
&\boldsymbol{\lambda} \sim \text{MVN}\left(\boldsymbol{\mu},\boldsymbol{\Sigma}\right)
\end{aligned}
\end{equation}
where $\boldsymbol{\mu}$ and $\boldsymbol{\Sigma}$ represent the mean vector and variance-covariance matrix of the lognormal distribution, respectively. In matrix $\boldsymbol{\Sigma}$, non-diagonal elements can be positive, negative or zero. Therefore, Equation \eqref{pleq1} specifies that $\boldsymbol{R}$ follows a multivariate Poisson-lognormal mixture distribution with parameters mean vector $\boldsymbol{\mu}$ and variance matrix $\boldsymbol{\Sigma}$. The probability mass function of this distribution is given by
\begin{equation}
\label{pleq2}
\begin{aligned}
P\left( \boldsymbol{R} \right) =
&\int_{\boldsymbol{\lambda}}{\prod_{s=1}^N\left(\frac{\exp\left(R_s\cdot\lambda_s\right)}{R_s!}\exp\left(-\exp\left(\lambda_s\right)\right)\right)} \\
&\ast\frac{1}{\sqrt{\left(2\pi\right)^N\left|\sigma\right|}}\exp\left(-\frac{1}{2}\left(\boldsymbol{\lambda}-\boldsymbol{\mu}\right)^{\prime}\boldsymbol{\Sigma}^{-1}\left(\boldsymbol{\lambda}-\boldsymbol{\mu}\right)\right)d\boldsymbol{\lambda}. 
\end{aligned}
\end{equation}

According to \citet{izsak2008}, we can provide some properties of the first-order partial derivatives of the probability density function for the multivariate Poisson-lognormal mixture distribution. These properties can play an important role in parameter estimation using maximum likelihood method. They are described as follows.
\begin{mylemma}
\label{pllemma1}
For the $N$-dimensional Poisson-lognormal mixture distribution with parameters $\boldsymbol{\mu}$ and $\boldsymbol{\Sigma}$, the first-order partial derivatives of the probability density function have the following recursion formulas: for $\forall s,t \in \{1,2,\dots,N\}$,
$$
\frac{\partial P_{\boldsymbol{r}}}{\partial\mu_s} = r_sP_{\boldsymbol{r}} - (r_s+1)P_{\boldsymbol{r}+\boldsymbol{e}_s},
$$
$$
\frac{\partial P_{\boldsymbol{r}}}{\partial \sigma _{st}}=
\begin{cases}
	0.5\left\{ r_sr_tP_{\boldsymbol{r}}-r_s\left( r_t+1 \right) P_{\boldsymbol{r}+\boldsymbol{e}_t}-r_t\left( r_s+1 \right) P_{\boldsymbol{r}+\boldsymbol{e}_s}+\left( r_s+1 \right) \left( r_t+1 \right) P_{\boldsymbol{r}+\boldsymbol{e}_s+\boldsymbol{e}_t} \right\}   \quad s\ne t  \\
	0.5\left\{ r_{s}^{2}P_{\boldsymbol{r}}-\left( r_s+1 \right) \left( 2r_s+1 \right) P_{\boldsymbol{r}+\boldsymbol{e}_s}+\left( r_s+1 \right) \left( r_s+2 \right) P_{\boldsymbol{r}+\boldsymbol{e}_s+\boldsymbol{e}_t} \right\}   \, \,\,\, \quad \quad \quad \quad \quad \quad s=t ,
\end{cases}
$$
where $\boldsymbol{e}_s$ denotes the $N$-dimensional unit vector with the $s$-th element being 1 and other elements being 0.
\end{mylemma}


Next, we provide the proof of the moment properties of the multivariate normal distribution and introduce the corresponding assumptions and lemmas. Firstly, we should have the following assumptions.
\begin{myhypo}
\label{plhypo2}
The parameter space $\Theta$ is a closed compact set.
\end{myhypo}

Next, to prove the moment properties of the multivariate normal distribution, we provide the following lemma.
\begin{mylemma}
\label{pllemma2}
For a random variable $\lambda$ following a one-dimensional lognormal distribution with parameters $\mu$ and $\sigma^2$, we have: for any given $k \in \mathbb{Z}^{+} $, 
$$
E\left( \lambda ^k \right) =\exp \left( k\mu +\frac{1}{2}k^2\sigma ^2 \right) .
$$
\end{mylemma}

Next, we provide the following lemma based on the previous content.
\begin{mylemma}
\label{pllemma3}
Suppose a random vector $\boldsymbol{\lambda}$ that follows a N-dimensional lognormal distribution with mean vector $\boldsymbol{\mu}$ and variance-covariance matrix $\boldsymbol{\Sigma}$, we have the following formula: for a given non-negative integer vector $(k_1, k_2, \dots, k_{\scrS N})$,
$$
E\left( \lambda _{1}^{k_1}\lambda _{2}^{k_2}\cdots \lambda_{\scrS N}^{k_{\scrS N}} \right) =\exp \left[ k_1\mu _1+k_2\mu _2+\cdots +k_{\scrS N} \mu_{\scrS N}+\frac{1}{2}\sum_{i=1}^N{\sum_{j=1}^N{k_i}}k_j\sigma _{ij} \right] .  
$$
\end{mylemma}

\begin{mylemma}
\label{pllemma4}
For a random variable $R$ that follows a Poisson distribution with parameter $\lambda$, its high-order moments have the following property:
$$
E R^{k+1}=\lambda E R^k+\lambda \frac{d E R^k}{d \lambda}, \, \, k = 0,1,2,\dots.
$$
Furthermore, we can conclude that the $(k+1)$-th moment of a Poisson distribution with parameter $\lambda$ is a $(k+1)$th degree polynomial in $\lambda$.
\end{mylemma}

\begin{mytheom}
\label{pltheorem1}
For a random vector $\boldsymbol{R}$ that follows a N-dimensional Poisson-lognormal mixture distribution with parameters mean vector $\boldsymbol{\mu}$ and variance-covariance $\boldsymbol{\Sigma}$, it holds that for any given positive integer $k$,
$$
E\boldsymbol{R}^k<\infty .
$$
\end{mytheom}

Finally, based on the previous theorem, we present the properties of the first and second moments of a random vector $\boldsymbol{R}$ that follows a multivariate Poisson-lognormal mixture distribution with parameters mean vector $\boldsymbol{\mu}$ and variance matrix $\boldsymbol{\Sigma}$.
\begin{myprop}
\label{plprop1}
For the first and second moments of a random vector $\boldsymbol{R}$ that follows a multivariate Poisson-lognormal mixture distribution with parameters $\boldsymbol{\mu}$ and $\boldsymbol{\Sigma}$, we have the following equations:
\begin{enumerate}[label=(\roman*)]
\item $\mu _{\scrS   R,i} =\exp \left( \mu _i+\frac{1}{2}\sigma _{ii} \right) $;
\item $ER_{i}^{2}=\exp \left( \mu _i+\frac{1}{2}\sigma _{ii} \right) +\exp \left( 2\mu _i+2\sigma _{ii} \right)$;
\item $ER_iR_j=\exp \left[ \mu _i+\mu _j+\frac{1}{2}\left( \sigma _{ii}+\sigma _{jj} \right) +\sigma _{ij} \right]$ ;
\item $\sigma _{\scrS   R,ii} =\exp \left( \mu _i+\frac{1}{2}\sigma _{ii} \right) +\exp \left( 2\mu _i+\sigma _{ii} \right) \left[ \exp \left( \sigma _{ii} \right) -1 \right] $;
\item $\sigma _{\scrS   R,ij} =\exp \left[ \mu _i+\mu _j+\frac{1}{2}\left( \sigma _{ii}+\sigma _{jj} \right) \right] \left[ \exp \left( \sigma _{ij} \right) -1 \right] $;
\item $corr\left( R_i,R_j \right) =\frac{\exp \left( \sigma _{ij} \right) -1}{\sqrt{\left\{ \exp \left( \sigma _{ii} \right) -1+\exp \left( -\left( \mu _i+\frac{1}{2}\sigma _{ii} \right) \right) \right\} \left\{ \exp \left( \sigma _{jj} \right) -1+\exp \left( -\left( \mu _j+\frac{1}{2}\sigma _{jj} \right) \right) \right\}}} $.
\end{enumerate}
\end{myprop}

According to Proposition \ref{plprop1}, we can observe that the multivariate Poisson-lognormal mixture distribution can handle overdispersion in multivariate integer-valued data. Therefore, the proposed multivariate integer autoregressive model \eqref{geneq1} based on this mixture distribution can be quite flexible in capturing the correlation structure of multivariate count sequences.

Under the definition of the innovation process mentioned above, we can define an first-order multivariate integer-valued autoregressive process with multivariate Poisson-lognormal processes denoted by MINAR(1)-PL. Based on Propositions \ref{genprop1} and \ref{plprop1}, we can derive the moment properties of the proposed model.
\begin{myprop}
\label{plprop2}
For the multivariate integer autoregressive model defined by \eqref{geneq1} and \eqref{pleq2}, we have the following moment properties:
\begin{enumerate}[label=(\roman*)]
\item $\mu _{\scrS   X,i}=\frac{\exp \left( \mu _i+\frac{1}{2}\sigma _{ii} \right)}{1-\alpha _i}$;
\item $\sigma _{\scrS   X,ii}=\frac{\alpha _i\exp \left( \mu _i+\frac{1}{2}\sigma _{ii} \right) +\left\{ \exp \left( \mu _i+\frac{1}{2}\sigma _{ii} \right) +\exp \left( 2\mu _i+\sigma _{ii} \right) \left\{ \exp \left( \sigma _{ii} \right) -1 \right\} \right\}}{1-\alpha _{i}^{2}}  $;
\item $\sigma _{\scrS   X,ij}=\frac{\exp \left( \mu _i+\mu _j+\frac{1}{2}\left( \sigma _{ii}+\sigma _{jj} \right) \right) \left\{ \exp \left( \sigma _{ij} \right) -1 \right\}}{1-\alpha _i\alpha _j}$;
\item $Cov\left( X_{i,t},X_{i,t+h} \right) =\alpha _{i}^{h}\sigma _{\scrS   X,i},  \,  h=0,1,2,\dots .
$;
\item $Cov\left( X_{i,t+h},X_{j,t} \right) =\frac{\alpha _{i}^{h}}{1-\alpha _i\alpha _j}\sigma _{\scrS   R,ij},  \,  h=0,1,2,\dots ,i\ne j$.
\end{enumerate}
\end{myprop}

For the process defined by equations \eqref{geneq1} and \eqref{pleq2}, we can provide the following properties for the higher-order moments:
\begin{myprop}
\label{plprop3}
Let $\boldsymbol{X}_t$ be the process defined by equations \eqref{geneq1} and \eqref{pleq2}, and assume that hypotheses \ref{genhypo1} and \ref{plhypo2} hold. Then, for any fixed positive integer $k$, $E(\boldsymbol{X}_t^k )<\infty$.
\end{myprop}

\subsection{The geometric-logitnormal MINAR(1) process}
Based on the definition of multivariate Poisson-lognormal mixture distribution, assuming that the elements of the random vector $\boldsymbol{R}=(R_1, R_2, \dots, R_{\scrS N})$ are mutually independent and follow geometric distributions with parameters $\frac{\exp\left(p_i\right)}{1+\exp\left(p_i\right)}$. Given $\boldsymbol{p}$, we can obtain the conditional probability mass function of the random vector $\boldsymbol{R}=(R_1,R_2,\dots,R_{\scrS N})$ as follows:
\begin{equation}
\label{gleq1}
\begin{aligned}
P\left( \boldsymbol{R}|\boldsymbol{p} \right) 
&=\prod_{s=1}^N{\left( 1-\frac{\exp \left( p_s \right)}{1+\exp \left( p_s \right)} \right) ^{R_s-1}\frac{\exp \left( p_s \right)}{1+\exp \left( p_s \right)}}\\
&=\prod_{s=1}^N{\frac{\exp \left( p_s \right)}{\left\{ 1+\exp \left( p_s \right) \right\} ^{R_s}}}.
\end{aligned}
\end{equation}
Furthermore, assuming that $\mathrm{logit} (\boldsymbol{p})$ follows a $N$-dimensional multivariate normal distribution with parameters mean vector $\boldsymbol{\mu}$ and covariance matrix $\boldsymbol{\Sigma}$, we can derive the joint distribution function of the random vector $\boldsymbol{R}$ as follows:
\begin{equation}
\label{gleq2}
\begin{aligned}
P\left( \boldsymbol{R } \right) =&\int_{\boldsymbol{\lambda }}{\prod_{s=1}^{N}{\frac{\exp \left( p_s \right)}{\left\{ 1+\exp \left( p_s \right) \right\} ^{R_s}}}}
\\
&\ast \frac{1}{\sqrt{\left( 2\pi \right) ^{N}\left| \sigma \right|}}\exp \left( -\frac{1}{2}\left( \boldsymbol{p}-\boldsymbol{\mu } \right) ^{\prime}\boldsymbol{\Sigma }^{-1}\left( \boldsymbol{p }-\boldsymbol{\mu } \right) \right) d\boldsymbol{p }.
\end{aligned}
\end{equation}

For simplicity, we can derive some properties of this distribution. Firstly, according to the definition of the multivariate geometric-logitnormal mixture distribution, we can obtain the recursive formula for the first-order partial derivatives of its probability density function.
\begin{mylemma}
\label{gllemma1}
For the $N$-dimensional geometric-logitnormal mixture distribution with parameters mean vector $\boldsymbol{\mu}$ and covariance of $\boldsymbol{\Sigma}$, the first-order partial derivatives of its probability density function satisfy the following recursive formulas: for $\forall s,t\in \left\{1,2,\dots,N\right\}$,
$$
\frac{\partial P_{\boldsymbol{r}}}{\partial \mu _s}=\left( r_s+1 \right) P_{\boldsymbol{r}+\boldsymbol{e}_{s}}-r_sP_{\boldsymbol{r}},
$$
$$
\frac{\partial P_{\boldsymbol{r}}}{\partial \sigma _{st}}=\begin{cases}
	0.5\left\{ r_sr_tP_{\boldsymbol{r}}-r_s\left( r_t+1 \right) P_{\boldsymbol{r}+\boldsymbol{e}_t}-r_t\left( r_s+1 \right) P_{\boldsymbol{r}+\boldsymbol{e}_s}+\left( r_s+1 \right) \left( r_t+1 \right) P_{\boldsymbol{r}+\boldsymbol{e}_s+\boldsymbol{e}_t} \right\} \quad s\ne t\\
	0.5\left\{ r_{s}^{2}P_{\boldsymbol{r}}-\left( r_s+1 \right) \left( 2r_s+1 \right) P_{\boldsymbol{r}+\boldsymbol{e}_s}+\left( r_s+1 \right) \left( r_s+2 \right) P_{\boldsymbol{r}+\boldsymbol{e}_s+\boldsymbol{e}_t} \right\}   \, \,\,\, \quad \quad \quad \quad \quad \quad s=t   ,
\end{cases}
$$
where $\boldsymbol{e}_s$ denotes the $N$-dimensional unit vector with the $s$-th element being 1 and other elements being 0.
\end{mylemma}


To demonstrate the properties of the multivariate geometric-logitnormal distribution, we need to make some assumptions and lemmas. First, we assume that the parameter space $\Theta$ is a closed and bounded set as follows, to ensure that the parameters of the multivariate-logitnormal mixture distribution have finite upper and lower bounds.
\begin{myhypo}
\label{glhypo1}
The parameter space $\Theta$ is a closed and dense set.
\end{myhypo}
Next, we present the following lemma.
\begin{mylemma}
\label{gllemma2}
For the one-dimensional logitnormal distribution with parameters mean vector $\mu$ and covariance matrix $\sigma^{2}$, we have the following result: for any given $k \in \mathbb{Z}^{+}$,
$$
E\left( \frac{1}{p^k} \right) =\sum_{i=0}^k{C_{k}^{i}\exp \left( -i\mu +\frac{1}{2}i^2\sigma ^2 \right)}.
$$
\end{mylemma}

\begin{mylemma}
\label{gllemma3}
For a random vector $(p_{1}, p_{2}, \dots, p_{\scrS N})$ that follows a N-dimensional logitnormal distribution with parameters mean vector $\boldsymbol{\mu}$ and covariance matrix $\boldsymbol{\Sigma}$, we have the following formula: for a given non-negative integer vector $(k_{1}, k_{2}, \dots, k_{\scrS N})$,
$$
E\left[ \left( \frac{1}{p_1} \right) ^{k_1}\left( \frac{1}{p_2} \right) ^{k_2}\cdots \left( \frac{1}{p_{\scrS N}} \right) ^{k_{\scrS N}} \right] =\sum_{i_1=0}^{k_1}{\sum_{i_2=0}^{k_2}{\cdots \sum_{i_{\scrS N}=0}^{k_{\scrS N}}{C_{k_1}^{i_1}C_{k_2}^{i_2}\cdots C_{k_{\scrS N}}^{i_{\scrS N}}}}}\cdot \exp \left[ -\sum_{p=1}^N{i_p\mu _p}+\frac{1}{2}\sum_{p=1}^N{\sum_{q=1}^N{i_p}}i_q\sigma _{pq} \right] .
$$
\end{mylemma}

\begin{mylemma}
\label{gllemma4}
For a random variable $R$ following a geometric distribution with parameter $p$, the high-order moments have the following property:
$$
E R^{k+1}= ER \cdot\sum_{m=0}^k{C_{k+1}^{m}ER^m} , \, \, k = 0,1,2,\dots.
$$
Furthermore, from the low-order moments of the geometric distribution less than k, we can obtain that the $(k+1)$-th moment of a Poisson distribution with parameter $p$ is a $(k+1)$-th order polynomial of $\frac{1}{p}$.
\end{mylemma}

\begin{mytheom}
\label{gltheo1}
For a random vector $\boldsymbol{R}$ following a multivariate geometric-logitnormal mixture distribution with parameters $\boldsymbol{\mu}$ and $\boldsymbol{\Sigma}$, for any given positive integer $k$, we have
$$
E\boldsymbol{R}^k<\infty .
$$
\end{mytheom}

\begin{myprop}
\label{glprop1}
For a random vector $\boldsymbol{R}$ following a multivariate geometric-logitnormal mixture distribution with parameters $\boldsymbol{\mu}$ and $\boldsymbol{\Sigma}$, we can obtain the following formulas for the first and second moments:
\begin{enumerate}[label=(\roman*)]
\item $ \mu_{\scrS R,i} =\exp \left( -\mu _i+\frac{1}{2}\sigma _{ii} \right) $;
\item $ER_{i}^{2}=\exp \left(- \mu _i+\frac{1}{2}\sigma _{ii} \right) + 2 \exp \left(- 2\mu _i+2\sigma _{ii} \right)$;
\item $ER_iR_j=\exp \left[ - \mu _i - \mu _j+\frac{1}{2}\left( \sigma _{ii}+\sigma _{jj} \right) +\sigma _{ij} \right]$ ;
\item $\sigma_{\scrS R, ii}   =\exp \left(- \mu _i+\frac{1}{2}\sigma _{ii} \right) +\exp \left(- 2\mu _i+\sigma _{ii} \right) \left[ 2\exp \left( \sigma _{ii} \right) -1 \right] $;
\item $\sigma_{\scrS R, ij} =\exp \left[ -\mu _i - \mu _j+\frac{1}{2}\left( \sigma _{ii}+\sigma _{jj} \right) \right] \left[ \exp \left( \sigma _{ij} \right) -1 \right] $;
\item $corr\left( R_i,R_j \right) =\frac{\exp \left( \sigma _{ij} \right) -1}{\sqrt{\left\{ 2\exp \left( \sigma _{ii} \right) -1+\exp \left( \mu _i-\frac{1}{2}\sigma _{ii} \right) \right\} \left\{ 2\exp \left( \sigma _{jj} \right) -1+\exp \left( \mu _j-\frac{1}{2}\sigma _{jj} \right) \right\}}} $.
\end{enumerate}
\end{myprop}

Under the definition of the innovation process, we can define an first-order multivariate integer-valued autoregressive process with multivariate geometric-logitnormal processes denoted by MINAR(1)-GL. Based on Propositions \ref{genprop1} and \ref{glprop1}, we can derive the moment properties of the proposed model.
\begin{myprop}
\label{glprop2}
For an multivariate integer autoregressive model defined by \eqref{geneq1} and \eqref{gleq2}, we have:
\begin{enumerate}[label=(\roman*)]
\item $\mu _{\scrS   X,i}=\frac{\exp \left( -\mu _i+\frac{1}{2}\sigma _{ii} \right)}{1-\alpha _i} $;
\item $\sigma _{\scrS   X,ii}=\frac{\alpha _i\exp \left( -\mu _i+\frac{1}{2}\sigma _{ii} \right) +\left\{ \exp \left( -\mu _i+\frac{1}{2}\sigma _{ii} \right) +\exp \left( -2\mu _i+\sigma _{ii} \right) \left[ 2\exp \left( \sigma _{ii} \right) -1 \right] \right\}}{1-\alpha _{i}^{2}} $;
\item $\sigma _{\scrS   X,ij}=\frac{\exp \left( \mu _i+\mu _j+\frac{1}{2}\left( \sigma _{ii}+\sigma _{jj} \right) \right) \left\{ \exp \left( \sigma _{ij} \right) -1 \right\}}{1-\alpha _i\alpha _j}$;
\item $Cov\left( X_{i,t},X_{i,t+h} \right) =\alpha _{i}^{h}\sigma _{\scrS   X,i}, \quad h=0,1,2,\dots $;
\item $Cov\left( X_{i,t+h},X_{j,t} \right) =\frac{\alpha _{i}^{h}}{1-\alpha _i\alpha _j}\sigma _{\scrS   R,ij};h=0,1,2,\dots ,i\ne j$.
\end{enumerate}
\end{myprop}

\begin{myprop}
\label{glprop3}
Let $\boldsymbol{X}_t$ be the process defined by Equation \eqref{geneq1} and \eqref{gleq2}, and assume that Hypotheses \ref{genhypo1} and \ref{glhypo1} hold. Then, for any fixed positive integer $k$, $E(\boldsymbol{X}_t^k )<\infty$.
\end{myprop}

\section{Estimation}

The conditional probability density function of the MINAR(1) model can be expressed as a combination of $N$ binomial probability density functions:
\begin{equation}
\label{estieq1}
f_{s}\left( k_s|X_{s,t-1} \right) =\left( \begin{array}{c}
	X_{s,t-1}\\
	k_{s}\\
\end{array} \right) \alpha _{s}^{k_{s}}\left( 1-\alpha _s \right) ^{X_{d, t-1} - k_{j}}, s=1,2,\dots , N , 
\end{equation}
and the distribution function of the innovation process $g\left( r_1, r_2,\dots , r_{\scrS N} \right) = P\left( R_1=r_1, R_2=r_2,\dots ,  R_{\scrS N}=r_{\scrS N} \right)$ can be represented in a convolution form. Therefore, the conditional density function of the model \eqref{geneq1} can be written as
\begin{equation}
\label{estieq2}
f\left( \boldsymbol{X}_t|\boldsymbol{X}_{t-1} , \boldsymbol{\theta}\right) =\sum_{k_1=0}^{m_1}{\cdots \sum_{k_{\scrS N}=0}^{m_{\scrS N}}{f_1\left( k_1|X_{1,t-1} \right) \cdots f_{\scrS N}\left( k_{\scrS N}|X_{N,t-1} \right) g\left( X_{1,t}-k_1, \dots , X_{N,t}-k_{\scrS N} \right)}},
\end{equation}
where $m_s=\min \left( X_{s,t},  X_{s,t-1} \right) $, $s=1,2,\dots,N$ and $\boldsymbol{\theta}$ is the vector of unknown parameters vector. The density function $g\left( k_1,k_2,\dots ,k_{\scrS N} \right)$ depends on the model specification and different model specifications have different expressions. The likelihood function of the model \eqref{geneq1} can be obtained as follows:
\begin{equation}
\label{estieq3}
L\left( \boldsymbol{\theta }|\boldsymbol{X} \right) =\prod_{t=2}^{T}{f\left( \boldsymbol{X}_t|\boldsymbol{X}_{t-1},\boldsymbol{\theta } \right)} .
\end{equation}
Then, the corresponding parameters can be estimated using maximum likelihood estimation. However, as the dimension of the model increases, the likelihood function \eqref{estieq3} involves multiple summations and integration, making the computation of maximum likelihood estimation extremely complex and more time-consuming. To simplify the computation, we propose to use the EM algorithm (\citet{dempster1977}) for parameter estimation in the model.

In the proposed model, the unknown parameter vector is denoted as $\boldsymbol{\theta}$.We assume that the complete data is given by 
$$
\left\{ \boldsymbol{X}_{t}, \boldsymbol{Z}_{t}, \boldsymbol{\eta}_{t} \right\} _{t=1}^{T},
$$
where $\boldsymbol{Z}_t\in \left\{ \left( Z_{1,t},Z_{2,t},\dots ,Z_{N,t} \right) :Z_{s,t}=0,1,2,\dots ,\min \left( X_{s,t},X_{s,t-1} \right) ,s=1,2,\dots ,N \right\}$, and $\boldsymbol{\eta}_{t}=\left( \eta_{1}, \eta_{2}, \dots, \eta_{\scrS N} \right)^{\prime}$ denote $\boldsymbol{\lambda}_{t}$ or $\boldsymbol{p}_{t}$.
Based on equations \eqref{geneq1}, \eqref{pleq2}, and \eqref{gleq2}, we can obtain the log-likelihood function for the complete sample at a given time $t$, which is defined as
\begin{equation}
\label{estieq4}
\begin{aligned}
	\mathcal{L} (\boldsymbol{\theta}|\boldsymbol{X}_{t}, \boldsymbol{X}_{t-1})=&	\log \left( f\left( \boldsymbol{X}_t,\boldsymbol{Z}_t,\boldsymbol{\eta }_t|\boldsymbol{X}_{t-1},\boldsymbol{\theta} \right) \right) \\
	=&\log \left( f\left( \boldsymbol{X}_t|\boldsymbol{X}_{t-1},\boldsymbol{Z}_t,\boldsymbol{\eta }_t,\boldsymbol{\theta} \right) \right) +\log \left( f(\boldsymbol{Z}_t|\boldsymbol{\eta }_t,\boldsymbol{X}_t,\boldsymbol{X}_{t-1}),\boldsymbol{\theta} \right)  +\log \left( f(\boldsymbol{\eta }_t|\boldsymbol{X}_{t-1}, \boldsymbol{\theta}) \right).
\end{aligned}
\end{equation}

Therefore, in the $E$ step of the EM algorithm we calculate the expectation of the log-likelihood function. Given the estimation $\boldsymbol{\theta}^{(k-1)}$ from the $(k-1)$-th iteration, the $Q$ function in the $k$-th iteration is defined as
\begin{equation}
\label{estieq5}
\begin{aligned}
	Q_t(\boldsymbol{\theta }|\boldsymbol{\theta }^{\left( k-1 \right)})=&E_{\left( \boldsymbol{Z}_t,\boldsymbol{\eta }_t \right) |\boldsymbol{X}_t,\boldsymbol{X}_{t-1},\boldsymbol{\theta }^{\left( k-1 \right)}}\left( \log \left( f\left( \boldsymbol{X}_t,\boldsymbol{Z}_t,  \boldsymbol{\eta }_t|\boldsymbol{X}_{t-1},\boldsymbol{\theta } \right) \right) \right)
	\\
	=&E_{\left( \boldsymbol{Z}_t,  \boldsymbol{\eta }_t \right) |\boldsymbol{X}_t,\boldsymbol{X}_{t-1},\boldsymbol{\theta }^{\left( k-1 \right)}}\left( \log \left( f\left( \boldsymbol{X}_t|\boldsymbol{X}_{t-1},\boldsymbol{Z}_t,\boldsymbol{\eta }_t,\boldsymbol{\theta } \right) \right) \right)
	\\
	&+E_{\left( \boldsymbol{Z}_t,\boldsymbol{\eta }_t \right) |\boldsymbol{X}_t,\boldsymbol{X}_{t-1},\boldsymbol{\theta }^{\left( k-1 \right)}}\left( \log \left( f(\boldsymbol{Z}_t| \boldsymbol{\eta }_t,\boldsymbol{X}_t,\boldsymbol{X}_{t-1},,\boldsymbol{\theta }) \right) \right)
	\\
	&+E_{\left( \boldsymbol{Z}_t,\boldsymbol{\eta }_t \right) |\boldsymbol{X}_t,\boldsymbol{X}_{t-1},\boldsymbol{\theta }^{\left( k-1 \right)}}\left( \log \left( f(\boldsymbol{\eta }_t|\boldsymbol{X}_{t-1},,\boldsymbol{\theta }) \right) \right)
	\\
	=&\sum_{t=2}^{T}{\sum_{\boldsymbol{Z}_t}{\int_{\boldsymbol{\eta }_t}{\log \left( f\left( \boldsymbol{X}_t|\boldsymbol{X}_{t-1},\boldsymbol{Z}_t,\boldsymbol{\eta }_t,\boldsymbol{\theta } \right) \right) *f_{post}\left( \boldsymbol{Z}_t,\boldsymbol{\eta }_t|\boldsymbol{X}_t,\boldsymbol{X}_{t-1},\boldsymbol{\theta }^{(k-1)} \right)}d\boldsymbol{\eta }_t}}
	\\
	&+\sum_{t=2}^{T}{\sum_{\boldsymbol{Z}_t}{\int_{\boldsymbol{\eta }_t}{\log \left( f(\boldsymbol{Z}_t|\boldsymbol{\eta }_t,\boldsymbol{X}_t,\boldsymbol{X}_{t-1},,\boldsymbol{\theta }) \right) *f_{post}\left( \boldsymbol{Z}_t,\boldsymbol{\eta }_t|\boldsymbol{X}_t,\boldsymbol{X}_{t-1},\boldsymbol{\theta }^{(k-1)} \right)}d\boldsymbol{\eta }_t}}
	\\
	&+\sum_{t=2}^{T}{\sum_{\boldsymbol{Z}_t}{\int_{\boldsymbol{\eta }_t}{\log \left( f(\boldsymbol{\eta }_t|\boldsymbol{X}_{t-1},,\boldsymbol{\theta }) \right) *f_{post}\left( \boldsymbol{Z}_t,\boldsymbol{\eta }_t|\boldsymbol{X}_t,\boldsymbol{X}_{t-1},\boldsymbol{\theta }^{(k-1)} \right)}d\boldsymbol{\eta }_t}},
\end{aligned}
\end{equation}
where $f_{post}\left( \boldsymbol{Z}_t,\boldsymbol{\eta }_t|\boldsymbol{X}_t,\boldsymbol{X}_{t-1},\boldsymbol{\theta }^{\left( k-1 \right)} \right)$ is the posterior probability density function, which can be calculated using the following formula:
\begin{footnotesize}
\begin{equation}
f_{post}\left( \boldsymbol{Z}_t,\boldsymbol{\eta }_t|\boldsymbol{X}_t,\boldsymbol{X}_{t-1},\boldsymbol{\theta }^{\left( k-1 \right)} \right) =\frac{f\left( \boldsymbol{X}_t|\boldsymbol{X}_{t-1},\boldsymbol{Z}_t,\boldsymbol{\eta }_t,\boldsymbol{\theta }^{\left( k-1 \right)} \right) f\left( \boldsymbol{Z}_t|\boldsymbol{\eta }_t,\boldsymbol{X}_t,\boldsymbol{X}_{t-1},\boldsymbol{\theta }^{\left( k-1 \right)} \right) f\left( \boldsymbol{\eta }_t|\boldsymbol{X}_{t-1},\boldsymbol{\theta }^{\left( k-1 \right)} \right)}{\sum_{\boldsymbol{Z}_t}{\int_{\boldsymbol{\eta }_t}{f\left( \boldsymbol{X}_t|\boldsymbol{X}_{t-1},\boldsymbol{Z}_t,\boldsymbol{\eta }_t,\boldsymbol{\theta }^{\left( k-1 \right)} \right) f\left( \boldsymbol{Z}_t|\boldsymbol{\eta }_t,\boldsymbol{X}_t,\boldsymbol{X}_{t-1},\boldsymbol{\theta }^{\left( k-1 \right)} \right) f\left( \boldsymbol{\eta }_t|\boldsymbol{X}_{t-1},\boldsymbol{\theta }^{\left( k-1 \right)} \right) d\boldsymbol{\eta }_t}}}.
\label{estieq6}
\end{equation}
\end{footnotesize}

The $Q$ function for the observed data $\left\{ \boldsymbol{X}_{t}, \boldsymbol{Z}_{t}, \boldsymbol{\eta}_{t} \right\} _{t=1}^{T}$ can be expressed as 
\begin{equation}
\label{estieq7}
Q(\boldsymbol{\theta }|\boldsymbol{\theta }^{\left( k-1 \right)}) = \sum_{t=2}^{n_{t}}{Q_{t}(\boldsymbol{\theta }|\boldsymbol{\theta }^{\left( k-1 \right)})}.
\end{equation}
Then, in the $M$ step of the EM algorithm, we maximize the $Q$ function \eqref{estieq6} with respect to the parameter $\boldsymbol{\theta}$ to obtain a new parameter estimate $\boldsymbol{\theta }^{(k)}$:
\begin{equation}
\label{estieq8}
\boldsymbol{\theta }^{\left( k \right)}=\underset{\boldsymbol{\theta }}{argmax} \,\, Q( \boldsymbol{\theta }|\boldsymbol{\theta }^{\left( k-1 \right)} ).
\end{equation}
Subsequently, we repeat the $E$-step and $M$-step until certain termination conditions are met to stop the iterations. The parameter estimation expressions under two different assumptions will be introduced separately below.

\subsection{The Poisson-lognormal innovation process}
According to equations \eqref{pleq2} and \eqref{estieq2}, the logarithm of the likelihood function is respectively given by:
\begin{subequations}
\label{estieq9}
\begin{gather}
\log \left( f\left( \boldsymbol{X}_t|\boldsymbol{X}_{t-1},\boldsymbol{Z}_t,\boldsymbol{\lambda }_t,\boldsymbol{\theta } \right) \right)=\sum_{s=1}^{N}{\log \left( \frac{\exp(\lambda_{s,t})}{\left( X_{s,t}-Z_{s,t} \right) !} \exp(-\exp(\lambda _{s,t})) \right)}, \label{23a}
\\
\log \left( f(\boldsymbol{Z}_t|\boldsymbol{\lambda }_t,\boldsymbol{X}_t,\boldsymbol{X}_{t-1},\boldsymbol{\theta }) \right) =\sum_{s=1}^N{\log \left( C_{\scrS   X_{s,t-1}}^{Z_{s,t}}\alpha _{s}^{Z_{s,t}}\left( 1-\alpha _s \right) ^{X_{s,t-1}-Z_{s,t}} \right)},  \label{23b} 
\\
\log \left( f(\boldsymbol{\lambda }_t|\boldsymbol{X}_{t-1},,\boldsymbol{\theta }) \right) =\frac{1}{\sqrt{\left( 2\pi \right) ^{N}\left| \boldsymbol{\Sigma } \right|}}\exp \left( -\frac{\left( \boldsymbol{\lambda }_t-\boldsymbol{\mu } \right) ^{\prime}\boldsymbol{\Sigma }^{-1}\left(  \boldsymbol{\lambda }_t-\boldsymbol{\mu } \right)}{2} \right),  \label{23c} 
\end{gather}
\end{subequations}
and the calculation formula for the posterior probability density function is
\begin{tiny}
\begin{equation}
\begin{aligned}
&f_{post}\left( \boldsymbol{Z}_t,\boldsymbol{\lambda }_t|\boldsymbol{X}_t,\boldsymbol{X}_{t-1},\boldsymbol{\theta }^{\left( k-1 \right)} \right)
\\
&=\frac{\prod_{s=1}^N{\frac{\exp\mathrm{(}\lambda _{s,t})}{\left( X_{s,t}-Z_{s,t} \right) !}\exp\mathrm{(}-\exp\mathrm{(}\lambda _{s,t}))}\cdot \prod_{s=1}^N{C_{\scrS   X_{s,t-1}}^{Z_{s,t}}\alpha _{s}^{Z_{s,t}}\left( 1-\alpha _s \right) ^{X_{s,t-1}-Z_{s,t}}}  \cdot   \frac{1}{\sqrt{\left( 2\pi \right) ^N\left| \mathbf{\Sigma } \right|}}\exp \left( -\frac{1}{2}\left( \boldsymbol{\lambda }_t-\boldsymbol{\mu } \right) ^{\prime}\mathbf{\Sigma }^{-1}\left( \boldsymbol{\lambda }_t-\boldsymbol{\mu } \right) \right)}{\sum_{\boldsymbol{Z}_t}{\int_{\boldsymbol{\lambda }_t}{\prod_{s=1}^N{\frac{\exp\mathrm{(}\lambda _{s,t})}{\left( X_{s,t}-Z_{s,t} \right) !}\exp\mathrm{(}-\exp\mathrm{(}\lambda _{s,t}))}\cdot \prod_{s=1}^N{C_{\scrS   X_{s,t-1}}^{Z_{s,t}}\alpha _{s}^{Z_{s,t}}\left( 1-\alpha _s \right) ^{X_{s,t-1}-Z_{s,t}}} \cdot \frac{1}{\sqrt{\left( 2\pi \right) ^N\left| \mathbf{\Sigma } \right|}}\exp \left( -\frac{1}{2}\left( \boldsymbol{\lambda }_t-\boldsymbol{\mu } \right) ^{\prime}\mathbf{\Sigma }^{-1}\left( \boldsymbol{\lambda }_t-\boldsymbol{\mu } \right) \right) d\boldsymbol{\lambda }_t}}},
\end{aligned}
\label{estieq10}
\end{equation}
\end{tiny}
where the integral can be calculated using the Gauss-Hermite method. In the E-step, the expected value of the log-likelihood can be obtained based on equations \eqref{estieq9} and \eqref{estieq10}, using the observed multivariate count data and existing parameter estimations.

In the M-step, due to the special form of the log-likelihood function and the complete data setting, we can obtain explicit solutions for parameter estimation under the multivariate Poisson-lognormal setting without the need for numerical algorithms like Newton-Raphson. By taking the derivatives of the $Q$ function with respect to parameters $\boldsymbol{\alpha}$, $\boldsymbol{\mu}$, and $\boldsymbol{\Sigma}$, and after some algebraic manipulation, we have
\begin{equation}
\label{estieq11}
\begin{aligned}
&\frac{\partial Q\left( \boldsymbol{\theta }|\boldsymbol{\theta }^{\left( k-1 \right)} \right)}{\partial \alpha _s}=\sum_{t=2}^T{\sum_{\boldsymbol{Z}_t}{\int_{\boldsymbol{\lambda }_t}{\frac{Z_{s,t}-\alpha _sX_{s,t-1}}{\alpha _s\left( 1-\alpha _s \right)}*f_{post}\left( \boldsymbol{Z}_t,\boldsymbol{\lambda }_t|\boldsymbol{X}_t,\boldsymbol{X}_{t-1},\boldsymbol{\theta }^{(k-1)} \right)}d\boldsymbol{\lambda }_t}},
\\
&\qquad \qquad \qquad s=1,2,\dots ,N,
\\
&\frac{\partial Q\left( \boldsymbol{\theta }|\boldsymbol{\theta }^{\left( k-1 \right)} \right)}{\partial \boldsymbol{\mu }}=\sum_{t=2}^T{\sum_{\boldsymbol{Z}_t}{\int_{\boldsymbol{\lambda }_t}{\left\{ \mathbf{\Sigma }^{-1}\left( \boldsymbol{\lambda }_t-\boldsymbol{\mu } \right) \right\} *f_{post}\left( \boldsymbol{Z}_t,\boldsymbol{\lambda }_t|\boldsymbol{X}_t,\boldsymbol{X}_{t-1},\boldsymbol{\theta }^{(k-1)} \right) d\boldsymbol{\lambda }_t}}},
\\
&\frac{\partial Q\left( \boldsymbol{\theta }|\boldsymbol{\theta }^{\left( k-1 \right)} \right)}{\partial \mathbf{\Sigma }}=\sum_{t=2}^T{\sum_{\boldsymbol{Z}_t}{\int_{\boldsymbol{\lambda }_t}{\left\{ \mathbf{\Sigma }^{-1}-\mathbf{\Sigma }^{-1}\left( \boldsymbol{\lambda }_t-\boldsymbol{\mu } \right) \left( \boldsymbol{\lambda }_t-\boldsymbol{\mu } \right) ^{\prime}\mathbf{\Sigma }^{-1} \right\} *f_{post}\left( \boldsymbol{Z}_t,\boldsymbol{\lambda }_t|\boldsymbol{X}_t,\boldsymbol{X}_{t-1},\boldsymbol{\theta }^{(k-1)} \right)}d\boldsymbol{\lambda }_t}}.
\end{aligned}
\end{equation}

By setting the equation in Eq. \eqref{estieq11} to be equal to 0, we can obtain
\begin{equation}
\label{estieq12}
\begin{aligned}
&\alpha _{s}^{\left( k \right)}=\frac{\sum_{t=2}^{T}{\sum_{\boldsymbol{Z}_t}{\int_{\boldsymbol{\lambda }_t}{Z_{s,t}*f_{post}\left( \boldsymbol{Z}_t,\boldsymbol{\lambda }_t|\boldsymbol{X}_t,\boldsymbol{X}_{t-1},\boldsymbol{\theta }^{(k-1)} \right)}d\boldsymbol{\lambda }_t}}}{\sum_{t=2}^{T}{X_{s,t-1}}},
\\
&\qquad \qquad \qquad s=1,2,\dots ,N,
\\
&\boldsymbol{\mu }^{\left( k \right)}=\frac{1}{T-1}\sum_{t=2}^{T}{\sum_{\boldsymbol{Z}_t}{\int_{\boldsymbol{\lambda }_t}{\boldsymbol{\lambda }_t*f_{post}\left( \boldsymbol{Z}_t,\boldsymbol{\lambda }_t|\boldsymbol{X}_t,\boldsymbol{X}_{t-1},\boldsymbol{\theta }^{(k-1)} \right) d\boldsymbol{\lambda }_t}}},
\\
&\boldsymbol{\Sigma }^{\left( k \right)}=\frac{1}{T-1}\sum_{t=2}^{T}{\sum_{\boldsymbol{Z}_t}{\int_{\boldsymbol{\lambda }_t}{\left\{ \left( \boldsymbol{\lambda }_t-\boldsymbol{\mu } \right) \left( \boldsymbol{\lambda }_t-\boldsymbol{\mu } \right) ^{\prime} \right\} *f_{post}\left( \boldsymbol{Z}_t,\boldsymbol{\lambda }_t|\boldsymbol{X}_t,\boldsymbol{X}_{t-1},\boldsymbol{\theta }^{(k-1)} \right)}d\boldsymbol{\lambda }_t}}.
\end{aligned}
\end{equation}

Repeat the E-step and M-step until certain convergence criteria are satisfied.

\subsection{The geometric-logitnormal innovation process}

According to equations \eqref{pleq2} and \eqref{estieq2}, the log-likelihood functions are given by
\begin{subequations}
\label{estieq13}
\begin{align}
&	
	\begin{aligned}
	\log \left( f\left( \boldsymbol{X}_t|\boldsymbol{X}_{t-1},\boldsymbol{Z}_t,\boldsymbol{p}_t,\boldsymbol{\theta } \right) \right) &=\sum_{s=1}^N{\log \left( \left( 1-\frac{\exp\mathrm{(}p_{s,t})}{1+\exp\mathrm{(}p_{s,t})} \right) ^{X_{s,t}-Z_{s,t}}\cdot \frac{\exp\mathrm{(}p_{s,t})}{1+\exp\mathrm{(}p_{s,t})} \right)}
	\\
	&=\sum_{s=1}^{N}{\log \left( \frac{\exp\mathrm{(}p _{s,t})}{\left( 1+\exp( p_{s,t}) \right) ^{X_{s,t}-Z_{s,t}+1}} \right)},
	\end{aligned} \label{27a}
\\
&
	\log \left( f(\boldsymbol{Z}_t|\boldsymbol{p }_t,\boldsymbol{X}_t,\boldsymbol{X}_{t-1},\boldsymbol{\theta }) \right) =\sum_{s=1}^N{\log \left( C_{\scrS   X_{s,t-1}}^{Z_{s,t}}\alpha _{s}^{Z_{s,t}}\left( 1-\alpha _s \right) ^{X_{s,t-1}-Z_{s,t}} \right)}, \label{27b} 
\\
&	
	\log \left( f(\boldsymbol{p }_t|\boldsymbol{X}_{t-1}, \boldsymbol{\theta } ) \right) =\frac{1}{  \sqrt{\left( 2\pi \right)^N \left| \boldsymbol{\Sigma} \right| }   }\exp \left( -\frac{\left( \boldsymbol{p}_t-\boldsymbol{\mu } \right) ^{\prime}  \boldsymbol{\Sigma }^{-1}  \left( \boldsymbol{p}_t-\boldsymbol{\mu } \right)}{2} \right) ,  \label{27c} 
\end{align}
\end{subequations}
and the formula for calculating the posterior probability density function is given by
\begin{tiny}
\begin{equation}
\label{estieq14}
\begin{aligned}
	&f_{post}\left( \boldsymbol{Z}_t,\boldsymbol{p}_t|\boldsymbol{X}_t,\boldsymbol{X}_{t-1},\boldsymbol{\theta }^{\left( k-1 \right)} \right)\\
	&=\frac{\prod_{s=1}^N{\frac{\exp\mathrm{(}p_{s,t})}{\left( 1+\exp\mathrm{(}p_{s,t}) \right) ^{X_{s,t}-Z_{s,t}+1}}}\cdot \prod_{s=1}^N{C_{\scrS   X_{s,t-1}}^{Z_{s,t}}\alpha _{s}^{Z_{s,t}}\left( 1-\alpha _s \right) ^{X_{s,t-1}-Z_{s,t}}}\cdot \frac{1}{\sqrt{\left( 2\pi \right) ^N\left| \mathbf{\Sigma } \right|}}\exp \left( -\frac{1}{2}\left( \boldsymbol{p}_t-\boldsymbol{\mu } \right) ^{\prime}\mathbf{\Sigma }^{-1}\left( \boldsymbol{p}_t-\boldsymbol{\mu } \right) \right)}{\sum_{\boldsymbol{Z}_t}{\int_{\boldsymbol{p}_t}{\prod_{s=1}^N{\frac{\exp\mathrm{(}p_{s,t})}{\left( 1+\exp\mathrm{(}p_{s,t}) \right) ^{X_{s,t}-Z_{s,t}+1}}}\cdot \prod_{s=1}^N{C_{\scrS   X_{s,t-1}}^{Z_{s,t}}\alpha _{s}^{Z_{s,t}}\left( 1-\alpha _s \right) ^{X_{s,t-1}-Z_{s,t}}}\cdot \frac{1}{\sqrt{\left( 2\pi \right) ^N\left| \mathbf{\Sigma } \right|}}\exp \left( -\frac{1}{2}\left( \boldsymbol{p}_t-\boldsymbol{\mu } \right) ^{\prime}\mathbf{\Sigma }^{-1}\left( \boldsymbol{p}_t-\boldsymbol{\mu } \right) \right) d\boldsymbol{p}_t}}}.
\end{aligned}
\end{equation}
\end{tiny}

In the E-step, the expectation of the log-likelihood can be computed based on equations \eqref{estieq13} and \eqref{estieq14}, using the observed integer-valued data and the current parameter estimation.
In the M-step, due to the specific form of the log-likelihood function and the setting of complete data, we can obtain explicit solutions for parameter estimation under the geometric-logitnormal setting without numerical algorithms like Newton-Raphson. By taking derivatives of the $Q$ function with respect to the parameters $\boldsymbol{\alpha}$, $\boldsymbol{\mu}$, $\boldsymbol{\Sigma}$ and setting them into zero, we have
\begin{equation}
\label{estieq15}
\begin{aligned}
&\frac{\partial Q( \boldsymbol{\theta }|\boldsymbol{\theta }^{\left( k-1 \right)} )}{\partial \alpha _s}=\sum_{t=2}^T{\sum_{\boldsymbol{Z}_t}{\int_{\boldsymbol{p }_t}{\frac{Z_{s,t}-\alpha _sX_{s,t-1}}{\alpha _s\left( 1-\alpha _s \right)}*f_{post}   ( \boldsymbol{Z}_t,\boldsymbol{p }_t|\boldsymbol{X}_t,\boldsymbol{X}_{t-1},\boldsymbol{\theta }^{(k-1)}  )}d\boldsymbol{p }_t}},
\\
&\qquad \qquad \qquad s=1,2,\dots ,  N,
\\
&\frac{\partial Q( \boldsymbol{\theta }|\boldsymbol{\theta }^{\left( k-1 \right)} )}{\partial \boldsymbol{\mu }}=\sum_{t=2}^T{\sum_{\boldsymbol{Z}_t}{\int_{\boldsymbol{p }_t}{\left\{ \mathbf{\Sigma }^{-1}\left( \boldsymbol{p }_t-\boldsymbol{\mu } \right) \right\} *f_{post}  ( \boldsymbol{Z}_t,\boldsymbol{p }_t|\boldsymbol{X}_t,\boldsymbol{X}_{t-1},\boldsymbol{\theta }^{(k-1)}   ) d\boldsymbol{p }_t}}},
\\
&\frac{\partial Q( \boldsymbol{\theta }|\boldsymbol{\theta }^{\left( k-1 \right)} )}{\partial \mathbf{\Sigma }}=\sum_{t=2}^{\prime}{\sum_{\boldsymbol{Z}_t}{\int_{\boldsymbol{p }_t}{\left\{ \mathbf{\Sigma }^{-1}-\mathbf{\Sigma }^{-1}\left( \boldsymbol{p }_t-\boldsymbol{\mu } \right)   ( \boldsymbol{p }_t-\boldsymbol{\mu }  ) ^{\prime}\mathbf{\Sigma }^{-1} \right\} *f_{post}  ( \boldsymbol{Z}_t,\boldsymbol{p }_t|\boldsymbol{X}_t,\boldsymbol{X}_{t-1},\boldsymbol{\theta }^{(k-1)}  )}d\boldsymbol{p }_t}}.
\end{aligned}
\end{equation}
Subsequently, we can calculate
\begin{equation}
\label{estieq16}
\begin{aligned}
&\alpha _{s}^{\left( k \right)}=\frac{\sum_{t=2}^{T}{\sum_{\boldsymbol{Z}_t}{\int_{\boldsymbol{p }_t}{Z_{s,t}*f_{post}   ( \boldsymbol{Z}_t,\boldsymbol{p }_t|\boldsymbol{X}_t,\boldsymbol{X}_{t-1},\boldsymbol{\theta }^{(k-1)}  )}d\boldsymbol{p }_t}}}{\sum_{t=2}^{T}{X_{s,t-1}}},
\\
&\qquad \qquad \qquad s=1,2,\dots ,  N,
\\
&\boldsymbol{\mu }^{\left( k \right)}=\frac{1}{T-1}\sum_{t=2}^{T}{\sum_{\boldsymbol{Z}_t}{\int_{\boldsymbol{p }_t}{\boldsymbol{p }_t*f_{post}   ( \boldsymbol{Z}_t,\boldsymbol{p }_t|\boldsymbol{X}_t,\boldsymbol{X}_{t-1},\boldsymbol{\theta }^{(k-1)}  ) d\boldsymbol{p }_t}}} ,  
\\
&\boldsymbol{\Sigma }^{\left( k \right)}=\frac{1}{T-1}\sum_{t=2}^{T}{\sum_{\boldsymbol{Z}_t}{\int_{\boldsymbol{p }_t}{\left\{ \left( \boldsymbol{p }_t-\boldsymbol{\mu } \right)  ( \boldsymbol{p }_t-\boldsymbol{\mu }  ) ^{\prime} \right\} *f_{post}  ( \boldsymbol{Z}_t,\boldsymbol{p }_t|\boldsymbol{X}_t,\boldsymbol{X}_{t-1},\boldsymbol{\theta }^{(k-1)}  )}     d\boldsymbol{p }_t}}.
\end{aligned}
\end{equation}

Repeat the E-step and M-step until certain convergence are met.

\section{Simulation}
This section aims to evaluate the finite sample performance of the proposed models under the 3-dimensional setting through a series of simulation experiments. Firstly, we set the true values of parameters as follows. For parameter $\boldsymbol{\alpha}$, we consider three scenarios: (A1) $\boldsymbol{\alpha}^{\prime}=\left(0.1, 0.3, 0.5\right)$, (A2) $\boldsymbol{\alpha}^{\prime}=\left(0.3, 0.3, 0.3\right)$, (A3) $\boldsymbol{\alpha}^{\prime}=\left(0.5, 0.5, 0.5\right)$. For parameter $\boldsymbol{\mu}$, we have two scenarios: (B1) $\boldsymbol{\mu}^{\prime}=\left(0.5, 0.5, 0.5\right)$, (B2) $\boldsymbol{\mu}^{\prime}=\left(1, 1, 1\right)$. As for parameter $\boldsymbol{\Sigma}$, we consider two cases: (C1) $\text{vec}\left(\mathbf{\Sigma}\right)^{\prime}=(0.64, 0.5709447, 0.5344570, 0.5709447, \allowbreak 0.64, 0.5919781, 0.5344570, 0.5919781, 0.64)$, (C2) $\text{vec}\left(\mathbf{\Sigma}\right)^{\prime}=  ( 0.640,  0.320, -0.192,  0.320, 0.640,  0.192, -0.192, \allowbreak  0.192,  0.640 )$. The sample size is set to be 50, 100, 300, respectively. The simulation under each setting repeats 300 times.

\subsection{The Multivariate Poisson-lognormal innovation process}
The simulation results can be found in Table \ref{tab1} and Table \ref{tab2}, which provide the empirical biases and standard deviations (SDs) of the parameters. The results in the tables show that the EM algorithm performs well, with estimated means close to the true values used to generate the data. In particular, the parameters of multivariate Poisson-lognormal innovation process yield good results across various parameter scenarios, even when the sample size is 50, with relatively small estimation biases and standard deviations. As the sample size increases, the standard deviations gradually decrease. In conclusion, the performance of the parameter estimation algorithm is good for all parameter scenarios, and the estimation biases and standard deviations decrease gradually with the increase in sample size.

\begin{table}[!htbp]
\caption{In scenario (B1), the estimated biases and standard deviations (in parentheses) of parameters $\boldsymbol{\theta}$ are presented. (multivariate Poisson-lognormal distribution)}
\centering
\label{tab1}
\begin{tiny}
\begin{tabular}{l|rrr|rrr}
\hline
\multicolumn{1}{c|}{}                      & \multicolumn{3}{c|}{(C1)}                                                                                      & \multicolumn{3}{c}{(C2)}                                                                                      \\ \cline{2-7} 
\multicolumn{1}{c|}{$\boldsymbol{\theta}$} & \multicolumn{1}{c|}{$n_{t}=50$}      & \multicolumn{1}{c|}{$n_{t}=100$}     & \multicolumn{1}{c|}{$n_{t}=300$} & \multicolumn{1}{c|}{$n_{t}=50$}      & \multicolumn{1}{c|}{$n_{t}=100$}     & \multicolumn{1}{c}{$n_{t}=300$} \\ \hline
\multicolumn{1}{c|}{(A1)(B1)}              & \multicolumn{1}{r|}{}                & \multicolumn{1}{r|}{}                &                                  & \multicolumn{1}{r|}{}                & \multicolumn{1}{r|}{}                &                                 \\
$\hat{\alpha}_1$                           & \multicolumn{1}{r|}{-0.0014(0.0512)} & \multicolumn{1}{r|}{0.0057(0.0385)}  & 0.0013(0.0212)                   & \multicolumn{1}{r|}{0.0075(0.0604)}  & \multicolumn{1}{r|}{-0.0005(0.0410)} & -0.0002(0.0213)                 \\
$\hat{\alpha}_2$                           & \multicolumn{1}{r|}{0.0039(0.0657)}   & \multicolumn{1}{r|}{0.0028(0.0482)}  & 0.0019(0.0246)                   & \multicolumn{1}{r|}{-0.0036(0.0691)} & \multicolumn{1}{r|}{-0.0039(0.0553)} & 0.0027(0.0258)                  \\
$\hat{\alpha}_3$                           & \multicolumn{1}{r|}{-0.0094(0.0605)} & \multicolumn{1}{r|}{-0.0002(0.0408)} & -0.001(0.0208)                   & \multicolumn{1}{r|}{-0.0076(0.0619)} & \multicolumn{1}{r|}{-0.0047(0.0444)} & -0.0003(0.0232)                 \\
$\hat{\mu}_1$                              & \multicolumn{1}{r|}{-0.0105(0.1621)} & \multicolumn{1}{r|}{-0.0172(0.1234)} & -0.0068(0.0615)                  & \multicolumn{1}{r|}{-0.0127(0.2068)} & \multicolumn{1}{r|}{-0.0057(0.1252)} & -0.0074(0.0641)                 \\
$\hat{\mu}_2$                              & \multicolumn{1}{r|}{-0.0033(0.1683)} & \multicolumn{1}{r|}{-0.0132(0.1196)} & -0.0064(0.0601)                  & \multicolumn{1}{r|}{-0.0202(0.2058)} & \multicolumn{1}{r|}{-0.0093(0.1412)} & -0.0091(0.0737)                 \\
$\hat{\mu}_3$                              & \multicolumn{1}{r|}{-0.0124(0.1917)} & \multicolumn{1}{r|}{-0.0093(0.1274)} & -0.0028(0.0630)                  & \multicolumn{1}{r|}{-0.0165(0.2280)} & \multicolumn{1}{r|}{-0.0043(0.1429)} & -0.0058(0.0812)                 \\
$\hat{\sigma}_{11}$                        & \multicolumn{1}{r|}{0.0000(0.2522)}  & \multicolumn{1}{r|}{0.0117(0.1632)}  & 0.006(0.0913)                    & \multicolumn{1}{r|}{0.0187(0.2465)}  & \multicolumn{1}{r|}{0.0078(0.1682)}  & -0.0066(0.0884)                 \\
$\hat{\sigma}_{12}$                        & \multicolumn{1}{r|}{-0.0355(0.2158)} & \multicolumn{1}{r|}{-0.0057(0.1433)} & -0.0015(0.0824)                  & \multicolumn{1}{r|}{-0.022(0.1971)}  & \multicolumn{1}{r|}{-0.0045(0.1276)} & -0.0064(0.0725)                 \\
$\hat{\sigma}_{13}$                        & \multicolumn{1}{r|}{-0.0484(0.2086)} & \multicolumn{1}{r|}{-0.0137(0.1400)} & -0.0108(0.0795)                  & \multicolumn{1}{r|}{0.0061(0.1785)}  & \multicolumn{1}{r|}{0.0122(0.1307)}  & 0.0040(0.0627)                  \\
$\hat{\sigma}_{22}$                        & \multicolumn{1}{r|}{-0.0169(0.2246)} & \multicolumn{1}{r|}{-0.0030(0.1597)} & 0.0034(0.084)                    & \multicolumn{1}{r|}{0.0029(0.2570)}  & \multicolumn{1}{r|}{-0.0007(0.1737)} & -0.0042(0.089)                  \\
$\hat{\sigma}_{23}$                        & \multicolumn{1}{r|}{-0.0352(0.2152)} & \multicolumn{1}{r|}{-0.0106(0.1514)} & -0.0079(0.0798)                  & \multicolumn{1}{r|}{-0.019(0.2053)}  & \multicolumn{1}{r|}{-0.0138(0.1363)} & 0.0007(0.0726)                  \\
$\hat{\sigma}_{33}$                        & \multicolumn{1}{r|}{-0.0043(0.2521)} & \multicolumn{1}{r|}{0.0163(0.1708)}  & 0.0026(0.083)                    & \multicolumn{1}{r|}{0.0139(0.2729)}  & \multicolumn{1}{r|}{-0.0017(0.1815)} & 0.0100(0.0936)                  \\ \hline
\multicolumn{1}{c|}{(A2)(B1)}              & \multicolumn{1}{r|}{}                & \multicolumn{1}{r|}{}                &                                  & \multicolumn{1}{r|}{}                & \multicolumn{1}{r|}{}                &                                 \\
$\hat{\alpha}_1$                           & \multicolumn{1}{r|}{-0.0062(0.0696)} & \multicolumn{1}{r|}{0.0015(0.0495)}  & 0.0029(0.0248)                   & \multicolumn{1}{r|}{-0.0031(0.0744)} & \multicolumn{1}{r|}{-0.0013(0.0565)} & -0.0015(0.0249)                 \\
$\hat{\alpha}_2$                           & \multicolumn{1}{r|}{-0.0033(0.0636)} & \multicolumn{1}{r|}{-0.0021(0.0484)} & 0.0018(0.0227)                   & \multicolumn{1}{r|}{-0.0119(0.0731)} & \multicolumn{1}{r|}{-0.0076(0.0487)} & -0.0008(0.0260)                 \\
$\hat{\alpha}_3$                           & \multicolumn{1}{r|}{-0.0056(0.0733)} & \multicolumn{1}{r|}{-0.0011(0.0484)} & 0.0006(0.0260)                   & \multicolumn{1}{r|}{-0.0013(0.0729)} & \multicolumn{1}{r|}{-0.0049(0.052)}  & -0.002(0.0253)                  \\
$\hat{\mu}_1$                              & \multicolumn{1}{r|}{-0.0143(0.1973)} & \multicolumn{1}{r|}{-0.0034(0.1280)} & -0.0099(0.0665)                  & \multicolumn{1}{r|}{-0.0039(0.2154)} & \multicolumn{1}{r|}{-0.0048(0.1389)} & -0.0015(0.0735)                 \\
$\hat{\mu}_2$                              & \multicolumn{1}{r|}{-0.0147(0.1800)} & \multicolumn{1}{r|}{-0.0023(0.1235)} & -0.0067(0.0676)                  & \multicolumn{1}{r|}{0.0131(0.2043)}  & \multicolumn{1}{r|}{0.0108(0.1469)}  & 0.0008(0.0729)                  \\
$\hat{\mu}_3$                              & \multicolumn{1}{r|}{-0.0022(0.1949)} & \multicolumn{1}{r|}{-0.0003(0.1258)} & -0.0036(0.0662)                  & \multicolumn{1}{r|}{-0.0170(0.2229)} & \multicolumn{1}{r|}{0.0035(0.1524)}  & 0.0047(0.0735)                  \\
$\hat{\sigma}_{11}$                        & \multicolumn{1}{r|}{-0.0112(0.2338)} & \multicolumn{1}{r|}{-0.0150(0.1650)} & -0.0038(0.0889)                  & \multicolumn{1}{r|}{-0.0206(0.2492)} & \multicolumn{1}{r|}{-0.0035(0.1787)} & 0.0061(0.0918)                  \\
$\hat{\sigma}_{12}$                        & \multicolumn{1}{r|}{-0.0361(0.2056)} & \multicolumn{1}{r|}{-0.0242(0.1449)} & -0.0098(0.0839)                  & \multicolumn{1}{r|}{-0.0376(0.1921)} & \multicolumn{1}{r|}{-0.0213(0.1340)} & -0.0023(0.0753)                 \\
$\hat{\sigma}_{13}$                        & \multicolumn{1}{r|}{-0.0528(0.1985)} & \multicolumn{1}{r|}{-0.0402(0.1346)} & -0.0194(0.0802)                  & \multicolumn{1}{r|}{0.0162(0.1857)}  & \multicolumn{1}{r|}{0.009(0.1346)}   & 0.0004(0.0724)                  \\
$\hat{\sigma}_{22}$                        & \multicolumn{1}{r|}{-0.0163(0.2279)} & \multicolumn{1}{r|}{-0.0070(0.1595)} & -0.0029(0.0893)                  & \multicolumn{1}{r|}{-0.0140(0.2680)} & \multicolumn{1}{r|}{-0.0059(0.1778)} & -0.0045(0.0902)                 \\
$\hat{\sigma}_{23}$                        & \multicolumn{1}{r|}{-0.0408(0.2163)} & \multicolumn{1}{r|}{-0.0240(0.1428)} & -0.0140(0.0858)                  & \multicolumn{1}{r|}{-0.0093(0.1979)} & \multicolumn{1}{r|}{-0.0029(0.1416)} & -0.0048(0.0630)                 \\
$\hat{\sigma}_{33}$                        & \multicolumn{1}{r|}{-0.0153(0.2367)} & \multicolumn{1}{r|}{-0.0090(0.1577)} & -0.0020(0.0908)                  & \multicolumn{1}{r|}{0.0074(0.2556)}  & \multicolumn{1}{r|}{-0.0034(0.1891)} & 0.0000(0.0905)                  \\ \hline
\multicolumn{1}{c|}{(A3)(B1)}              & \multicolumn{1}{r|}{}                & \multicolumn{1}{r|}{}                &                                  & \multicolumn{1}{r|}{}                & \multicolumn{1}{r|}{}                &                                 \\
$\hat{\alpha}_1$                           & \multicolumn{1}{r|}{-0.0042(0.0664)} & \multicolumn{1}{r|}{-0.0030(0.0403)} & 0.0001(0.0228)                   & \multicolumn{1}{r|}{-0.002(0.0626)}  & \multicolumn{1}{r|}{0.0003(0.0432)}  & -0.0005(0.0261)                 \\
$\hat{\alpha}_2$                           & \multicolumn{1}{r|}{-0.0032(0.0577)} & \multicolumn{1}{r|}{-0.0016(0.0400)} & -0.0027(0.0212)                  & \multicolumn{1}{r|}{-0.0091(0.0613)} & \multicolumn{1}{r|}{-0.0062(0.0444)} & 0.0024(0.0237)                  \\
$\hat{\alpha}_3$                           & \multicolumn{1}{r|}{0.0025(0.0606)}  & \multicolumn{1}{r|}{0.0015(0.0378)}  & 0.0001(0.0213)                   & \multicolumn{1}{r|}{-0.0094(0.0670)} & \multicolumn{1}{r|}{-0.0049(0.0469)} & 0.0020(0.0231)                  \\
$\hat{\mu}_1$                              & \multicolumn{1}{r|}{-0.0118(0.2134)} & \multicolumn{1}{r|}{-0.0021(0.1437)} & -0.0009(0.0716)                  & \multicolumn{1}{r|}{-0.0083(0.2510)} & \multicolumn{1}{r|}{-0.0041(0.1599)} & -0.0035(0.0814)                 \\
$\hat{\mu}_2$                              & \multicolumn{1}{r|}{-0.0111(0.2007)} & \multicolumn{1}{r|}{-0.0002(0.1226)} & 0.0009(0.0703)                   & \multicolumn{1}{r|}{0.0171(0.2473)}  & \multicolumn{1}{r|}{0.0033(0.1570)}  & 0.0003(0.0763)                  \\
$\hat{\mu}_3$                              & \multicolumn{1}{r|}{-0.0264(0.2074)} & \multicolumn{1}{r|}{-0.0011(0.1223)} & 0.0003(0.0720)                   & \multicolumn{1}{r|}{-0.0076(0.2514)} & \multicolumn{1}{r|}{-0.0083(0.1717)} & -0.0068(0.0759)                 \\
$\hat{\sigma}_{11}$                        & \multicolumn{1}{r|}{-0.0085(0.2592)} & \multicolumn{1}{r|}{-0.0006(0.1816)} & -0.0039(0.0889)                  & \multicolumn{1}{r|}{-0.0016(0.3013)} & \multicolumn{1}{r|}{0.0119(0.2042)}  & 0.0000(0.0916)                  \\
$\hat{\sigma}_{12}$                        & \multicolumn{1}{r|}{-0.0388(0.2224)} & \multicolumn{1}{r|}{-0.0120(0.1622)} & -0.0095(0.0851)                  & \multicolumn{1}{r|}{-0.0348(0.2159)} & \multicolumn{1}{r|}{-0.0005(0.1636)} & -0.0076(0.0747)                 \\
$\hat{\sigma}_{13}$                        & \multicolumn{1}{r|}{-0.0367(0.2207)} & \multicolumn{1}{r|}{-0.0229(0.1459)} & -0.0199(0.0827)                  & \multicolumn{1}{r|}{0.0274(0.2094)}  & \multicolumn{1}{r|}{0.0151(0.1289)}  & -0.0044(0.0690)                 \\
$\hat{\sigma}_{22}$                        & \multicolumn{1}{r|}{-0.0066(0.2585)} & \multicolumn{1}{r|}{-0.0060(0.1699)} & -0.0034(0.0917)                  & \multicolumn{1}{r|}{-0.0102(0.2603)} & \multicolumn{1}{r|}{0.0223(0.1908)}  & -0.0068(0.0953)                 \\
$\hat{\sigma}_{23}$                        & \multicolumn{1}{r|}{-0.0204(0.2424)} & \multicolumn{1}{r|}{-0.0188(0.1541)} & -0.0151(0.0889)                  & \multicolumn{1}{r|}{-0.0247(0.2175)} & \multicolumn{1}{r|}{0.0069(0.1433)}  & -0.0028(0.0763)                 \\
$\hat{\sigma}_{33}$                        & \multicolumn{1}{r|}{0.0294(0.2772)}  & \multicolumn{1}{r|}{-0.0027(0.1612)} & -0.0052(0.0932)                  & \multicolumn{1}{r|}{-0.0038(0.3241)} & \multicolumn{1}{r|}{0.0153(0.2008)}  & 0.0060(0.0910)                  \\ \hline
\end{tabular}
\end{tiny}
\end{table}

\begin{table}[!htbp]
\caption{In scenario (B2), the estimated biases and standard deviations (in parentheses) of parameters $\boldsymbol{\theta}$ are presented. (multivariate Poisson-lognormal distribution)}
\centering
\label{tab2}
\begin{tiny}
\begin{tabular}{l|rrr|rrr}
\hline
\multicolumn{1}{c|}{}                      & \multicolumn{3}{c|}{(C1)}                                                                                      & \multicolumn{3}{c}{(C2)}                                                                                      \\ \cline{2-7} 
\multicolumn{1}{c|}{$\boldsymbol{\theta}$} & \multicolumn{1}{c|}{$n_{t}=50$}      & \multicolumn{1}{c|}{$n_{t}=100$}     & \multicolumn{1}{c|}{$n_{t}=300$} & \multicolumn{1}{c|}{$n_{t}=50$}      & \multicolumn{1}{c|}{$n_{t}=100$}     & \multicolumn{1}{c}{$n_{t}=300$} \\ \hline
\multicolumn{1}{c|}{(A1)(B2)}              & \multicolumn{1}{r|}{}                & \multicolumn{1}{r|}{}                &                                  & \multicolumn{1}{r|}{}                & \multicolumn{1}{r|}{}                &                                 \\
$\hat{\alpha}_1$                           & \multicolumn{1}{r|}{0.0120(0.0557)}  & \multicolumn{1}{r|}{0.0060(0.0333)}  & 0.0014(0.0153)                   & \multicolumn{1}{r|}{0.0091(0.0593)}  & \multicolumn{1}{r|}{0.0040(0.0323)}  & 0.0010(0.0142)                  \\
$\hat{\alpha}_2$                           & \multicolumn{1}{r|}{-0.0007(0.0579)} & \multicolumn{1}{r|}{0.0013(0.0404)}  & 0.0039(0.0199)                   & \multicolumn{1}{r|}{-0.0017(0.0603)} & \multicolumn{1}{r|}{-0.0037(0.0417)} & 0.0012(0.0212)                  \\
$\hat{\alpha}_3$                           & \multicolumn{1}{r|}{0.0012(0.0526)}  & \multicolumn{1}{r|}{0.0011(0.0360)}  & 0.0024(0.0176)                   & \multicolumn{1}{r|}{-0.005(0.0578)}  & \multicolumn{1}{r|}{-0.0035(0.0373)} & -0.0009(0.0183)                 \\
$\hat{\mu}_1$                              & \multicolumn{1}{r|}{-0.0411(0.1827)} & \multicolumn{1}{r|}{-0.0199(0.1047)} & -0.0117(0.0567)                  & \multicolumn{1}{r|}{-0.0344(0.1825)} & \multicolumn{1}{r|}{-0.0018(0.1155)} & -0.0025(0.0590)                 \\
$\hat{\mu}_2$                              & \multicolumn{1}{r|}{-0.0310(0.1944)} & \multicolumn{1}{r|}{-0.0239(0.1117)} & -0.0125(0.0598)                  & \multicolumn{1}{r|}{-0.0154(0.1864)} & \multicolumn{1}{r|}{0.0118(0.1237)}  & -0.0035(0.0669)                 \\
$\hat{\mu}_3$                              & \multicolumn{1}{r|}{-0.0254(0.2114)} & \multicolumn{1}{r|}{-0.0275(0.1157)} & -0.0141(0.0614)                  & \multicolumn{1}{r|}{0.0044(0.2104)}  & \multicolumn{1}{r|}{0.0002(0.1424)}  & -0.0031(0.0652)                 \\
$\hat{\sigma}_{11}$                        & \multicolumn{1}{r|}{0.0168(0.2258)}  & \multicolumn{1}{r|}{-0.0005(0.1615)} & 0.0031(0.0775)                   & \multicolumn{1}{r|}{0.0043(0.2290)}  & \multicolumn{1}{r|}{-0.0154(0.1570)} & -0.0048(0.0791)                 \\
$\hat{\sigma}_{12}$                        & \multicolumn{1}{r|}{-0.0059(0.2129)} & \multicolumn{1}{r|}{-0.0073(0.1520)} & -0.0063(0.0732)                  & \multicolumn{1}{r|}{-0.0114(0.1819)} & \multicolumn{1}{r|}{-0.0166(0.1267)} & -0.0063(0.0653)                 \\
$\hat{\sigma}_{13}$                        & \multicolumn{1}{r|}{-0.0237(0.2071)} & \multicolumn{1}{r|}{-0.0128(0.1488)} & -0.0167(0.0743)                  & \multicolumn{1}{r|}{0.0171(0.1534)}  & \multicolumn{1}{r|}{0.0001(0.1103)}  & -0.0009(0.0584)                 \\
$\hat{\sigma}_{22}$                        & \multicolumn{1}{r|}{0.0091(0.2581)}  & \multicolumn{1}{r|}{0.0111(0.1708)}  & 0.0015(0.0848)                   & \multicolumn{1}{r|}{-0.0045(0.2403)} & \multicolumn{1}{r|}{-0.0129(0.1606)} & -0.0046(0.0865)                 \\
$\hat{\sigma}_{23}$                        & \multicolumn{1}{r|}{-0.0209(0.2326)} & \multicolumn{1}{r|}{-0.0002(0.1597)} & -0.0095(0.0840)                  & \multicolumn{1}{r|}{-0.0117(0.1707)} & \multicolumn{1}{r|}{-0.0067(0.1175)} & -0.0031(0.0687)                 \\
$\hat{\sigma}_{33}$                        & \multicolumn{1}{r|}{0.0041(0.2635)}  & \multicolumn{1}{r|}{0.0214(0.1715)}  & 0.0021(0.0903)                   & \multicolumn{1}{r|}{-0.0048(0.2640)} & \multicolumn{1}{r|}{0.0047(0.1779)}  & -0.0022(0.0811)                 \\ \hline
\multicolumn{1}{c|}{(A2)(B2)}              & \multicolumn{1}{r|}{}                & \multicolumn{1}{r|}{}                &                                  & \multicolumn{1}{r|}{}                & \multicolumn{1}{r|}{}                &                                 \\
$\hat{\alpha}_1$                           & \multicolumn{1}{r|}{0.0036(0.0588)}  & \multicolumn{1}{r|}{0.0032(0.0420)}  & 0.0032(0.0194)                   & \multicolumn{1}{r|}{0.0022(0.0693)}  & \multicolumn{1}{r|}{0.0025(0.0432)}  & 0.0007(0.0189)                  \\
$\hat{\alpha}_2$                           & \multicolumn{1}{r|}{0.0025(0.0586)}  & \multicolumn{1}{r|}{0.0064(0.0398)}  & 0.0024(0.0190)                   & \multicolumn{1}{r|}{-0.0011(0.0645)} & \multicolumn{1}{r|}{0.0038(0.0459)}  & 0.0004(0.0193)                  \\
$\hat{\alpha}_3$                           & \multicolumn{1}{r|}{0.0028(0.0670)}  & \multicolumn{1}{r|}{0.0014(0.0414)}  & 0.0028(0.0204)                   & \multicolumn{1}{r|}{-0.0012(0.0601)} & \multicolumn{1}{r|}{0.0019(0.0424)}  & 0.0012(0.0194)                  \\
$\hat{\mu}_1$                              & \multicolumn{1}{r|}{-0.0343(0.1834)} & \multicolumn{1}{r|}{-0.0255(0.1265)} & -0.0066(0.0619)                  & \multicolumn{1}{r|}{-0.0235(0.2005)} & \multicolumn{1}{r|}{-0.0113(0.1264)} & -0.0031(0.0646)                 \\
$\hat{\mu}_2$                              & \multicolumn{1}{r|}{-0.0211(0.1679)} & \multicolumn{1}{r|}{-0.0173(0.1230)} & -0.0065(0.0588)                  & \multicolumn{1}{r|}{-0.0015(0.1912)} & \multicolumn{1}{r|}{-0.0140(0.1315)} & -0.0028(0.0647)                 \\
$\hat{\mu}_3$                              & \multicolumn{1}{r|}{-0.0260(0.1820)} & \multicolumn{1}{r|}{-0.0115(0.1193)} & -0.0056(0.0608)                  & \multicolumn{1}{r|}{0.0076(0.1832)}  & \multicolumn{1}{r|}{-0.0198(0.1354)} & 0.0027(0.0615)                  \\
$\hat{\sigma}_{11}$                        & \multicolumn{1}{r|}{0.0147(0.2462)}  & \multicolumn{1}{r|}{0.0180(0.1668)}  & -0.0003(0.0786)                  & \multicolumn{1}{r|}{0.0163(0.2506)}  & \multicolumn{1}{r|}{0.0096(0.1685)}  & 0.0058(0.0759)                  \\
$\hat{\sigma}_{12}$                        & \multicolumn{1}{r|}{-0.0026(0.2056)} & \multicolumn{1}{r|}{0.0040(0.1524)}  & -0.0081(0.0719)                  & \multicolumn{1}{r|}{-0.0197(0.1799)} & \multicolumn{1}{r|}{-0.0090(0.1233)} & 0.0000(0.0600)                  \\
$\hat{\sigma}_{13}$                        & \multicolumn{1}{r|}{-0.0101(0.1989)} & \multicolumn{1}{r|}{-0.0106(0.1516)} & -0.0180(0.0660)                  & \multicolumn{1}{r|}{-0.0110(0.1703)} & \multicolumn{1}{r|}{-0.0074(0.1176)} & 0.0007(0.0618)                  \\
$\hat{\sigma}_{22}$                        & \multicolumn{1}{r|}{0.0197(0.2383)}  & \multicolumn{1}{r|}{0.0137(0.1666)}  & -0.0010(0.0765)                  & \multicolumn{1}{r|}{-0.0113(0.2346)} & \multicolumn{1}{r|}{0.0029(0.1658)}  & 0.0036(0.0806)                  \\
$\hat{\sigma}_{23}$                        & \multicolumn{1}{r|}{-0.0016(0.2143)} & \multicolumn{1}{r|}{-0.0053(0.1632)} & -0.0125(0.0693)                  & \multicolumn{1}{r|}{-0.0153(0.1668)} & \multicolumn{1}{r|}{-0.0059(0.1248)} & -0.0002(0.0648)                 \\
$\hat{\sigma}_{33}$                        & \multicolumn{1}{r|}{0.0207(0.2467)}  & \multicolumn{1}{r|}{0.0073(0.1783)}  & -0.0015(0.0724)                  & \multicolumn{1}{r|}{-0.0054(0.2401)} & \multicolumn{1}{r|}{0.0075(0.1611)}  & 0.0020(0.0816)                  \\ \hline
\multicolumn{1}{c|}{(A3)(B2)}              & \multicolumn{1}{r|}{}                & \multicolumn{1}{r|}{}                &                                  & \multicolumn{1}{r|}{}                & \multicolumn{1}{r|}{}                &                                 \\
$\hat{\alpha}_1$                           & \multicolumn{1}{r|}{-0.0013(0.0554)} & \multicolumn{1}{r|}{-0.0034(0.0355)} & 0.0011(0.0175)                   & \multicolumn{1}{r|}{-0.0008(0.0594)} & \multicolumn{1}{r|}{-0.0014(0.0387)} & 0.0004(0.0186)                  \\
$\hat{\alpha}_2$                           & \multicolumn{1}{r|}{-0.0017(0.0524)} & \multicolumn{1}{r|}{-0.0011(0.0369)} & 0.0004(0.0170)                   & \multicolumn{1}{r|}{-0.0073(0.0610)} & \multicolumn{1}{r|}{-0.0006(0.0384)} & 0.0002(0.0195)                  \\
$\hat{\alpha}_3$                           & \multicolumn{1}{r|}{0.0039(0.0509)}  & \multicolumn{1}{r|}{0.0014(0.0356)}  & -0.0004(0.0182)                  & \multicolumn{1}{r|}{-0.0034(0.0605)} & \multicolumn{1}{r|}{0.0010(0.0395)}  & -0.0012(0.0188)                 \\
$\hat{\mu}_1$                              & \multicolumn{1}{r|}{-0.0209(0.2008)} & \multicolumn{1}{r|}{-0.0051(0.1330)} & -0.0076(0.0680)                  & \multicolumn{1}{r|}{-0.0221(0.2348)} & \multicolumn{1}{r|}{-0.0037(0.1350)} & 0.0019(0.0667)                  \\
$\hat{\mu}_2$                              & \multicolumn{1}{r|}{-0.0178(0.1753)} & \multicolumn{1}{r|}{-0.0109(0.1278)} & -0.0093(0.0652)                  & \multicolumn{1}{r|}{0.0132(0.2328)}  & \multicolumn{1}{r|}{-0.0174(0.1420)} & -0.0015(0.0701)                 \\
$\hat{\mu}_3$                              & \multicolumn{1}{r|}{-0.0237(0.1784)} & \multicolumn{1}{r|}{-0.0174(0.1390)} & -0.0107(0.0662)                  & \multicolumn{1}{r|}{-0.0096(0.2576)} & \multicolumn{1}{r|}{-0.0187(0.1588)} & 0.0026(0.0687)                  \\
$\hat{\sigma}_{11}$                        & \multicolumn{1}{r|}{0.0198(0.2531)}  & \multicolumn{1}{r|}{0.0031(0.1783)}  & 0.0048(0.0792)                   & \multicolumn{1}{r|}{0.0060(0.2667)}  & \multicolumn{1}{r|}{-0.0067(0.1617)} & -0.0009(0.0887)                 \\
$\hat{\sigma}_{12}$                        & \multicolumn{1}{r|}{-0.0158(0.2083)} & \multicolumn{1}{r|}{-0.0113(0.1517)} & -0.0035(0.0780)                  & \multicolumn{1}{r|}{-0.0253(0.1884)} & \multicolumn{1}{r|}{0.0014(0.1304)}  & -0.0059(0.0717)                 \\
$\hat{\sigma}_{13}$                        & \multicolumn{1}{r|}{-0.0294(0.1999)} & \multicolumn{1}{r|}{-0.0153(0.1489)} & -0.0119(0.0757)                  & \multicolumn{1}{r|}{-0.0115(0.1823)} & \multicolumn{1}{r|}{0.0034(0.1193)}  & 0.0009(0.0612)                  \\
$\hat{\sigma}_{22}$                        & \multicolumn{1}{r|}{0.0049(0.2285)}  & \multicolumn{1}{r|}{-0.0021(0.1591)} & 0.0023(0.0870)                   & \multicolumn{1}{r|}{-0.0002(0.2577)} & \multicolumn{1}{r|}{0.0229(0.1763)}  & -0.0070(0.0889)                 \\
$\hat{\sigma}_{23}$                        & \multicolumn{1}{r|}{-0.0171(0.2195)} & \multicolumn{1}{r|}{-0.0106(0.1543)} & -0.0065(0.0833)                  & \multicolumn{1}{r|}{0.0086(0.1766)}  & \multicolumn{1}{r|}{-0.0015(0.1304)} & -0.0077(0.0664)                 \\
$\hat{\sigma}_{33}$                        & \multicolumn{1}{r|}{0.0097(0.2521)}  & \multicolumn{1}{r|}{0.0138(0.1806)}  & 0.0072(0.0881)                   & \multicolumn{1}{r|}{0.0217(0.3030)}  & \multicolumn{1}{r|}{0.0200(0.1918)}  & -0.0027(0.0829)                 \\ \hline
\end{tabular}
\end{tiny}
\end{table}

\subsection{The Multivariate geometric-logitnormal innovation process}
The outcomes of the simulation are displayed in Table \ref{tab3} and Table \ref{tab4}, offering insights into the empirical biases and standard deviations of the parameters. The EM algorithm demonstrate commendable and effective performance in the experiments, with the estimated means close to the true parameter values that were used to generate the data. In particular, the parameters related to the multivariate geometric-logitnormal innovation process are accurately estimated across various scenarios, even when the sample size was as small as 50. As the sample size expands, there is a gradual reduction in the empirical standard deviations. To sum up, the parameter estimation algorithm exhibit robust performance across all parameter scenarios. Furthermore, an increase in sample size leads to a steady decrease in both the biases and standard deviations of the parameter estimation results.

\begin{table}[!htbp]
\caption{In scenario (B1), the estimated biases and standard deviations (in parentheses) of parameters $\boldsymbol{\theta}$ are presented. (multivariate geometric-logitnormal distribution)}
\centering
\label{tab3}
\begin{tiny}
\begin{tabular}{l|rrr|rrr}
\hline
                      & \multicolumn{3}{l|}{(C1)}                                                                                      & \multicolumn{3}{l}{(C2)}                                                                                      \\ \cline{2-7} 
$\boldsymbol{\theta}$ & \multicolumn{1}{l|}{$n_{t}=50$}      & \multicolumn{1}{l|}{$n_{t}=100$}     & \multicolumn{1}{l|}{$n_{t}=300$} & \multicolumn{1}{l|}{$n_{t}=50$}      & \multicolumn{1}{l|}{$n_{t}=100$}     & \multicolumn{1}{l}{$n_{t}=300$} \\ \hline
(A1)(B1)              & \multicolumn{1}{r|}{}                & \multicolumn{1}{r|}{}                &                                  & \multicolumn{1}{r|}{}                & \multicolumn{1}{r|}{}                &                                 \\
$\hat{\alpha}_1$      & \multicolumn{1}{r|}{-0.0016(0.0551)} & \multicolumn{1}{r|}{0.0048(0.0413)}  & 0.0005(0.0207)                   & \multicolumn{1}{r|}{0.0099(0.0601)}  & \multicolumn{1}{r|}{0.0059(0.0437)}  & 0.0014(0.0213)                  \\
$\hat{\alpha}_2$      & \multicolumn{1}{r|}{0.0001(0.0589)}  & \multicolumn{1}{r|}{-0.0011(0.0432)} & 0.0009(0.0238)                   & \multicolumn{1}{r|}{-0.0006(0.0647)} & \multicolumn{1}{r|}{0.0029(0.0446)}  & 0.0001(0.0252)                  \\
$\hat{\alpha}_3$      & \multicolumn{1}{r|}{-0.0051(0.0515)} & \multicolumn{1}{r|}{-0.0005(0.0364)} & -0.0012(0.0211)                  & \multicolumn{1}{r|}{0.0036(0.0526)}  & \multicolumn{1}{r|}{-0.0047(0.0387)} & -0.0024(0.0196)                 \\
$\hat{\mu}_1$         & \multicolumn{1}{r|}{-0.0363(0.2089)} & \multicolumn{1}{r|}{-0.0122(0.1622)} & -0.0043(0.0732)                  & \multicolumn{1}{r|}{-0.0667(0.2629)} & \multicolumn{1}{r|}{-0.0272(0.1665)} & -0.0114(0.0966)                 \\
$\hat{\mu}_2$         & \multicolumn{1}{r|}{-0.0302(0.1889)} & \multicolumn{1}{r|}{-0.0145(0.1432)} & -0.0038(0.0699)                  & \multicolumn{1}{r|}{-0.0604(0.2506)} & \multicolumn{1}{r|}{-0.0253(0.1743)} & -0.0103(0.0967)                 \\
$\hat{\mu}_3$         & \multicolumn{1}{r|}{-0.0334(0.2034)} & \multicolumn{1}{r|}{-0.0175(0.1449)} & -0.0039(0.0699)                  & \multicolumn{1}{r|}{-0.0473(0.2451)} & \multicolumn{1}{r|}{0.0019(0.1887)}  & -0.0073(0.0897)                 \\
$\hat{\sigma}_{11}$   & \multicolumn{1}{r|}{0.0302(0.3695)}  & \multicolumn{1}{r|}{0.0047(0.2377)}  & -0.0015(0.1001)                  & \multicolumn{1}{r|}{0.1109(0.4597)}  & \multicolumn{1}{r|}{0.0448(0.2479)}  & 0.0105(0.1295)                  \\
$\hat{\sigma}_{12}$   & \multicolumn{1}{r|}{-0.0237(0.3158)} & \multicolumn{1}{r|}{-0.0107(0.1981)} & -0.0029(0.0969)                  & \multicolumn{1}{r|}{-0.0287(0.3334)} & \multicolumn{1}{r|}{-0.0001(0.2177)} & 0.0076(0.1154)                  \\
$\hat{\sigma}_{13}$   & \multicolumn{1}{r|}{-0.0442(0.3095)} & \multicolumn{1}{r|}{-0.0152(0.1876)} & -0.0029(0.0919)                  & \multicolumn{1}{r|}{-0.0021(0.2828)} & \multicolumn{1}{r|}{-0.0142(0.2034)} & 0.0002(0.1114)                  \\
$\hat{\sigma}_{22}$   & \multicolumn{1}{r|}{0.0057(0.3423)}  & \multicolumn{1}{r|}{-0.0112(0.2036)} & -0.0036(0.1018)                  & \multicolumn{1}{r|}{0.0833(0.4571)}  & \multicolumn{1}{r|}{0.0441(0.2923)}  & 0.0101(0.1357)                  \\
$\hat{\sigma}_{23}$   & \multicolumn{1}{r|}{0.0008(0.3207)}  & \multicolumn{1}{r|}{-0.0113(0.1976)} & -0.0032(0.0983)                  & \multicolumn{1}{r|}{-0.0192(0.3166)} & \multicolumn{1}{r|}{-0.0256(0.2241)} & -0.0067(0.0985)                 \\
$\hat{\sigma}_{33}$   & \multicolumn{1}{r|}{0.0211(0.3383)}  & \multicolumn{1}{r|}{-0.0057(0.2082)} & -0.0022(0.0992)                  & \multicolumn{1}{r|}{0.0132(0.3678)}  & \multicolumn{1}{r|}{0.0284(0.3036)}  & -0.0035(0.1335)                 \\ \hline
(A2)(B1)              & \multicolumn{1}{l|}{}                & \multicolumn{1}{l|}{}                & \multicolumn{1}{l|}{}            & \multicolumn{1}{l|}{}                & \multicolumn{1}{l|}{}                & \multicolumn{1}{l}{}            \\
$\hat{\alpha}_1$      & \multicolumn{1}{r|}{0.0089(0.0618)}  & \multicolumn{1}{r|}{0.0019(0.0459)}  & 0.0003(0.0242)                   & \multicolumn{1}{r|}{0.0079(0.0662)}  & \multicolumn{1}{r|}{-0.0004(0.0455)} & 0.0003(0.0253)                  \\
$\hat{\alpha}_2$      & \multicolumn{1}{r|}{0.0048(0.0635)}  & \multicolumn{1}{r|}{-0.0007(0.0422)} & -0.0003(0.0247)                  & \multicolumn{1}{r|}{0.0034(0.0682)}  & \multicolumn{1}{r|}{0.0027(0.0433)}  & 0.0007(0.0245)                  \\
$\hat{\alpha}_3$      & \multicolumn{1}{r|}{0.0039(0.0605)}  & \multicolumn{1}{r|}{0.0055(0.0405)}  & 0.0001(0.0235)                   & \multicolumn{1}{r|}{-0.0019(0.0633)} & \multicolumn{1}{r|}{-0.0027(0.0441)} & -0.0005(0.0252)                 \\
$\hat{\mu}_1$         & \multicolumn{1}{r|}{-0.0439(0.2152)} & \multicolumn{1}{r|}{-0.0245(0.1453)} & 0.0073(0.0705)                   & \multicolumn{1}{r|}{-0.0567(0.2662)} & \multicolumn{1}{r|}{-0.0267(0.1862)} & -0.0095(0.0927)                 \\
$\hat{\mu}_2$         & \multicolumn{1}{r|}{-0.0448(0.1973)} & \multicolumn{1}{r|}{-0.0258(0.1432)} & 0.0071(0.0721)                   & \multicolumn{1}{r|}{-0.0559(0.2636)} & \multicolumn{1}{r|}{-0.0269(0.1762)} & -0.0124(0.0972)                 \\
$\hat{\mu}_3$         & \multicolumn{1}{r|}{-0.0438(0.2318)} & \multicolumn{1}{r|}{-0.0279(0.1528)} & 0.0059(0.0734)                   & \multicolumn{1}{r|}{-0.0633(0.2891)} & \multicolumn{1}{r|}{-0.0328(0.2049)} & -0.0067(0.0975)                 \\
$\hat{\sigma}_{11}$   & \multicolumn{1}{r|}{0.0093(0.3247)}  & \multicolumn{1}{r|}{0.0304(0.2187)}  & -0.0052(0.1063)                  & \multicolumn{1}{r|}{0.1012(0.4452)}  & \multicolumn{1}{r|}{0.0549(0.3256)}  & 0.0233(0.1287)                  \\
$\hat{\sigma}_{12}$   & \multicolumn{1}{r|}{-0.0248(0.2828)} & \multicolumn{1}{r|}{0.0199(0.1918)}  & -0.0075(0.1016)                  & \multicolumn{1}{r|}{-0.0209(0.3464)} & \multicolumn{1}{r|}{-0.0072(0.2375)} & 0.0112(0.1098)                  \\
$\hat{\sigma}_{13}$   & \multicolumn{1}{r|}{-0.0437(0.3147)} & \multicolumn{1}{r|}{0.0203(0.1919)}  & -0.0085(0.0983)                  & \multicolumn{1}{r|}{0.0118(0.3166)}  & \multicolumn{1}{r|}{-0.0255(0.2364)} & -0.0074(0.1057)                 \\
$\hat{\sigma}_{22}$   & \multicolumn{1}{r|}{-0.0026(0.3132)} & \multicolumn{1}{r|}{0.0247(0.2011)}  & -0.0083(0.1056)                  & \multicolumn{1}{r|}{0.0767(0.4641)}  & \multicolumn{1}{r|}{0.0406(0.2742)}  & 0.0224(0.1391)                  \\
$\hat{\sigma}_{23}$   & \multicolumn{1}{r|}{-0.0043(0.3253)} & \multicolumn{1}{r|}{0.0279(0.2065)}  & -0.0093(0.1026)                  & \multicolumn{1}{r|}{-0.0216(0.3484)} & \multicolumn{1}{r|}{-0.0281(0.2204)} & 0.0068(0.1168)                  \\
$\hat{\sigma}_{33}$   & \multicolumn{1}{r|}{0.0267(0.3783)}  & \multicolumn{1}{r|}{0.0378(0.2366)}  & -0.0097(0.1034)                  & \multicolumn{1}{r|}{0.1238(0.5248)}  & \multicolumn{1}{r|}{0.0791(0.3412)}  & 0.0246(0.1453)                  \\ \hline
(A3)(B1)              & \multicolumn{1}{l|}{}                & \multicolumn{1}{l|}{}                & \multicolumn{1}{l|}{}            & \multicolumn{1}{l|}{}                & \multicolumn{1}{l|}{}                & \multicolumn{1}{l}{}            \\
$\hat{\alpha}_1$      & \multicolumn{1}{r|}{0.0033(0.0523)}  & \multicolumn{1}{r|}{0.0034(0.0362)}  & 0.0014(0.0203)                   & \multicolumn{1}{r|}{-0.0032(0.0536)} & \multicolumn{1}{r|}{0.0021(0.0385)}  & -0.0001(0.0218)                 \\
$\hat{\alpha}_2$      & \multicolumn{1}{r|}{-0.0049(0.0502)} & \multicolumn{1}{r|}{-0.0009(0.0378)} & 0.0006(0.0205)                   & \multicolumn{1}{r|}{0.0007(0.0525)}  & \multicolumn{1}{r|}{0.0021(0.0372)}  & 0.0011(0.0207)                  \\
$\hat{\alpha}_3$      & \multicolumn{1}{r|}{-0.0043(0.0497)} & \multicolumn{1}{r|}{0.0024(0.0335)}  & 0.0006(0.0207)                   & \multicolumn{1}{r|}{0.0008(0.0553)}  & \multicolumn{1}{r|}{-0.0004(0.0364)} & 0.0008(0.0215)                  \\
$\hat{\mu}_1$         & \multicolumn{1}{r|}{-0.0393(0.2222)} & \multicolumn{1}{r|}{-0.0191(0.1493)} & -0.0018(0.0724)                  & \multicolumn{1}{r|}{-0.0461(0.2861)} & \multicolumn{1}{r|}{-0.0389(0.2048)} & -0.0166(0.0916)                 \\
$\hat{\mu}_2$         & \multicolumn{1}{r|}{-0.0157(0.2206)} & \multicolumn{1}{r|}{-0.0151(0.1412)} & -0.0065(0.0712)                  & \multicolumn{1}{r|}{-0.0532(0.2891)} & \multicolumn{1}{r|}{-0.0368(0.1967)} & -0.0114(0.0951)                 \\
$\hat{\mu}_3$         & \multicolumn{1}{r|}{-0.0142(0.2411)} & \multicolumn{1}{r|}{-0.0042(0.1526)} & -0.0099(0.0721)                  & \multicolumn{1}{r|}{-0.0482(0.2964)} & \multicolumn{1}{r|}{-0.0181(0.1881)} & -0.0014(0.1006)                 \\
$\hat{\sigma}_{11}$   & \multicolumn{1}{r|}{0.0106(0.3746)}  & \multicolumn{1}{r|}{-0.0033(0.2308)} & 0.0046(0.1156)                   & \multicolumn{1}{r|}{0.0775(0.4159)}  & \multicolumn{1}{r|}{0.0179(0.3147)}  & 0.0089(0.1288)                  \\
$\hat{\sigma}_{12}$   & \multicolumn{1}{r|}{-0.0389(0.2844)} & \multicolumn{1}{r|}{-0.0094(0.2086)} & 0.0039(0.1151)                   & \multicolumn{1}{r|}{-0.0645(0.3256)} & \multicolumn{1}{r|}{-0.0352(0.2343)} & -0.0015(0.1141)                 \\
$\hat{\sigma}_{13}$   & \multicolumn{1}{r|}{-0.0828(0.2764)} & \multicolumn{1}{r|}{-0.0098(0.2056)} & 0.0016(0.1089)                   & \multicolumn{1}{r|}{-0.0054(0.2886)} & \multicolumn{1}{r|}{0.0145(0.2001)}  & 0.0022(0.0989)                  \\
$\hat{\sigma}_{22}$   & \multicolumn{1}{r|}{-0.0195(0.2873)} & \multicolumn{1}{r|}{-0.0021(0.2146)} & 0.0047(0.1214)                   & \multicolumn{1}{r|}{0.0343(0.4296)}  & \multicolumn{1}{r|}{0.0478(0.3027)}  & 0.0077(0.1329)                  \\
$\hat{\sigma}_{23}$   & \multicolumn{1}{r|}{-0.0485(0.2807)} & \multicolumn{1}{r|}{0.0028(0.2102)}  & 0.0023(0.1156)                   & \multicolumn{1}{r|}{-0.0392(0.3293)} & \multicolumn{1}{r|}{0.0094(0.2104)}  & 0.0003(0.1045)                  \\
$\hat{\sigma}_{33}$   & \multicolumn{1}{r|}{-0.0378(0.3058)} & \multicolumn{1}{r|}{0.0161(0.2247)}  & 0.0005(0.1145)                   & \multicolumn{1}{r|}{0.0728(0.5148)}  & \multicolumn{1}{r|}{0.0275(0.2742)}  & 0.0047(0.1287)                  \\ \hline
\end{tabular}
\end{tiny}
\end{table}

\begin{table}[!htbp]
\caption{In scenario (B2), the estimated biases and standard deviations (in parentheses) of parameters $\boldsymbol{\theta}$ are presented. (multivariate geometric-logitnormal distribution)}
\centering
\label{tab4}
\begin{tiny}
\begin{tabular}{l|rrr|rrr}
\hline
\multicolumn{1}{c|}{}                      & \multicolumn{3}{c|}{(C1)}                                                                                      & \multicolumn{3}{c}{(C2)}                                                                                      \\ \cline{2-7} 
\multicolumn{1}{c|}{$\boldsymbol{\theta}$} & \multicolumn{1}{c|}{$n_{t}=50$}      & \multicolumn{1}{c|}{$n_{t}=100$}     & \multicolumn{1}{c|}{$n_{t}=300$} & \multicolumn{1}{c|}{$n_{t}=50$}      & \multicolumn{1}{c|}{$n_{t}=100$}     & \multicolumn{1}{c}{$n_{t}=300$} \\ \hline
\multicolumn{1}{c|}{(A1)(B1)}              & \multicolumn{1}{r|}{}                & \multicolumn{1}{r|}{}                &                                  & \multicolumn{1}{r|}{}                & \multicolumn{1}{r|}{}                &                                 \\
$\hat{\alpha}_1$                           & \multicolumn{1}{r|}{0.0066(0.0487)}  & \multicolumn{1}{r|}{0.0024(0.0328)}  & 0.0004(0.0167)                   & \multicolumn{1}{r|}{0.0064(0.0524)}  & \multicolumn{1}{r|}{0.0064(0.0337)}  & 0.0014(0.0195)                  \\
$\hat{\alpha}_2$                           & \multicolumn{1}{r|}{0.0043(0.0484)}  & \multicolumn{1}{r|}{-0.0039(0.0351)} & -0.0011(0.0185)                  & \multicolumn{1}{r|}{-0.0006(0.0521)} & \multicolumn{1}{r|}{0.0016(0.0357)}  & 0.0006(0.0213)                  \\
$\hat{\alpha}_3$                           & \multicolumn{1}{r|}{0.0018(0.0442)}  & \multicolumn{1}{r|}{-0.0021(0.0312)} & -0.0014(0.0182)                  & \multicolumn{1}{r|}{0.0012(0.0483)}  & \multicolumn{1}{r|}{-0.0003(0.0322)} & -0.0012(0.0185)                 \\
$\hat{\mu}_1$                              & \multicolumn{1}{r|}{-0.0406(0.1984)} & \multicolumn{1}{r|}{-0.0079(0.1305)} & 0.0007(0.0767)                   & \multicolumn{1}{r|}{-0.0334(0.2427)} & \multicolumn{1}{r|}{-0.0179(0.1767)} & -0.0099(0.0832)                 \\
$\hat{\mu}_2$                              & \multicolumn{1}{r|}{-0.0325(0.2012)} & \multicolumn{1}{r|}{-0.0124(0.1381)} & -0.0004(0.0726)                  & \multicolumn{1}{r|}{-0.0976(0.2554)} & \multicolumn{1}{r|}{-0.0205(0.1814)} & -0.0107(0.0871)                 \\
$\hat{\mu}_3$                              & \multicolumn{1}{r|}{-0.0346(0.2129)} & \multicolumn{1}{r|}{-0.0049(0.1479)} & -0.0007(0.0731)                  & \multicolumn{1}{r|}{-0.0595(0.2578)} & \multicolumn{1}{r|}{-0.024(0.1954)}  & -0.0062(0.0922)                 \\
$\hat{\sigma}_{11}$                        & \multicolumn{1}{r|}{0.0258(0.3278)}  & \multicolumn{1}{r|}{0.0052(0.2487)}  & 0.0007(0.1048)                   & \multicolumn{1}{r|}{0.0733(0.4188)}  & \multicolumn{1}{r|}{0.0304(0.2505)}  & 0.0262(0.1242)                  \\
$\hat{\sigma}_{12}$                        & \multicolumn{1}{r|}{-0.0089(0.2845)} & \multicolumn{1}{r|}{-0.0124(0.1999)} & 0.0011(0.1021)                   & \multicolumn{1}{r|}{-0.0591(0.2879)} & \multicolumn{1}{r|}{-0.0199(0.1862)} & 0.0066(0.1065)                  \\
$\hat{\sigma}_{13}$                        & \multicolumn{1}{r|}{-0.0219(0.2823)} & \multicolumn{1}{r|}{-0.0179(0.1971)} & 0.0009(0.0972)                   & \multicolumn{1}{r|}{-0.0015(0.2698)} & \multicolumn{1}{r|}{-0.0098(0.1985)} & -0.0113(0.0966)                 \\
$\hat{\sigma}_{22}$                        & \multicolumn{1}{r|}{0.0375(0.3313)}  & \multicolumn{1}{r|}{-0.0126(0.2022)} & 0.0029(0.1055)                   & \multicolumn{1}{r|}{0.0613(0.4102)}  & \multicolumn{1}{r|}{0.0255(0.2627)}  & 0.0099(0.1268)                  \\
$\hat{\sigma}_{23}$                        & \multicolumn{1}{r|}{0.0253(0.3183)}  & \multicolumn{1}{r|}{-0.0107(0.2032)} & 0.0029(0.1016)                   & \multicolumn{1}{r|}{-0.0263(0.2952)} & \multicolumn{1}{r|}{-0.0111(0.1983)} & 0.0018(0.0994)                  \\
$\hat{\sigma}_{33}$                        & \multicolumn{1}{r|}{0.0421(0.3603)}  & \multicolumn{1}{r|}{0.0005(0.2205)}  & 0.0034(0.1019)                   & \multicolumn{1}{r|}{0.0479(0.4413)}  & \multicolumn{1}{r|}{0.0489(0.3088)}  & 0.0205(0.1429)                  \\ \hline
\multicolumn{1}{c|}{(A2)(B1)}              & \multicolumn{1}{l|}{}                & \multicolumn{1}{l|}{}                & \multicolumn{1}{l|}{}            & \multicolumn{1}{l|}{}                & \multicolumn{1}{l|}{}                & \multicolumn{1}{l}{}            \\
$\hat{\alpha}_1$                           & \multicolumn{1}{r|}{0.0031(0.0496)}  & \multicolumn{1}{r|}{0.0028(0.0357)}  & -0.0004(0.0204)                  & \multicolumn{1}{r|}{0.0074(0.0539)}  & \multicolumn{1}{r|}{0.0031(0.037)}   & 0.0011(0.0205)                  \\
$\hat{\alpha}_2$                           & \multicolumn{1}{r|}{0.0031(0.0537)}  & \multicolumn{1}{r|}{0.0029(0.0364)}  & -0.0006(0.0202)                  & \multicolumn{1}{r|}{0.0025(0.0563)}  & \multicolumn{1}{r|}{0.0026(0.036)}   & -0.0004(0.0207)                 \\
$\hat{\alpha}_3$                           & \multicolumn{1}{r|}{-0.0002(0.0533)} & \multicolumn{1}{r|}{-0.0009(0.0361)} & 0.0001(0.0182)                   & \multicolumn{1}{r|}{0.0048(0.0566)}  & \multicolumn{1}{r|}{0.0018(0.043)}   & 0.0036(0.0207)                  \\
$\hat{\mu}_1$                              & \multicolumn{1}{r|}{-0.0237(0.2199)} & \multicolumn{1}{r|}{0.0039(0.1404)}  & -0.0052(0.0773)                  & \multicolumn{1}{r|}{-0.0591(0.2361)} & \multicolumn{1}{r|}{-0.0158(0.1655)} & -0.0029(0.0843)                 \\
$\hat{\mu}_2$                              & \multicolumn{1}{r|}{-0.0207(0.2096)} & \multicolumn{1}{r|}{0.0047(0.1241)}  & -0.0054(0.0736)                  & \multicolumn{1}{r|}{-0.0559(0.2563)} & \multicolumn{1}{r|}{-0.0247(0.1773)} & -0.0113(0.0841)                 \\
$\hat{\mu}_3$                              & \multicolumn{1}{r|}{-0.0183(0.2293)} & \multicolumn{1}{r|}{0.0091(0.1241)}  & -0.0027(0.0742)                  & \multicolumn{1}{r|}{-0.0371(0.2504)} & \multicolumn{1}{r|}{-0.0305(0.1759)} & -0.0118(0.0874)                 \\
$\hat{\sigma}_{11}$                        & \multicolumn{1}{r|}{0.0292(0.3828)}  & \multicolumn{1}{r|}{0.0011(0.2119)}  & -0.0034(0.1029)                  & \multicolumn{1}{r|}{0.0443(0.3802)}  & \multicolumn{1}{r|}{0.0225(0.2536)}  & 0.0119(0.1274)                  \\
$\hat{\sigma}_{12}$                        & \multicolumn{1}{r|}{-0.0325(0.2827)} & \multicolumn{1}{r|}{-0.0149(0.1938)} & -0.0015(0.1023)                  & \multicolumn{1}{r|}{-0.0396(0.3212)} & \multicolumn{1}{r|}{-0.0117(0.1941)} & -0.0009(0.1114)                 \\
$\hat{\sigma}_{13}$                        & \multicolumn{1}{r|}{-0.0492(0.2784)} & \multicolumn{1}{r|}{-0.0175(0.1957)} & 0.0002(0.0995)                   & \multicolumn{1}{r|}{0.0171(0.2795)}  & \multicolumn{1}{r|}{0.0041(0.1845)}  & -0.0087(0.0923)                 \\
$\hat{\sigma}_{22}$                        & \multicolumn{1}{r|}{-0.0167(0.3001)} & \multicolumn{1}{r|}{-0.0143(0.1966)} & 0.0015(0.1088)                   & \multicolumn{1}{r|}{0.0705(0.3982)}  & \multicolumn{1}{r|}{0.0345(0.2528)}  & 0.0073(0.1319)                  \\
$\hat{\sigma}_{23}$                        & \multicolumn{1}{r|}{-0.0271(0.2872)} & \multicolumn{1}{r|}{-0.0146(0.1955)} & 0.0036(0.1069)                   & \multicolumn{1}{r|}{-0.0037(0.2687)} & \multicolumn{1}{r|}{-0.0172(0.1927)} & 0.0032(0.0952)                  \\
$\hat{\sigma}_{33}$                        & \multicolumn{1}{r|}{-0.0113(0.3087)} & \multicolumn{1}{r|}{-0.0076(0.2096)} & 0.0063(0.1088)                   & \multicolumn{1}{r|}{0.0237(0.3787)}  & \multicolumn{1}{r|}{0.0245(0.2703)}  & 0.0231(0.1273)                  \\ \hline
\multicolumn{1}{c|}{(A3)(B1)}              & \multicolumn{1}{l|}{}                & \multicolumn{1}{l|}{}                & \multicolumn{1}{l|}{}            & \multicolumn{1}{l|}{}                & \multicolumn{1}{l|}{}                & \multicolumn{1}{l}{}            \\
$\hat{\alpha}_1$                           & \multicolumn{1}{r|}{-0.0009(0.0446)} & \multicolumn{1}{r|}{-0.0001(0.0304)} & -0.0011(0.0169)                  & \multicolumn{1}{r|}{0.0059(0.0482)}  & \multicolumn{1}{r|}{-0.0013(0.0311)} & 0.0014(0.0181)                  \\
$\hat{\alpha}_2$                           & \multicolumn{1}{r|}{-0.0006(0.0451)} & \multicolumn{1}{r|}{-0.0015(0.0332)} & -0.0006(0.0167)                  & \multicolumn{1}{r|}{0.0047(0.0473)}  & \multicolumn{1}{r|}{0.0027(0.0334)}  & 0.0024(0.0176)                  \\
$\hat{\alpha}_3$                           & \multicolumn{1}{r|}{-0.0003(0.0457)} & \multicolumn{1}{r|}{-0.0016(0.0303)} & -0.0003(0.0155)                  & \multicolumn{1}{r|}{-0.0021(0.0448)} & \multicolumn{1}{r|}{0.0017(0.0304)}  & 0.0012(0.0177)                  \\
$\hat{\mu}_1$                              & \multicolumn{1}{r|}{0.0081(0.2172)}  & \multicolumn{1}{r|}{-0.0147(0.1384)} & 0.0025(0.0706)                   & \multicolumn{1}{r|}{-0.0873(0.2769)} & \multicolumn{1}{r|}{-0.0089(0.1687)} & -0.0077(0.0933)                 \\
$\hat{\mu}_2$                              & \multicolumn{1}{r|}{-0.0199(0.1831)} & \multicolumn{1}{r|}{-0.0117(0.1367)} & 0.0045(0.0672)                   & \multicolumn{1}{r|}{-0.0804(0.2641)} & \multicolumn{1}{r|}{-0.0247(0.1709)} & -0.0129(0.0932)                 \\
$\hat{\mu}_3$                              & \multicolumn{1}{r|}{-0.0257(0.1971)} & \multicolumn{1}{r|}{-0.0115(0.1419)} & 0.0044(0.0689)                   & \multicolumn{1}{r|}{-0.0625(0.2761)} & \multicolumn{1}{r|}{-0.0133(0.1855)} & -0.0054(0.0917)                 \\
$\hat{\sigma}_{11}$                        & \multicolumn{1}{r|}{-0.0373(0.3021)} & \multicolumn{1}{r|}{0.0028(0.2176)}  & 0.0019(0.1074)                   & \multicolumn{1}{r|}{0.0509(0.4209)}  & \multicolumn{1}{r|}{0.0403(0.2852)}  & 0.0144(0.1378)                  \\
$\hat{\sigma}_{12}$                        & \multicolumn{1}{r|}{-0.0669(0.2845)} & \multicolumn{1}{r|}{-0.0098(0.1869)} & 0.0003(0.1058)                   & \multicolumn{1}{r|}{-0.0574(0.3214)} & \multicolumn{1}{r|}{-0.0102(0.2211)} & 0.0069(0.1107)                  \\
$\hat{\sigma}_{13}$                        & \multicolumn{1}{r|}{-0.0739(0.2884)} & \multicolumn{1}{r|}{-0.0196(0.1888)} & -0.0003(0.1022)                  & \multicolumn{1}{r|}{-0.0049(0.2841)} & \multicolumn{1}{r|}{0.0083(0.1939)}  & 0.0021(0.0958)                  \\
$\hat{\sigma}_{22}$                        & \multicolumn{1}{r|}{-0.0255(0.3006)} & \multicolumn{1}{r|}{-0.0021(0.1983)} & 0.0001(0.1122)                   & \multicolumn{1}{r|}{0.0648(0.4576)}  & \multicolumn{1}{r|}{0.0474(0.2809)}  & 0.0051(0.1231)                  \\
$\hat{\sigma}_{23}$                        & \multicolumn{1}{r|}{-0.0189(0.2932)} & \multicolumn{1}{r|}{-0.0051(0.1985)} & -0.0004(0.1095)                  & \multicolumn{1}{r|}{-0.0031(0.3164)} & \multicolumn{1}{r|}{-0.0153(0.2086)} & -0.0058(0.1008)                 \\
$\hat{\sigma}_{33}$                        & \multicolumn{1}{r|}{0.0149(0.3384)}  & \multicolumn{1}{r|}{0.0006(0.2122)}  & -0.0003(0.1096)                  & \multicolumn{1}{r|}{0.0763(0.4292)}  & \multicolumn{1}{r|}{0.0116(0.2477)}  & -0.0029(0.1255)                 \\ \hline
\end{tabular}
\end{tiny}
\end{table}

\section{Application}
\begin{figure}[!htbp]
\centering
\subfigure[The sample path and autocorrelation plot for sequence A]
{
	\centering
	\includegraphics[scale=0.5]{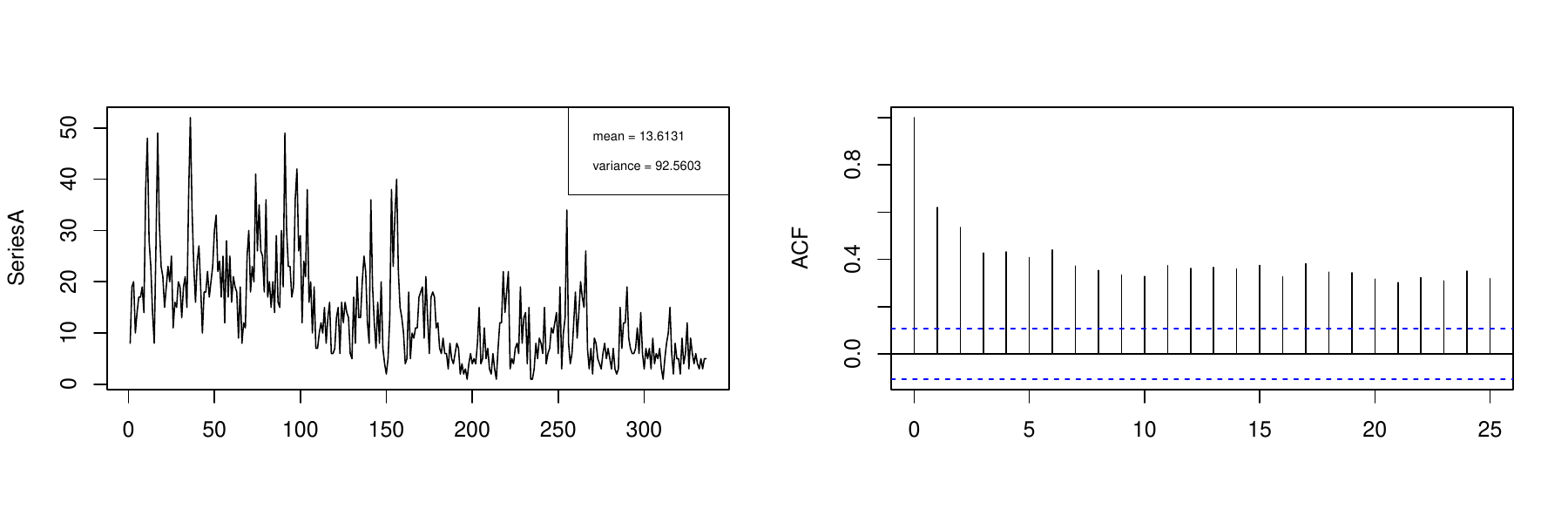}
}

\subfigure[The sample path and autocorrelation plot for sequence B]
{
	\centering
	\includegraphics[scale=0.5]{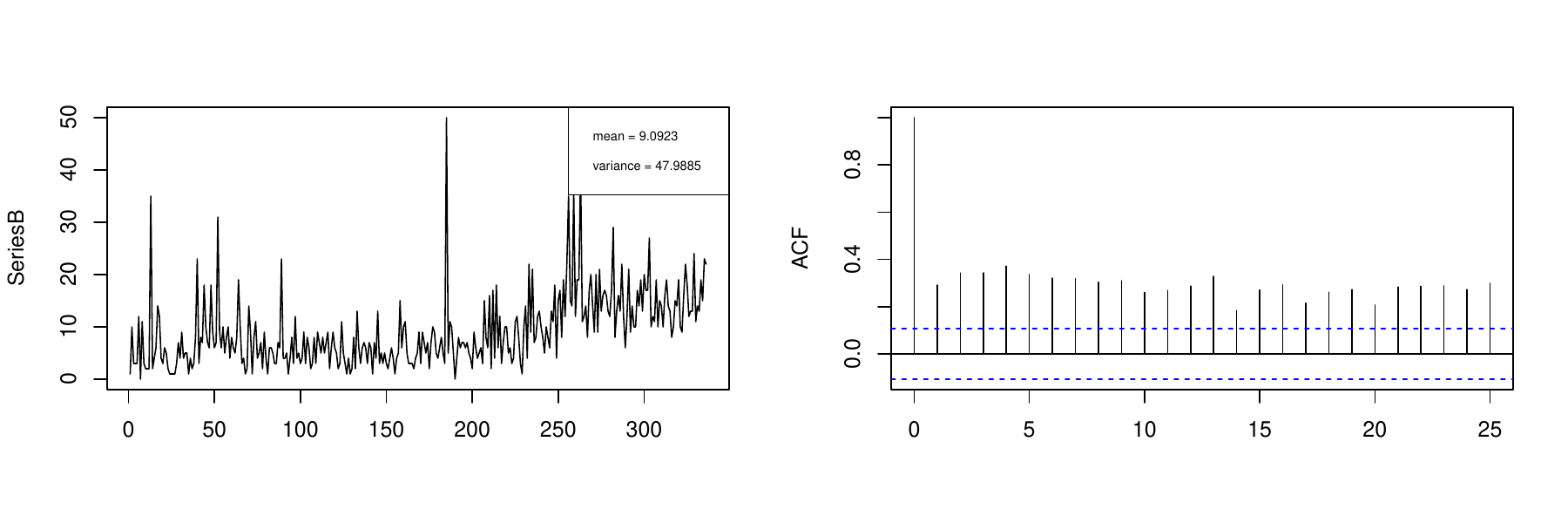}
}

\subfigure[The sample path and autocorrelation plot for sequence C]
{
	\centering
	\includegraphics[scale=0.5]{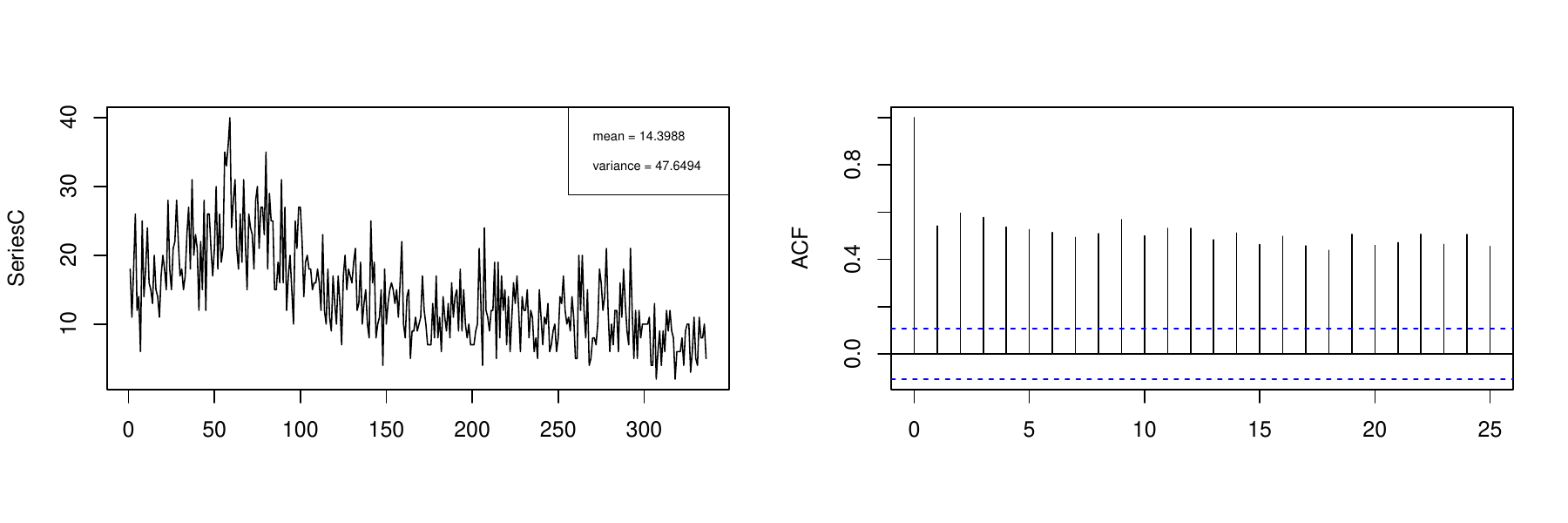}
}

\caption{Figure of the related data.}
\label{fig01}
\end{figure}

In this section, we use the two types of models proposed above to model and fit a set of crime data. The data is sourced from the Bureau of Crime Statistics and Research (BOCSAR) in New Wales, Australia (\url{https://www.bocsar.nsw.gov.au/}). It records monthly crime data from January 1995 to December 2022, including categories such as sexual crimes, robberies, thefts, etc., with a total of 336 observations. Within these categories, there are various subcategories of crimes. In this study, we focus on the three subcategories within the theft category, specifically named by "Break and enter non-dwelling," "Fraud," and "Other theft," which are denoted as A, B, and C, respectively. We denote the monthly observations as $\left\{ X_{1,t},X_{2,t},X_{3,t} \right\}_{t=1}^{336} $. Figure \ref{fig01} displays the sample paths and autocorrelation plots for the three crime categories, along with the sample means and variances in the upper-right corner. From the plotting, it can be observed that the data exhibits obvious overdispersion. The correlation coefficients between different sequences are as follows: A-B is -0.1988, A-C is 0.5314, and B-C is -0.2671. Therefore, the correlation among  sequences cannot be ignored, indicating that a multivariate model is more suitable for this dataset.

For this dataset, \citet{chen2017bayesian} fitted the INGARCHX model to the data from January 1995 to December 2014 in the Ballina area of New South Wales. They selected crime data including sexual offenses, drug offenses, motor vehicle theft, and domestic violence-related assault to examine the effect of temperature on these different types of criminal activities. They found that temperature does indeed have an impact on criminal activity. In addition, \citet{yang2021random} fitted the RCTINAR(1)-X model to the sexual offenses data in the Ballina area. Apart from temperature, they also included drug offenses data as a covariate and found that drug offenses are related to sexual offenses.

Next, the MINAR(1)-PL, MINAR(1)-GL models and other integer time series models are used to fit the dataset. The results are compared by using Log-Likelihood (Log-Like), Akaike Information Criterion (AIC), and Bayesian Information Criterion (BIC). 
\begin{itemize}
\item The independent Poisson-INAR(1) model (\citet{alosh1987});
\item The independent Geometric-INAR(1) model (\citet{alzaid1988first});
\item The multivariate Poisson-INAR(1) model (\citet{pedeli2013}).
\end{itemize}
The results are shown in Table \ref{TABapplic1}. From the table, we can see that the MINAR(1)-PL model has the lowest AIC and BIC values, and the highest log-likelihood value. This indicates that the MINAR(1)-PL model is more suitable for the given set of actual data compared to other models. It is worth noting that the independent Geometric-INAR(1) model has better fitting results than MINAR(1)-GL model. Additionally, by examining Figure \ref{fig01} and considering existing studies, we found that the dataset exhibits overdispersion, a possible change point, an increasing trend, and the influence of other variables. Therefore, the MINAR(1)-PL model can be further improved in the near future.

\begin{table}[h]
\caption{The fitting results of the crime data}
\centering
\label{TABapplic1}
\begin{tabular}{l|c|c|c}
\hline
\multicolumn{1}{c|}{Model} & Log-Like    & AIC       & BIC       \\ \hline
The independent Poisson-INAR(1) model             & -4032.4581 & 8070.9161 & 8099.8188 \\ \hline
The independent Geometric-INAR(1) model           & -3207.6418 & 6421.2836 & 6450.1863 \\ \hline
MINAR(1)-PL                                      & -3162.4801 & 6336.9602 & 6394.7656 \\ \hline
MINAR(1)-GL                                      & -3208.0052 & 6428.0104 & 6485.8158 \\ \hline
The multivariate Poisson-INAR(1) model             & -3981.2527 & 7971.5055 & 8014.8595 \\ \hline
\end{tabular}
\end{table}

\section{Conclusion}
In this study, a new multivariate integer-valued autoregressive model with two classes of multivariate mixture distributions is proposed and the corresponding parameter estimation method is given. The model extends existing multivariate integer-valued autoregressive models by considering both the complex correlation among multivariate random vectors and the dependence between multivariate time series. This makes out model more flexible and applicable in a wider range of scenarios. The effectiveness of proposed method is verified through numerical simulations and case study of a real data. However, the proposed model still has some limitations, such as not addressing change-point issues and considering only the impact of correlation among the elements of innovation process. Future research is needed to explore these issues.

\section*{Disclosure statement}
No potential conflict of interest was reported by the author(s).

\section*{Funding}
This work is supported by the National Natural Science Foundation of China [grant numbers 51578471] and the Fundamental Research Funds for the Central
Universities [grant numbers 2682021ZTPY078].

\bibliographystyle{plainnat}
\bibliography{ref2}

\appendix
\section*{Appendix}
\subsection*{A.1. Proof of Theorem \ref{gentheo3}}
We introduce a random sequence $\{\boldsymbol{X}_t^{(n)}\}_{n \in \mathbb{Z}}$ as follows :
$$
\boldsymbol{X}_t^{(n)}= \begin{cases}\boldsymbol{0}, & n<0 
\\ 
\boldsymbol{R}_t, & n=0 
\\ 
\boldsymbol{A} \circ \boldsymbol{X}_{t-1}^{(n-1)}+\boldsymbol{R}_t, & n>0 .
\end{cases}
$$
For any $n$, $\boldsymbol{R}_t$ is independent of $\boldsymbol{A} \circ \boldsymbol{X}_{t-1}^{(n-1)}$ and $\boldsymbol{X}_s^{(n)}$ for $s<t$.

For a random vector $\boldsymbol{X}=\left(X_1, X_2, \dots, X_{\scrS N}\right)^{\prime}$, we define a Hilbert space $L^2(\Omega, \mathcal{F}, P)=$ $\{ \boldsymbol{X} \mid E(\boldsymbol{X X}^{\prime}) < \infty \}$, where the measure between two random vectors $\boldsymbol{X}$ and $\boldsymbol{Y}$ is defined as $d(\boldsymbol{X}, \boldsymbol{Y})=E ( \boldsymbol{X} \boldsymbol{Y}^{\prime}  )$. Let $\boldsymbol{X}_t^{(n)}$ have first and second moments denoted as $\boldsymbol{\mu}^{(n)}$ and $\boldsymbol{\Gamma}^{(n)}$, and let $\boldsymbol{R}_t$ have first and second moments denoted as $\boldsymbol{\mu}_{\boldsymbol{R}}$ and $\boldsymbol{\Sigma}_{\boldsymbol{R}}$. We now prove that as $n \rightarrow \infty$, $\boldsymbol{X}_t^{(n)} \stackrel{L^2}{\rightarrow} \boldsymbol{X}_t$.

$(\mathrm{A} 1)$ Existence and uniqueness.

Step 1: For $n>0$, $\boldsymbol{X}_t^{(n)} \in L^2(\Omega, \mathcal{F}, P)$.
For the mean vector $\boldsymbol{\mu}^{(n)}$, we have
$$
\boldsymbol{\mu}^{(n)}=E\left(\boldsymbol{X}_t^{(n)}\right)=E( \boldsymbol{A} \circ \boldsymbol{X}_{t-1}^{(n-1)}+\boldsymbol{R}_t )=\boldsymbol{A} E( \boldsymbol{X}_{t-1}^{(n-1)} ) +\boldsymbol{\mu}_{\boldsymbol{R}}=\sum_{i=0}^n \boldsymbol{A}^i \boldsymbol{\mu}_{\boldsymbol{R}}
$$
Since all $N$ eigenvalues $\alpha_i \in(0,1), i=1, 2, \dots, N$ of $\boldsymbol{A}$ are inside the unit circle, and we have $\lim_{n \rightarrow \infty} \sum_{i=0}^n \boldsymbol{A}^i=(\boldsymbol{I}-\boldsymbol{A})^{-1}$, the mean vector $\boldsymbol{\mu}^{n)}$ is finite.

For the second moment $\boldsymbol{\Gamma}^{(n)}$ of $\boldsymbol{X}_t^{(n)}$, we have
$$
\begin{aligned}
	\boldsymbol{\Gamma} ^{(n)}=&E[ \boldsymbol{X}_{t}^{(n)} ( \boldsymbol{X}_{t}^{(n)}  )^{\prime} ]\\
	=&\boldsymbol{A}E [ \boldsymbol{X}_{t-1}^{(n-1)} ( \boldsymbol{X}_{t-1}^{(n-1)}  ) ^{\prime} ] \boldsymbol{A}^{\prime}+\mathrm{diag}  ( \boldsymbol{\Lambda }\boldsymbol{\mu }^{(n-1)} )
	\\
	&+\boldsymbol{A}\boldsymbol{\mu }^{(n-1)} ( \boldsymbol{\mu }_{\boldsymbol{R}} )^{\prime} + \boldsymbol{\mu }_{\boldsymbol{R}} ( \boldsymbol{\mu }^{(n-1)} ) ^{\prime}\boldsymbol{A}^{\prime} + \boldsymbol{\Sigma }_{\boldsymbol{R}}
\end{aligned}
$$
where $\boldsymbol{\Lambda}=\operatorname{diag}\left(\alpha_i\left(1-\alpha_i\right)\right), \, i=1, 2, \dots, N$.

Next, we will prove that $\boldsymbol{\Gamma}^{(n)}$ is bounded. From $\boldsymbol{\mu}^{(n)} \leq (\boldsymbol{I}-\boldsymbol{A})^{-1} \boldsymbol{\mu}_{\scrS R}$, we have
\begin{equation}
\label{th31}
\boldsymbol{\Gamma}^{(n)} \leq \boldsymbol{A} E [\boldsymbol{X}_{t-1}^{(n-1)} (\boldsymbol{X}_{t-1}^{(n-1)} )^{\prime} ] \boldsymbol{A}^{\prime}+\boldsymbol{M},
\end{equation}
where 
$$
\boldsymbol{M}=\left[ m_{ij} \right] =\mathrm{diag}\left( \boldsymbol{\Lambda }(\boldsymbol{I}-\boldsymbol{A})^{-1}\boldsymbol{\mu }_{\boldsymbol{R}} \right) +\boldsymbol{A}(\boldsymbol{I}-\boldsymbol{A})^{-1}\boldsymbol{\mu }_{\boldsymbol{R}}\left( \boldsymbol{\mu }_{\boldsymbol{R}} \right) ^{\prime}+\boldsymbol{\mu }_{\boldsymbol{R}}\left( \boldsymbol{\mu }_{\boldsymbol{R}} \right) ^{\prime}(\boldsymbol{I}-\boldsymbol{A})^{-1}\boldsymbol{A}^{\prime}+\boldsymbol{\Sigma }_{\boldsymbol{R}}.
$$
By iterating formula \eqref{th31} for $n$ times, we obtain
$$
\mathbf{\Gamma }^{(n)}\le \sum\limits_{k=0}^{n-1}{\boldsymbol{A}^k}\boldsymbol{M}(\boldsymbol{A}^{\prime})^k+\boldsymbol{A}^n\mathbf{\Sigma }_{\boldsymbol{R}}(\boldsymbol{A}^{\prime})^n.
$$
Furthermore, since $\lim_{n\rightarrow \infty} \boldsymbol{A}^n  \mathbf{\Sigma }_{\boldsymbol{R}}(\boldsymbol{A}^{\prime})^n=0$, and we have $\sum_{k=0}^{n-1} \boldsymbol{A}^k\boldsymbol{M}(\boldsymbol{A}^{\prime})^k  \le   \max_{1\le i,j\le N}   \{ m_{ij} \} \sum_{k=0}^{n-1}   \boldsymbol{A}^k(\boldsymbol{A}^{\prime})^k   \le \max_{1\le i,j\le N} \{ m_{ij} \} (\boldsymbol{I}-\boldsymbol{AA}^{\prime})^{-1}$, it follows that $\boldsymbol{\Gamma }^{\left( n \right)}<\infty$. This implies that $\boldsymbol{X}_t^{(n)} \in L^2(\Omega, \mathcal{F}, P)$.

Step 2: The sequence $\{ \boldsymbol{X}_t^{(n)} \}_{n \in \mathbb{Z}}$ is a Cauchy sequence.

Let $\boldsymbol{U}_{t, k}^{(n)}=\boldsymbol{X}_t^{(n)}-\boldsymbol{X}_t^{(n-k)}, k=1,2, \cdots, n$.
\begin{enumerate}[label=(\roman*)]
\item The sequence $\{ \boldsymbol{X}_t^{(n)} \}_{n \in \mathbb{Z}}$ is a non-decreasing sequence.

When $n=1$, $\boldsymbol{X}_t^{(1)}=\boldsymbol{A} \circ \boldsymbol{X}_{t-1}^{(0)}+\boldsymbol{R}_t=\boldsymbol{A} \circ \boldsymbol{R}_{t-1}+\boldsymbol{R}_t \geq \boldsymbol{R}_t=\boldsymbol{X}_t^{(0)}$. Assuming that for all $t \in \mathbb{Z}$ and $k=1,2, \cdots, n$, we have $\boldsymbol{X}_t^{(k)} \geq \boldsymbol{X}_t^{(k-1)}$. Next, we will prove that $\boldsymbol{X}_t^{(n+1)} \geq \boldsymbol{X}_t^{(n)}$. Considering the $s$-th component, we have
$$
( \boldsymbol{X}_t^{(n+1)}-\boldsymbol{X}_t^{(n)} )_{s }
= 
(\boldsymbol{A} \circ \boldsymbol{X}_{t-1}^{(n)}-\boldsymbol{A} \circ \boldsymbol{X}_{t-1}^{(n-1)})_{s}
=
\sum_{j=X_{s,t-1}^{(n-1)}+1}^{X_{s,t-1}^{(n)}}{Z_{s,j}}
\geq 
0.
$$
Therefore, $\{ \boldsymbol{X}_t^{(n)} \}_{n \in \mathbb{Z}}$ is a non-decreasing sequence which implies $\boldsymbol{U}_{t, k}^{(n)} \geq \boldsymbol{0}$.
\item $\lim _{n \rightarrow \infty} E [\boldsymbol{U}_{t, k}^{(n)} (\boldsymbol{U}_{t, k}^{(n)} )^{\prime} ]=0$.

For the expectation of $\boldsymbol{U}_{t, k}^{(n)}$, we have
$$
E\left(\boldsymbol{U}_{t, k}^{(n)}\right)=\boldsymbol{A} E\left(\boldsymbol{U}_{t-1, k}^{(n-1)}\right)=\boldsymbol{A}^n \boldsymbol{\mu}_{\boldsymbol{R}} \rightarrow \boldsymbol{0}, n \rightarrow \infty.
$$
Furthermore, we have
$$
\begin{aligned}
& E\left[\boldsymbol{U}_{t, k}^{(n)}\left(\boldsymbol{U}_{t, k}^{(n)}\right)^{\prime}\right] \\
& =\boldsymbol{A} E\left[\boldsymbol{U}_{t-1, k}^{(n-1)}\left(\boldsymbol{U}_{t-1, k}^{(n-1)}\right)^{\prime}\right] \boldsymbol{A}^{\prime}+\boldsymbol{\Lambda} \operatorname{diag}\left(E\left(\boldsymbol{U}_{t-1, k}^{(n-1)}\right)\right) \\
& =\boldsymbol{A}^n \boldsymbol{\Sigma}_{\boldsymbol{R}}\left(\boldsymbol{A}^{\prime}\right)^n+\sum_{j=0}^{n-1} \boldsymbol{A}^j \boldsymbol{\Lambda} \operatorname{diag}\left(\boldsymbol{A}^{n-j-1} \boldsymbol{\mu}_{\boldsymbol{R}}\right)\left(\boldsymbol{A}^j\right)^{\prime} \\
& \leq \boldsymbol{A}^n \boldsymbol{\Sigma}_{\boldsymbol{R}}\left(\boldsymbol{A}^{\prime}\right)^n+\boldsymbol{A}^{n-1} \boldsymbol{\Lambda} \operatorname{diag}\left((\boldsymbol{I}-\boldsymbol{A})^{-1} \boldsymbol{\mu}_{\boldsymbol{R}}\right) \rightarrow \boldsymbol{0}, n \rightarrow \infty,
\end{aligned}
$$
which implies that $E\left[\boldsymbol{U}_{t, k}^{(n)}\left(\boldsymbol{U}_{t, k}^{(n)}\right)^{\prime}\right] \rightarrow 0$ as $n \rightarrow \infty$. Therefore, $\left\{\boldsymbol{X}_t^{(n)}\right\}_{n \in N}$ is a Cauchy sequence.
\end{enumerate}

Step3: The sequence $\{ \boldsymbol{X}_t \}$ is unique.

Assuming the existence of another process $\{\boldsymbol{X}_{t}^*\}$, consider the $s$-th component such that
$$
X_{s, t}^{(n)} \stackrel{L^2}{\longrightarrow} X_{s, t}^{*}.
$$
By the Hölder inequality, we can obtain
$$
E\left(\left| X_{s, t}-X_{s, t}^*\right|\right) \leq\left(E\left(\left|X_{s, t}-X_{s, t}^{(n)}\right|^2\right)\right)^{\frac{1}{2}}\left(E\left(\left|X_{s, t}^*-X_{s, t}^{(n)}\right|^2\right)\right)^{\frac{1}{2}} \rightarrow 0, n \rightarrow \infty, i=1, 2, \dots, N .
$$
Therefore, $E\left(\left|X_{s, t}-X_{s, t}^*\right|\right)=0$, which implies $X_{s, t}=X_{s, t}^*$ a.s.

Step4: The sequence $\{ \boldsymbol{X}_t \}$ satisfies equation \eqref{geneq1}.

Considering the $s$-th component, from \citet{du1991}, we have
$$
\begin{aligned}
	E\left( \left| X_{s,t}^{(n)}-\alpha _s\circ X_{s,t-1}-R_{s,t} \right|^2 \right) &=E\left( \left| \alpha _s\circ X_{s,t-1}^{(n-1)}-\alpha _s\circ X_{s,t-1} \right|^2 \right)\\
	&=\alpha _s\left( 1-\alpha _s \right) E\left| X_{i,t-1}^{(n-1)}-X_{i,t-1} \right|+\alpha _{s}^{2}E\left( X_{i,t-1}^{(n-1)}-X_{i,t-1} \right) ^2.
\end{aligned}
$$
Since $X_{s, t}^{(n)} \stackrel{L^2}{\longrightarrow} X_{s, t}$ and $X_{s, t}=\lim _{n \rightarrow \infty} X_{s, t}^{(n)}$, we have $E\left(\left|X_{i, t-1}^{(n-1)}-X_{i, t-1}\right|^2\right) \rightarrow 0$ and $E\left(\left|X_{i, t-1}^{(n-1)}-X_{i, t-1}\right|\right) \allowbreak \rightarrow 0$, $n \rightarrow \infty$. Hence,
$$
X_{s, t}^{(n)} \stackrel{L^2}{\longrightarrow} \alpha_{s} \circ X_{s, t-1}+R_{s, t}.
$$
Note that
$$
X_{s, t}^{(n)} \stackrel{L^2}{\longrightarrow} X_{s, t},
$$
and $X_{s,t}$ is unique. Therefore, we have
$$
X_{s, t}=\alpha_{s} \circ X_{s, t-1}+R_{s, t}.
$$
From $X_{s, t}^{(n)} \stackrel{L^2}{\longrightarrow} X_{s, t}$, we have
$$
\lim _{n \rightarrow \infty} \operatorname{Cov}\left(X_{s, t}^{(n)}, R_{s t}\right)=\operatorname{Cov}\left(X_{s, t}, R_{s, t}\right).
$$

Because for any $n$ and $s<t$, we have
$$
\operatorname{Cov}\left(\boldsymbol{X}_s^{(n)}, \boldsymbol{R}_t\right)=0,
$$
it follows that
$$
\operatorname{Cov}\left(\boldsymbol{X}_s, \boldsymbol{R}_t\right)=0.
$$
Note that $X_{s,t}$ is unique for $s=1,2,\dots,N$. Therefore, $\left\{\boldsymbol{X}_t\right\}$ is unique. Thus, there exists a unique process $\left\{\boldsymbol{X}_t\right\}$ that satisfies \eqref{geneq1}.

$(\mathrm{A} 2)$ Stationarity.

From
$$
\begin{aligned}
	\boldsymbol{X}_{t}^{(n)}&\overset{d}{=}\boldsymbol{A}\circ \boldsymbol{X}_{t-1}^{(n-1)}+\boldsymbol{R}_t\\
	&\overset{d}{=}\boldsymbol{A}\circ \left( \boldsymbol{A}\circ \boldsymbol{X}_{t-2}^{(n-2)}+\boldsymbol{R}_{t-1} \right) +\boldsymbol{R}_t\\
	&\overset{d}{=}\boldsymbol{A}^2\circ \boldsymbol{X}_{t-2}^{(n-2)}+\boldsymbol{A}\circ \boldsymbol{R}_{t-1}+\boldsymbol{R}_t,
\end{aligned}
$$
we iterate the above equation $n$ times. Notice that $\{ \boldsymbol{R}_{t} \}$ is an i.i.d. sequence, then
$$
\boldsymbol{X}_{t}^{(n)}=\boldsymbol{A}\circ \boldsymbol{A}\circ \cdots \circ \boldsymbol{X}_{t-n}^{(0)}+\sum_{j=}^{n-1}{\boldsymbol{A}}\circ \boldsymbol{A}\circ \cdots \circ \boldsymbol{R}_{t-j}+\boldsymbol{R}_t
$$
which has the same distribution as
$$
\boldsymbol{A}\circ \boldsymbol{A}\circ \cdots \circ \boldsymbol{R}_0+\sum_{j=1}^{n-1}{\boldsymbol{A}}\circ \boldsymbol{A}\circ \cdots \circ \boldsymbol{R}_{n-j}+\boldsymbol{R}_{\scrS N}.
$$ 
Therefore, the distribution of $\boldsymbol{X}_t^{(n)}$ depends only on $n$ and does not depend on $t$. Therefore, for each $j$ and $n$, we prove that
$$
\left(\boldsymbol{X}_0^{(n)}, \boldsymbol{X}_1^{(n)}, \cdots, \boldsymbol{X}_j^{(n)}\right)
$$
and
$$
\left(\boldsymbol{X}_{\scrS N}^{(n)}, \boldsymbol{X}_{N+1}^{(n)}, \cdots, \boldsymbol{X}_{N+j}^{(n)}\right)
$$
have the same distribution. Therefore, according to \citet{latour1997}, we know that there exists a unique integer-valued stationary process $\left\{\boldsymbol{X}_t\right\}$ such that $\boldsymbol{X}_t=\boldsymbol{A}_t \circ \boldsymbol{X}_{t-1}+\boldsymbol{R}_t$.

$(\mathrm{A} 3)$ Ergodicity.

Let $\left\{\boldsymbol{Z}_t\right\}$ be all the counting sequences of $\boldsymbol{A} \circ \boldsymbol{X}_{t-1}$. From \eqref{geneq1}, we have
$$
\sigma \left( \boldsymbol{X}_t,\boldsymbol{X}_{t+1},\cdots \right) \subset \sigma \left( \boldsymbol{R}_t,\boldsymbol{Z}_t,\boldsymbol{R}_{t+1},\boldsymbol{Z}_{t+1},\cdots \right).
$$
Therefore,
$$
\cap _{t=1}^{+\infty}\sigma \left( \boldsymbol{X}_t,\boldsymbol{X}_{t+1},\cdots \right) \subset \cap _{t=1}^{+\infty}\sigma \left( \boldsymbol{R}_t,\boldsymbol{Z}_t,\boldsymbol{R}_{t+1},\boldsymbol{Z}_{t+1},\cdots \right).
$$
Since $\left\{ \left( \boldsymbol{R}_t,\boldsymbol{Z}_t \right) \right\}$ is an independent sequence, according to Kolmogorov's zero-one law, for any $\boldsymbol{B}\in \cap _{t=1}^{+\infty}\sigma ( \boldsymbol{R}_t,\boldsymbol{Z}_t,  \allowbreak \boldsymbol{R}_{t+1},\boldsymbol{Z}_{t+1},\cdots )$, we have $P(\boldsymbol{B})=0$ or $P(\boldsymbol{B})=1$. Therefore, the tail events of $\left\{\boldsymbol{X}_t\right\}$'s $\sigma$-field only contain measurable sets with probability 0 or 1. According to \citet{wang1982book}, $\left\{\boldsymbol{X}_t\right\}$ is ergodicity.

\subsection*{A.2. Proof of Lemma \ref{pllemma1}}
$$
\begin{aligned}
	P(\boldsymbol{R}=\boldsymbol{r})&=\int_{(0,+\infty )^N}{\prod_{s=1}^N{\frac{\lambda _{i}^{r_s-1}}{r_s!}e^{-\lambda _s}}\cdot \frac{1}{\sqrt{2\pi \left| \boldsymbol{\Sigma } \right|}}\cdot \exp \left[ -\frac{1}{2}(\log \boldsymbol{\lambda }-\boldsymbol{\mu })^{\top}\boldsymbol{\Sigma }^{-1}(\log \boldsymbol{\lambda }-\boldsymbol{\mu }) \right] d\boldsymbol{\lambda }}\\
	&=\frac{1}{\sqrt{2\pi \left| \boldsymbol{\Sigma } \right|}\prod_{s=1}^N{r_s!}}\int_{(0,+\infty )^N}{\prod_{s=1}^N{\lambda _{i}^{r_s-1}e^{-\lambda _s}}\cdot \exp \left[ -\frac{1}{2}(\log \boldsymbol{\lambda }-\boldsymbol{\mu })^{\top}\boldsymbol{\Sigma }^{-1}(\log \boldsymbol{\lambda }-\boldsymbol{\mu }) \right] d\boldsymbol{\lambda }},  
\end{aligned}
$$

$$
\begin{aligned}
\frac{\partial P_{\boldsymbol{r}}}{\partial \mu _s}=\frac{1}{\sqrt{2\pi \left| \boldsymbol{\Sigma } \right|}\prod_{s=1}^N{r_s!}}&\int_{(0,+\infty )^N}{\prod_{s=1}^N{\lambda _{i}^{r_s-1}e^{-\lambda _s}}\cdot \exp \left[ -\frac{1}{2}(\log \boldsymbol{\lambda }-\boldsymbol{\mu })^{\top}\boldsymbol{\Sigma }^{-1}(\log \boldsymbol{\lambda }-\boldsymbol{\mu }) \right]}
\\
&\cdot \left[ \frac{1}{2}\boldsymbol{e}_{s}^{\top}\boldsymbol{\Sigma }^{-1}(\log \boldsymbol{\lambda }-\boldsymbol{\mu })+\frac{1}{2}(\log \boldsymbol{\lambda }-\boldsymbol{\mu })^{\top}\boldsymbol{\Sigma }^{-1}\boldsymbol{e}_s \right] d\boldsymbol{\lambda },
\end{aligned}
$$
where $\boldsymbol{e}_s$ is an N-dimensional unit vector with the s-th element equal to 1.

Let us assume that
$$
f_{\boldsymbol{r}}\left( \boldsymbol{\lambda } \right)=\prod_{s=1}^N{\lambda _{i}^{r_s-1}e^{-\lambda _s}}\cdot \exp \left[ -\frac{1}{2}(\log \boldsymbol{\lambda }-\boldsymbol{\mu })^{\top}\boldsymbol{\Sigma }^{-1}(\log \boldsymbol{\lambda }-\boldsymbol{\mu }) \right],
$$
$$
\begin{aligned}
g_{\boldsymbol{r}}\left( \boldsymbol{\lambda } \right)=&\prod_{s=1}^N{\lambda _{i}^{r_s-1}e^{-\lambda _s}}\cdot \exp \left[ -\frac{1}{2}(\log \boldsymbol{\lambda }-\boldsymbol{\mu })^{\top}\boldsymbol{\Sigma }^{-1}(\log \boldsymbol{\lambda }-\boldsymbol{\mu }) \right]\\
&\cdot \left[ \frac{1}{2}\boldsymbol{e}_{s}^{\top}\boldsymbol{\Sigma }^{-1}(\log \boldsymbol{\lambda }-\boldsymbol{\mu })+\frac{1}{2}(\log \boldsymbol{\lambda }-\boldsymbol{\mu })^{\top}\boldsymbol{\Sigma }^{-1}\boldsymbol{e}_s \right] d\boldsymbol{\lambda }.
\end{aligned}
$$
Therefore, we have
$$
P(\boldsymbol{R}=\boldsymbol{r})=\frac{1}{\sqrt{2\pi \left| \boldsymbol{\Sigma } \right|}\prod_{s=1}^N{r_s!}}\int_{(0,+\infty )^N}{f_{\boldsymbol{r}}\left( \boldsymbol{\lambda } \right) d\boldsymbol{\lambda }},
$$
$$
\frac{\partial P_{\boldsymbol{r}}}{\partial \mu _s}=\frac{1}{\sqrt{2\pi \left| \boldsymbol{\Sigma } \right|}\prod_{s=1}^N{r_s!}}\int_{(0,+\infty )^N}{g_{\boldsymbol{r}}\left( \boldsymbol{\lambda } \right) d\boldsymbol{\lambda }}.
$$
Differentiating $f_{\boldsymbol{r}}(\boldsymbol{\lambda })$ with respect to the variables yields
$$
\begin{aligned}
\frac{df_{\boldsymbol{r}}(\boldsymbol{\lambda })}{d\lambda _s}&=\frac{\left( r_s-1 \right)}{\lambda _s}\prod_{s=1}^N{\lambda _{s}^{r_s-1}}\cdot e^{-\lambda _s}\cdot \exp \left[ -\frac{1}{2}(\log \boldsymbol{\lambda }-\boldsymbol{\mu })^{\top}\boldsymbol{\Sigma }^{-1}(\log \boldsymbol{\lambda }-\boldsymbol{\mu }) \right] 
\\
&\quad \quad -\prod_{s=1}^N{\lambda _{s}^{r_s-1}}\cdot e^{-\lambda _s}\cdot \exp \left[ -\frac{1}{2}(\log \boldsymbol{\lambda }-\boldsymbol{\mu })^{\top}\boldsymbol{\Sigma }^{-1}(\log \boldsymbol{\lambda }-\boldsymbol{\mu }) \right] 
\\
&\quad \quad+\prod_{s=1}^n{\lambda _{s}^{r_s-1}}\cdot e^{-r_s}\cdot \exp \left[ -\frac{1}{2}(\log \vec{x}-\vec{u})^{\top}\sigma ^{-1}(\log \vec{x}-\vec{u}) \right] 
\\
&\quad\quad\quad \quad \, \cdot \left[ -\frac{1}{2}\frac{1}{\lambda _s}\boldsymbol{e}_{s}^{\top}\boldsymbol{\Sigma }^{-1}(\log \boldsymbol{\lambda }-\boldsymbol{\mu })-\frac{1}{2}(\log \boldsymbol{\lambda }-\boldsymbol{\mu })^{\top}\boldsymbol{\Sigma }^{-1}\frac{1}{\lambda _s}\boldsymbol{e}_s \right] 
\\
&=-f_{\boldsymbol{r}}(\boldsymbol{\lambda })+\frac{\left( r_s-1 \right)}{\lambda _s}f_{\boldsymbol{r}}(\boldsymbol{\lambda })-\frac{1}{\lambda _s}g_{\boldsymbol{r}}(\boldsymbol{\lambda }).
\end{aligned}
$$
Furthermore, from the definition of $f_{\boldsymbol{r}}\left( \boldsymbol{\lambda } \right)$, we have
$$
f_{\boldsymbol{r}}\left( \boldsymbol{\lambda } \right)=\lambda_s f_{\boldsymbol{r}-\boldsymbol{e}_s}\left( \boldsymbol{\lambda } \right).
$$
Furthermore, we can derive
$$
\begin{aligned}
g_{\boldsymbol{r}}(\boldsymbol{\lambda })&=-\lambda _s\frac{df_{\boldsymbol{r}}(\boldsymbol{\lambda })}{d\lambda _s}-\lambda _sf_{\boldsymbol{r}}(\boldsymbol{\lambda })+\left( r_s-1 \right) f_{\boldsymbol{r}}(\boldsymbol{\lambda })
\\
&=-\lambda _s\frac{df_{\boldsymbol{r}}(\boldsymbol{\lambda })}{d\lambda _s}-f_{\boldsymbol{r}+\boldsymbol{e}_s}(\boldsymbol{\lambda })+\left( r_s-1 \right) f_{\boldsymbol{r}}(\boldsymbol{\lambda }).
\end{aligned}
$$
Therefore, according to \citet{izsak2008}, by the method of integration by parts, we have
$$
\begin{aligned}
\int_{(0,+\infty )^N}{g_{\boldsymbol{r}}(\boldsymbol{\lambda })d\boldsymbol{\lambda }}
&=\int_{(0,+\infty )^N}{f_{\boldsymbol{r}}(\boldsymbol{\lambda })d\boldsymbol{\lambda }}
\\
&\quad -\int_{(0,+\infty )^N}{f_{\boldsymbol{r}+\boldsymbol{e}_s}(\boldsymbol{\lambda })d\boldsymbol{\lambda }}+\left( r_s-1 \right) \int_{(0,+\infty )^N}{f_{\boldsymbol{r}}(\boldsymbol{\lambda })d\boldsymbol{\lambda }}.
\end{aligned}
$$
Therefore, we can obtain
$$
\frac{\partial P_{\boldsymbol{r}}}{\partial \mu _s}=r_sP_{\boldsymbol{r}}-\left( r_s+1 \right) P_{\boldsymbol{r+\boldsymbol{e}_s}}.
$$

Furthermore, we can also obtain
$$
\begin{aligned}
\frac{\partial ^2 P_{\boldsymbol{r}}}{\partial \mu _s\partial \mu _t}&=\frac{1}{\sqrt{2\pi \left| \boldsymbol{\Sigma } \right|}\prod_{s=1}^N{r_s!}}\int_{(0,+\infty )^N}{f_{\boldsymbol{r}}\left( \boldsymbol{\lambda } \right) \cdot \left[ \frac{1}{2}\boldsymbol{e}_{s}^{\top}\boldsymbol{\Sigma }^{-1}(\log \boldsymbol{\lambda }-\boldsymbol{\mu })+\frac{1}{2}(\log \boldsymbol{\lambda }-\boldsymbol{\mu })^{\top}\boldsymbol{\Sigma }^{-1}\boldsymbol{e}_s \right]}
\\
&\qquad \qquad\qquad\qquad\qquad\qquad\qquad \quad\, \cdot \left[ \frac{1}{2}\boldsymbol{e}_{t}^{\top}\boldsymbol{\Sigma }^{-1}(\log \boldsymbol{\lambda }-\boldsymbol{\mu })+\frac{1}{2}(\log \boldsymbol{\lambda }-\boldsymbol{\mu })^{\top}\boldsymbol{\Sigma }^{-1}\boldsymbol{e}_t \right] d\boldsymbol{\lambda }
\\
&\quad -\frac{1}{\sqrt{2\pi \left| \boldsymbol{\Sigma } \right|}\prod_{s=1}^N{r_s!}}\,\int_{(0,+\infty )^N}{f_{\boldsymbol{r}}\left( \boldsymbol{\lambda } \right) \cdot \left[ \frac{1}{2}\boldsymbol{e}_{s}^{\top}\boldsymbol{\Sigma }^{-1}\boldsymbol{e}_t+\frac{1}{2}\boldsymbol{e}_{t}^{\top}\boldsymbol{\Sigma }^{-1}\boldsymbol{e}_s \right] d\boldsymbol{\lambda }}
\\
&=\frac{1}{\sqrt{2\pi \left| \boldsymbol{\Sigma } \right|}\prod_{s=1}^N{r_s!}}\int_{(0,+\infty )^N}{f_{\boldsymbol{r}}\left( \boldsymbol{\lambda } \right) \cdot \left[ (\log \boldsymbol{\lambda }-\boldsymbol{\mu })^{\top}\boldsymbol{\Sigma }^{-1}\boldsymbol{e}_s\boldsymbol{e}_{t}^{\top}\boldsymbol{\Sigma }^{-1}(\log \boldsymbol{\lambda }-\boldsymbol{\mu }) \right] d\boldsymbol{\lambda }}
\\
&\quad -\frac{1}{\sqrt{2\pi \left| \boldsymbol{\Sigma } \right|}\prod_{s=1}^N{r_s!}}\,\int_{(0,+\infty )^N}{f_{\boldsymbol{r}}\left( \boldsymbol{\lambda } \right) \cdot \left[ \boldsymbol{e}_{t}^{\top}\boldsymbol{\Sigma }^{-1}\boldsymbol{e}_s \right] d\boldsymbol{\lambda }}
\\
&=\frac{1}{\sqrt{2\pi \left| \boldsymbol{\Sigma } \right|}\prod_{s=1}^N{r_s!}}\int_{(0,+\infty )^N}{f_{\boldsymbol{r}}\left( \boldsymbol{\lambda } \right) \cdot \left[ (\log \boldsymbol{\lambda }-\boldsymbol{\mu })^{\top}\boldsymbol{\Sigma }^{-1}\frac{\partial \boldsymbol{\Sigma }}{\partial \sigma _{st}}\boldsymbol{\Sigma }^{-1}(\log \boldsymbol{\lambda }-\boldsymbol{\mu }) \right] d\boldsymbol{\lambda }}
\\
&\quad -\frac{1}{\sqrt{2\pi \left| \boldsymbol{\Sigma } \right|}\prod_{s=1}^N{r_s!}}\,\int_{(0,+\infty )^N}{f_{\boldsymbol{r}}\left( \boldsymbol{\lambda } \right) \cdot tr\left( \boldsymbol{\Sigma }^{-1}\frac{\partial \boldsymbol{\Sigma }}{\partial \sigma _{st}} \right) d\boldsymbol{\lambda }}\\
&=\frac{1}{\sqrt{2\pi \left| \boldsymbol{\Sigma } \right|}\prod_{s=1}^N{r_s!}}\int_{(0,+\infty )^N}{f_{\boldsymbol{r}}\left( \boldsymbol{\lambda } \right) \cdot \left[ (\log \boldsymbol{\lambda }-\boldsymbol{\mu })^{\top}\boldsymbol{\Sigma }^{-1}\frac{\partial \boldsymbol{\Sigma }}{\partial \sigma _{st}}\boldsymbol{\Sigma }^{-1}(\log \boldsymbol{\lambda }-\boldsymbol{\mu }) \right] d\boldsymbol{\lambda }}
\\
&\quad -\frac{1}{\sqrt{2\pi \left| \boldsymbol{\Sigma } \right|}\prod_{s=1}^N{r_s!}}\,tr\left( \boldsymbol{\Sigma }^{-1}\frac{\partial \boldsymbol{\Sigma }}{\partial \sigma _{st}} \right) \int_{(0,+\infty )^N}{f_{\boldsymbol{r}}\left( \boldsymbol{\lambda } \right) d\boldsymbol{\lambda }},
\end{aligned}
$$
$$
\begin{aligned}
\frac{\partial P_{\boldsymbol{r}}}{\partial \sigma _{st}}&=\frac{1}{\sqrt{2\pi \left| \boldsymbol{\Sigma } \right|}\prod_{s=1}^N{r_s!}}\int_{(0,+\infty )^N}{f_{\boldsymbol{r}}\left( \boldsymbol{\lambda } \right) \cdot \left[ \frac{1}{2}(\log \boldsymbol{\lambda }-\boldsymbol{\mu })^{\top}\boldsymbol{\Sigma }^{-1}\frac{\partial \boldsymbol{\Sigma }}{\partial \sigma _{st}}\boldsymbol{\Sigma }^{-1}(\log \boldsymbol{\lambda }-\boldsymbol{\mu }) \right] d\boldsymbol{\lambda }}
\\
&\quad -\frac{1}{2}\frac{1}{\sqrt{2\pi \left| \boldsymbol{\Sigma } \right|}\prod_{s=1}^N{r_s!}}\,tr\left( \boldsymbol{\Sigma }^{-1}\frac{\partial \boldsymbol{\Sigma }}{\partial \sigma _{st}} \right) \int_{(0,+\infty )^N}{f_{\boldsymbol{r}}\left( \boldsymbol{\lambda } \right) d\boldsymbol{\lambda }}
\\
&=\frac{1}{2}\frac{\partial ^2P_{\boldsymbol{r}}}{\partial \mu_s\partial \mu_t}.
\end{aligned}
$$
Therefore, we have
$$
\frac{\partial P_{\boldsymbol{r}}}{\partial \sigma _{st}}=\frac{1}{2}\frac{\partial ^2P_{\boldsymbol{r}}}{\partial \mu _s\partial \mu _s}.
$$
Furthermore, if $s\ne t$, we have
$$
\begin{aligned}
\frac{\partial ^2P_{\boldsymbol{r}}}{\partial \mu _s\partial \mu _t}&=r_s\frac{\partial P_{\boldsymbol{r}}}{\partial \mu _t}-\left( r_s+1 \right) \frac{\partial P_{\boldsymbol{r}+\boldsymbol{e}_s}}{\partial \mu _t}
\\
&=r_s\left[ r_tP_{\boldsymbol{r}}-\left( r_t+1 \right) P_{\boldsymbol{r}+\boldsymbol{e}_t} \right] -\left( r_s+1 \right) \left[ r_tP_{\boldsymbol{r}+\boldsymbol{e}_s}-\left( r_t+1 \right) P_{\boldsymbol{r}+\boldsymbol{e}_s+\boldsymbol{e}_t} \right] 
\\
&=r_{s}r_{t}P_{\boldsymbol{r}}-r_s\left( r_t+1 \right) P_{\boldsymbol{r}+\boldsymbol{e}_t}-r_t\left( r_s+1 \right) P_{\boldsymbol{r}+\boldsymbol{e}_s}+\left( r_s+1 \right) \left( r_t+1 \right) P_{\boldsymbol{r}+\boldsymbol{e}_s+\boldsymbol{e}_t} ;
\end{aligned}
$$
if $s=t$, 
$$
\begin{aligned}
	\frac{\partial ^2P_{\boldsymbol{r}}}{\partial \mu _s\partial \mu _s}&=r_s\frac{\partial P_{\boldsymbol{r}}}{\partial \mu _s}-\left( r_s+1 \right) \frac{\partial P_{\boldsymbol{r}+\boldsymbol{e}_s}}{\partial \mu _s}\\
	&=r_s\left[ r_sP_{\boldsymbol{r}}-\left( r_s+1 \right) P_{\boldsymbol{r}+\boldsymbol{e}_s} \right] -\left( r_s+1 \right) \left[ \left( r_s+1 \right) P_{\boldsymbol{r}+\boldsymbol{e}_s}-\left( r_s+2 \right) P_{\boldsymbol{r}+2\boldsymbol{e}_s} \right]\\
	&=r_{s}^{2}P_{\boldsymbol{r}}-\left( r_s+1 \right) \left( 2r_s+1 \right) P_{\boldsymbol{r}+\boldsymbol{e}_s}+\left( r_s+1 \right) \left( r_s+2 \right) P_{\boldsymbol{r}+2\boldsymbol{e}_s}  .
\end{aligned}
$$
So, 
$$
\frac{\partial P_{\boldsymbol{r}}}{\partial \sigma _{st}}=\begin{cases}
	0.5\left\{ r_sr_tP_{\boldsymbol{r}}-r_s\left( r_t+1 \right) P_{\boldsymbol{r}+\boldsymbol{e}_t}-r_t\left( r_s+1 \right) P_{\boldsymbol{r}+\boldsymbol{e}_s}+\left( r_s+1 \right) \left( r_t+1 \right) P_{\boldsymbol{r}+\boldsymbol{e}_s+\boldsymbol{e}_t} \right\},    \quad s\ne t
	\\
	0.5\left\{ r_{s}^{2}P_{\boldsymbol{r}}-\left( r_s+1 \right) \left( 2r_s+1 \right) P_{\boldsymbol{r}+\boldsymbol{e}_s}+\left( r_s+1 \right) \left( r_s+2 \right) P_{\boldsymbol{r}+\boldsymbol{e}_s+\boldsymbol{e}_t} \right\},   \, \,\,\, \quad \quad \quad \quad \quad s=t.
\end{cases}
$$

\subsection*{A.3. Proof of Lemma \ref{pllemma2}}
Suppose that $\lambda$ follow lognormal distribution with parameters $\mu$ and $\sigma^{2}$, we have that for any given $k \in \mathbb{Z}^{+}$,
$$
\begin{aligned}
E\left( \lambda ^k \right) & =\int_0^{+\infty} \lambda^k\cdot \frac{1}{\sqrt{2\pi \sigma ^2}}\cdot \frac{1}{\lambda}\cdot \exp \left[ -\frac{1}{2}\left( \frac{\log \lambda -\mu}{\sigma ^2} \right) ^2 \right] d\lambda 
\\
&\overset{y=\log \lambda}{=}\int_{-\infty}^{+\infty} \exp(ky)\frac{1}{\sqrt{2\pi \sigma ^2}}\cdot \exp \left[ -\frac{1}{2}\left( \frac{y-\mu}{\sigma ^2} \right) ^2 \right] dy
\\
&\overset{x=\frac{y-\mu}{\sigma}}{=}  \int_{-\infty}^{+\infty}\exp(k\sigma x+k\mu )\frac{1}{\sqrt{2\pi}}\exp \left[ -\frac{1}{2}x^2 \right] dx
\\
&  =\exp\mathrm{(}k\mu )\cdot \int_{-\infty}^{+\infty} \frac{1}{\sqrt{2\pi}}\cdot \exp \left[ -\frac{1}{2}\left( x^2+2k\sigma x+k^2\sigma ^2 \right) +\frac{1}{2}k^2\sigma ^2 \right] dx
\\
&  =\exp\mathrm{(}k\mu )\cdot \exp \left( \frac{1}{2}k^2\sigma ^2 \right) 
\\
&  =\exp\mathrm{(}k\mu +\frac{1}{2}k^2\sigma ^2).
\end{aligned}
$$

\subsection*{A.4. Proof of Lemma \ref{pllemma3}}
Suppose that $\boldsymbol{\lambda}$ follow multivariate lognormal distribution with parameters $\boldsymbol{\mu}$ and $\mathbf{\Sigma}$, we have that for given non-negative integer vector $(k_1,  k_2,  \dots, k_{\scrS N})$,
$$
\begin{aligned}
E\left( \lambda _{1}^{k_1}\lambda _{2}^{k_2}\cdots \lambda _{\scrS N}^{k_{\scrS N}} \right) &   =\int_{(0,+\infty )^N}{\lambda _{1}^{k_1}\lambda _{2}^{k_2}\cdots \lambda _{\scrS N}^{k_{\scrS N}}}\cdot \frac{1}{\sqrt{2\pi \left| \boldsymbol{\Sigma } \right|}}\cdot \frac{1}{\prod_{s=1}^N{\lambda _i}}\cdot \exp \left[ -\frac{1}{2}(\log \boldsymbol{\lambda }-\boldsymbol{\mu })^{\top}\boldsymbol{\Sigma }^{-1}(\log \boldsymbol{\lambda }-\boldsymbol{\mu }) \right] d\boldsymbol{\lambda }
\\
&   \text{Let} \, \boldsymbol{y}=\log \boldsymbol{\lambda }
\\
& =\int_{(-\infty ,+\infty )^N}{\exp \left( k_1y_1+k_2y_2+\cdots +k_{\scrS N}y_{\scrS N} \right)}\cdot \frac{1}{\sqrt{2\pi \left| \boldsymbol{\Sigma } \right|}}\cdot \exp \left[ -\frac{1}{2}(\boldsymbol{y}-\boldsymbol{\mu })^{\top}\boldsymbol{\Sigma }^{-1}(\boldsymbol{y}-\boldsymbol{\mu }) \right] d\boldsymbol{y}
\\
&   \text{Let}\,  \boldsymbol{A}=k_1\boldsymbol{e}_1+k_2\boldsymbol{e}_2+\cdots +k_{\scrS N}\boldsymbol{e}_{\scrS N}
\\
&   =\int_{(-\infty ,+\infty )^N}{\exp \left( \boldsymbol{A}^{\prime}\boldsymbol{y} \right)}\cdot \frac{1}{\sqrt{2\pi \left| \boldsymbol{\Sigma } \right|}}\cdot \exp \left[ -\frac{1}{2}(\boldsymbol{y}-\boldsymbol{\mu })^{\top}\boldsymbol{\Sigma }^{-1}(\boldsymbol{y}-\boldsymbol{\mu }) \right] d\boldsymbol{y}
\\
&   \text{Let}\,  \boldsymbol{y}=\boldsymbol{U\varLambda }^{\frac{1}{2}}\boldsymbol{x}+\boldsymbol{\mu }
\\
&   =\int_{(-\infty ,+\infty )^N}{\exp \left( \boldsymbol{A}^{\prime}\boldsymbol{U\varLambda }^{\frac{1}{2}}\boldsymbol{x}+\boldsymbol{A}^{\prime}\boldsymbol{\mu } \right)}\cdot \frac{1}{\sqrt{2\pi}}\cdot \exp \left[ -\frac{1}{2}\boldsymbol{x}^{\prime}\boldsymbol{x} \right] d\boldsymbol{x}
\\
&   =\exp \left( \boldsymbol{A}^{\prime}\boldsymbol{\mu } \right) \cdot \int_{-\infty}^{+\infty}{\frac{1}{\sqrt{2\pi}}\cdot \exp}\left[ -\frac{1}{2}\boldsymbol{x}^{\top}\boldsymbol{x}+\boldsymbol{A}^{\prime}\boldsymbol{U\varLambda }^{\frac{1}{2}}\boldsymbol{x} \right] d\boldsymbol{x}
\\
&   =\exp \left( \boldsymbol{A}^{\prime}\boldsymbol{\mu } \right) \cdot \int_{-\infty}^{+\infty}{\frac{1}{\sqrt{2\pi}}}\cdot \exp \left[ -\frac{1}{2}\boldsymbol{x}^{\top}\boldsymbol{x}+\frac{1}{2}\boldsymbol{A}^{\prime}\boldsymbol{U\varLambda }^{\frac{1}{2}}\boldsymbol{x}+\frac{1}{2}\boldsymbol{x}^{\prime}\boldsymbol{\varLambda }^{\frac{1}{2}}\boldsymbol{UA} \right] d\boldsymbol{x}
\\
& =\exp \left( \boldsymbol{A}^{\prime}\boldsymbol{\mu } \right) \cdot \int_{-\infty}^{+\infty}{  \frac{1}{\sqrt{2\pi}}}\cdot \exp \left[ -\frac{1}{2}\left( \boldsymbol{x}-\boldsymbol{A}^{\prime}\boldsymbol{U\varLambda }^{\frac{1}{2}}\boldsymbol{x} \right) ^{\top}\left( \boldsymbol{x}-\boldsymbol{A}^{\prime}\boldsymbol{U\varLambda }^{\frac{1}{2}}\boldsymbol{x} \right) \right] 
\\
&\qquad\qquad\qquad\qquad -\exp \left[ \frac{1}{2}\boldsymbol{A}^{\prime}\boldsymbol{U\varLambda }^{\frac{1}{2}}\boldsymbol{\varLambda }^{\frac{1}{2}}\boldsymbol{UA} \right] d\boldsymbol{x}
\\
&   =\exp \left( \boldsymbol{A}^{\prime}\boldsymbol{\mu }+\frac{1}{2}\boldsymbol{A}^{\prime}\boldsymbol{\Sigma A} \right) 
\\
&   =\exp \left( k_1\mu _1+k_2\mu _2+\cdots +k_{\scrS N}\mu _{\scrS N}+\frac{1}{2}\sum_{i=1}^{\scrS N} {\sum_{j=1}^{\scrS N} {k_i}}k_j\sigma _{ij} \right),
\end{aligned}
$$
where $\boldsymbol{\Sigma}=\boldsymbol{U} \boldsymbol{\varLambda} \boldsymbol{U}^{\prime}$, where $\boldsymbol{U}$ is an orthogonal matrix and $\boldsymbol{\varLambda}$ is a symmetric matrix.

\subsection*{A.5. Proof of Lemma \ref{pllemma4}}
For $\forall  k \in \left\{ 0 \right\}  \cup  \mathbb{Z}^{+}$
$$
\begin{aligned}
ER^{k+1}&  =\sum_{\scrS   R=0}^{\infty}{r^{k+1}}\cdot \frac{\lambda ^r}{r!}e^{-\lambda}
\\
&  =\lambda \sum_{\scrS   R=1}^{\infty}{r^k}\cdot \frac{\lambda ^{r-1}}{\left( r-1 \right) !}e^{-\lambda}
\\
&   =\lambda \sum_{\scrS   R=1}^{\infty}{\left[ \left( r-1 \right) +1 \right] ^k}\cdot \frac{\lambda ^{r-1}}{\left( r-1 \right) !}e^{-\lambda}
\\
&  =\lambda \sum_{\scrS   R=1}^{\infty}{(}r-1)^k\cdot \frac{\lambda ^{r-1}}{\left( r-1 \right) !}e^{-\lambda}+\lambda \sum_{\scrS   R=1}^{\infty}{\sum_{m=0}^{k-1}{C_{k}^{m}}}(r-1)^m\cdot \frac{\lambda ^{r-1}}{\left( r-1 \right) !}e^{-\lambda}
\\
&   =\lambda ER^k+\lambda \sum_{\scrS   R=1}^{\infty}{\sum_{m=0}^{k-1}{C_{k}^{m}}}(r-1)^m\cdot \frac{\lambda ^{r-1}}{\left( r-1 \right) !}e^{-\lambda}.
\end{aligned}
$$
Furthermore, $ER^k=\sum_{\scrS   R=0}^{\infty}{r^k}\cdot \frac{\lambda ^r}{r!}e^{-\lambda}$, from which we can conclude that
$$
\begin{aligned}
	\frac{dER^k}{d\lambda}&=\sum_{\scrS   R=0}^{\infty}{r^k}\cdot \frac{\lambda ^{r-1}}{(r-1)!}e^{-\lambda}-\sum_{\scrS   R=0}^{\infty}{r^k}\cdot \frac{\lambda ^r}{r!}e^{-\lambda}\\
	&=\sum_{\scrS   R=1}^{\infty}{[}(r-1)+1]^k\cdot \frac{\lambda ^{r-1}}{(r-1)!}e^{-\lambda}-ER^k\\
	&=ER^k+\sum_{\scrS   R=1}^{\infty}{\sum_{m=0}^{k-1}{C_{k}^{m}}}(r-1)^m\cdot \frac{\lambda ^{r-1}}{(r-1)!}e^{-\lambda}-E\varepsilon ^{k.}\\
	&=\sum_{m=0}^{k-1}{\sum_{\scrS   R=1}^{\infty}{C_{k}^{m}}}(r-1)^m\cdot \frac{\lambda ^{r-1}}{(r-1)!}e^{-\lambda}\\
	&=\sum_{m=0}^{k-1}{C_{k}^{m}}\sum_{\scrS   R=1}^{\infty}{(}r-1)^m\cdot \frac{\lambda ^{r-1}}{(r-1)!}e^{-\lambda}\\
	&=\sum_{m=0}^{k-1}{C_{k}^{m}}\cdot ER^m   .
\end{aligned}
$$
Finally, we obtain
$$
E R^{k+1}=\lambda E R^k+\lambda \frac{d E R^k}{d \lambda}
$$
which can be proved using mathematical induction.

\subsection*{A.6. Proof of Theorem \ref{pltheorem1}}
Without loss of generality, let $\forall \left( k_1,k_2,\dots ,k_{\scrS N} \right) \in \{ \left( k_1,k_2,\dots ,k_{\scrS N} \right) :k_1+k_2+\cdots +k_{\scrS N}=k, k_s\in \left\{ 0,1,2,\dots ,k \right\} , s\in \left\{ 1,2,\dots ,N \right\} \} $
$$
\begin{aligned}
E\boldsymbol{R}_{k_1,k_2,\dots ,k_{\scrS N}}^{k}&=E\left( R^{k_1}R^{k_2}\cdots R^{k_{\scrS N}} \right) 
\\
&=E\left[ E\left( R^{k_1}R^{k_2}\cdots R^{k_{\scrS N}}|\boldsymbol{\lambda } \right) \right] 
\end{aligned}
$$
and 
$$
E\left( R^{k_1}R^{k_2}\cdots R^{k_{\scrS N}}|\boldsymbol{\lambda } \right) =E\left( R^{k_1}|\boldsymbol{\lambda } \right) E\left( R^{k_2}|\boldsymbol{\lambda } \right) \cdots E\left( R^{k_{\scrS N}}|\boldsymbol{\lambda } \right) .
$$
Therefore, we can conclude from the above equation that $E\left( R^{k_1}R^{k_2}\cdots R^{k_{\scrS N}}|\boldsymbol{\lambda } \right)$ is a polynomial of at most order $k$ with respect to $\lambda_{1}$, $\lambda_{2}$, $\dots$, $\lambda_{\scrS N}$.
Furthermore, for a given $\left( k_1,k_2,\dots ,k_{\scrS N} \right)$, we have
$$
E\left( \lambda _{1}^{k_1}\lambda _{2}^{k_2}\cdots \lambda _{\scrS N}^{k_{\scrS N}} \right) =\exp \left[ ku_1+\cdots +j_{\scrS N}u_{\scrS N}+\frac{1}{2}\sum_{i=1}^n{\sum_{j=1}^n{k_i}}k_j\sigma _{ij} \right] < \infty.
$$

\subsection*{Proof of Proposition \ref{plprop3}}

The proof steps are similar to \citet{liu2021}. Assuming Hypothesis \ref{genhypo1} holds, we can represent the multivariate distribution of process \eqref{geneq1} using the distribution of the multivariate increment process as follows:
$$
\boldsymbol{X}_t\overset{d}{=}\boldsymbol{A}^t\circ \boldsymbol{X}_0+\sum_{i=0}^{t-1}{\boldsymbol{A}^i}\circ \boldsymbol{R}_{t-i}. 
$$
Therefore, we have
$$
\boldsymbol{X}_t\overset{d}{=}\sum_{i=0}^{\infty}{\boldsymbol{A}^i}\circ \boldsymbol{R}_{t-i},
$$
which implies that process \eqref{geneq1} has a unique stationary solution. From Hypothesis \ref{plhypo2} and Theorem \ref{pltheorem1}, we know that $E( \boldsymbol{R}_{t}^{k} ) <\infty $, hence for any positive integer $k$, we have $E(\boldsymbol{X}_t^k )<\infty$.

\subsection*{A.7. Proof of Lemma \ref{gllemma1}}
$$
\begin{aligned}
	P(\boldsymbol{R}=\boldsymbol{r})&=\int_{(0, 1)^N}{\prod_{s=1}^N{p_s\left( 1-p_s \right) ^{r_s}}\cdot \frac{1}{\sqrt{2\pi \left| \mathbf{\Sigma } \right|}}\cdot \frac{1}{\prod_{s=1}^N{p_s\left( 1-p_s \right)}}\cdot \exp \left[ -\frac{1}{2}(\mathrm{logit} \boldsymbol{p}-\boldsymbol{\mu })^{\top}\mathbf{\Sigma }^{-1}(\mathrm{logit} \boldsymbol{p}-\boldsymbol{\mu }) \right] d\boldsymbol{p}}\\
	&=\frac{1}{\sqrt{2\pi \left| \mathbf{\Sigma } \right|}}\int_{(0,  1)^N}{\prod_{s=1}^N{\left( 1-p_s \right) ^{r_s-1}}\cdot \exp \left[ -\frac{1}{2}(\mathrm{logit} \boldsymbol{p}-\boldsymbol{\mu })^{\top}\mathbf{\Sigma }^{-1}(\mathrm{logit} \boldsymbol{p}-\boldsymbol{\mu }) \right] d\boldsymbol{p}}\\
\end{aligned}
$$
$$
\begin{aligned}
	\frac{\partial P_{\boldsymbol{r}}}{\partial \mu _s}=\frac{1}{\sqrt{2\pi \left| \mathbf{\Sigma } \right|}}
	&\int_{(0, 1)^N}{\prod_{s=1}^N{p_s\left( 1-p_s \right) ^{r_s-1}}\cdot \exp \left[ -\frac{1}{2}(\mathrm{logit}   \boldsymbol{p}-\boldsymbol{\mu })^{\top}\mathbf{\Sigma }^{-1}(\mathrm{logit}   \boldsymbol{p} - \boldsymbol{\mu }) \right]}
	\\
	&\cdot \left[ \frac{1}{2}\boldsymbol{e}_{s}^{\top}\mathbf{\Sigma }^{-1}(\mathrm{logit} \boldsymbol{p}-\boldsymbol{\mu })
	+
	\frac{1}{2}(\mathrm{logit} \boldsymbol{p}-\boldsymbol{\mu })^{\top}\mathbf{\Sigma }^{-1}\boldsymbol{e}_s \right] d\boldsymbol{p}.
\end{aligned}
$$
Here, $\boldsymbol{e}_s$ is an N-dimensional unit vector with the $s$-th element equal to 1.

Let us assume that
$$
f_{\boldsymbol{r}}\left( \boldsymbol{\lambda } \right) =\prod_{s=1}^N{p_s\left( 1-p_s \right) ^{r_s}}\cdot \exp \left[ -\frac{1}{2}(\mathrm{logit}   \boldsymbol{p}-\boldsymbol{\mu })^{\top}\mathbf{\Sigma }^{-1}(\mathrm{logit}  \boldsymbol{p}-\boldsymbol{\mu }) \right] ,
$$
$$
\begin{aligned}
	g_{\boldsymbol{r}}\left( \boldsymbol{\lambda } \right) &=\prod_{s=1}^N{p_s\left( 1-p_s \right) ^{r_s}}\cdot \exp \left[ -\frac{1}{2}(\mathrm{logit}   \boldsymbol{p} - \boldsymbol{\mu })^{\top}\mathbf{\Sigma }^{-1}(\mathrm{logit}   \boldsymbol{p} -\boldsymbol{\mu }) \right]\\
	&\cdot \left[ \frac{1}{2}\boldsymbol{e}_{s}^{\top}\mathbf{\Sigma }^{-1}(\mathrm{logit}   \boldsymbol{p} -\boldsymbol{\mu })+\frac{1}{2}(\mathrm{logit}   \boldsymbol{p} -\boldsymbol{\mu })^{\top}\mathbf{\Sigma }^{-1}\boldsymbol{e}_s \right] .
\end{aligned}
$$
Therefore, we have
$$
P(\boldsymbol{R}=\boldsymbol{r})=\frac{1}{\sqrt{2\pi \left| \mathbf{\Sigma } \right|}}\int_{(0, 1 )^N}{f_{\boldsymbol{r}}\left( \boldsymbol{p } \right) d\boldsymbol{p }},
$$
$$
\frac{\partial P_{\boldsymbol{r}}}{\partial \mu _s}=\frac{1}{\sqrt{2\pi \left| \mathbf{\Sigma } \right|}}\int_{(0, 1  )^N}{g_{\boldsymbol{r}}\left( \boldsymbol{p} \right) d \boldsymbol{p} }.
$$
Differentiating $f_{\boldsymbol{r}}(\boldsymbol{p})$ with respect to $p_s$, we obtain
$$
\begin{aligned}
	\frac{\partial f_{\boldsymbol{r}}(\boldsymbol{p})}{\partial p_s}&=-\frac{\left( r_s-1 \right)}{1-p_s}\prod_{s=1}^N{p_s\left( 1-p_s \right) ^{r_s}}\cdot \exp \left[ -\frac{1}{2}(\mathrm{logit}\boldsymbol{p}-\boldsymbol{\mu })^{\top}\mathbf{\Sigma }^{-1}(\mathrm{logit}\boldsymbol{p}-\boldsymbol{\mu }) \right]\\
	&\quad \quad -\frac{1}{p_s\left( 1-p_s \right)}\prod_{s=1}^n{\lambda _{s}^{r_s-1}}\cdot e^{-r_s}\cdot \exp \left[ -\frac{1}{2}(\mathrm{logit}\boldsymbol{p}-\boldsymbol{\mu })^{\top}\Sigma ^{-1}(\mathrm{logit}\boldsymbol{p}-\boldsymbol{\mu }) \right]\\
	&\quad \quad \quad \quad \,\cdot \left[ \frac{1}{2}\boldsymbol{e}_{s}^{\top}\mathbf{\Sigma }^{-1}(\mathrm{logit}\boldsymbol{p}-\boldsymbol{\mu })+\frac{1}{2}(\mathrm{logit}\boldsymbol{p}-\boldsymbol{\mu })^{\top}\mathbf{\Sigma }^{-1}\boldsymbol{e}_s \right]\\
	&=-\frac{\left( r_s-1 \right)}{1-p_s}f_{\boldsymbol{r}}(\boldsymbol{p})-\frac{1}{p_s\left( 1-p_s \right)}g_{\boldsymbol{r}}(\boldsymbol{\lambda }).
\end{aligned}
$$

Furthermore, from the definition of $f_{\boldsymbol{r}}(\boldsymbol{p})$, we have
$$
p_sf_{\boldsymbol{r}}\left( \boldsymbol{p} \right) =f_{\boldsymbol{r}}\left( \boldsymbol{p} \right) -f_{\boldsymbol{r}+\boldsymbol{e}_s}\left( \boldsymbol{p} \right) .
$$

Furthermore, we can obtain that
$$
\begin{aligned}
	g_{\boldsymbol{r}}(\boldsymbol{p})&=-\left( r_s-1 \right) p_sf_{\boldsymbol{r}}(\boldsymbol{p})-p_s\left( 1-p_s \right) \frac{\partial f_{\boldsymbol{r}}(\boldsymbol{p})}{\partial p_s}\\
	&=-\left( r_s-1 \right) f_{\boldsymbol{r}}(\boldsymbol{p})+\left( r_s-1 \right) f_{\boldsymbol{r}+\boldsymbol{e}_s}(\boldsymbol{p})-p_s\left( 1-p_s \right) \frac{\partial f_{\boldsymbol{r}}(\boldsymbol{p})}{\partial p_s}.\\
\end{aligned}
$$

Therefore, according to \citet{izsak2008}, we can obtain the following by using the method of integration by parts
$$
\begin{aligned}
\frac{1}{\sqrt{2\pi \left| \mathbf{\Sigma } \right|}}\int_{(0,1)^N}{g_{\boldsymbol{r}}(\boldsymbol{p})d\boldsymbol{p}}&=-\left( r_s-1 \right) P_{\boldsymbol{r}}+\left( r_s-1 \right) P_{\boldsymbol{r}+\boldsymbol{e}_s}
\\
&\quad -\frac{1}{\sqrt{2\pi \left| \mathbf{\Sigma } \right|}}\int_{(0,1)^N}{p_s\left( 1-p_s \right) \frac{\partial f_{\boldsymbol{r}}(\boldsymbol{p})}{\partial p_s}d\boldsymbol{p}}
\\
&=-\left( r_s-1 \right) P_{\boldsymbol{r}}+\left( r_s-1 \right) P_{\boldsymbol{r}+\boldsymbol{e}_s}
\\
&\quad +\frac{1}{\sqrt{2\pi \left| \mathbf{\Sigma } \right|}}\int_{(0,1)^N}{\left( 1-2p_s \right) f_{\boldsymbol{r}}(\boldsymbol{p})d\boldsymbol{p}}
\\
&=-\left( r_s-1 \right) P_{\boldsymbol{r}}+\left( r_s-1 \right) P_{\boldsymbol{r}+\boldsymbol{e}_s}+\frac{1}{\sqrt{2\pi \left| \mathbf{\Sigma } \right|}}\int_{(0,1)^N}{f_{\boldsymbol{r}}(\boldsymbol{p})d\boldsymbol{p}}
\\
&\quad -2\frac{1}{\sqrt{2\pi \left| \mathbf{\Sigma } \right|}}\int_{(0,1)^N}{p_sf_{\boldsymbol{r}}(\boldsymbol{p})d\boldsymbol{p}}
\\
&=-\left( r_s-2 \right) P_{\boldsymbol{r}}+\left( r_s-1 \right) P_{\boldsymbol{r}+\boldsymbol{e}_s}-2P_{\boldsymbol{r}}+2P_{\boldsymbol{r}+\boldsymbol{e}_s}
\\
&=\left( r_s+1 \right) P_{\boldsymbol{r}+\boldsymbol{e}_s}-r_sP_{\boldsymbol{r}} .
\end{aligned}
$$

Therefore, we can get 
$$
\frac{\partial P_{\boldsymbol{r}}}{\partial \mu _s}=\left( r_s+1 \right) P_{\boldsymbol{r}+\boldsymbol{e}_s}-r_sP_{\boldsymbol{r}}  .
$$

In addition, we can also derive 
$$
\begin{aligned}
	\frac{\partial ^2P_{\boldsymbol{r}}}{\partial \mu _s\partial \mu _t}&=\frac{1}{\sqrt{2\pi \left| \mathbf{\Sigma } \right|}}\int_{(0,1)^N}{f_{\boldsymbol{r}}\left( \boldsymbol{p} \right) \cdot \left[ \frac{1}{2}\boldsymbol{e}_{s}^{\top}\mathbf{\Sigma }^{-1}(\mathrm{logit}\boldsymbol{p}-\boldsymbol{\mu })+\frac{1}{2}(\mathrm{logit}\boldsymbol{p}-\boldsymbol{\mu })^{\top}\mathbf{\Sigma }^{-1}\boldsymbol{e}_s \right]}\\
	&\qquad \qquad \qquad \qquad \qquad \quad \,\cdot \left[ \frac{1}{2}\boldsymbol{e}_{t}^{\top}\mathbf{\Sigma }^{-1}(\mathrm{logit}\boldsymbol{p}-\boldsymbol{\mu })+\frac{1}{2}(\mathrm{logit}\boldsymbol{p}-\boldsymbol{\mu })^{\top}\mathbf{\Sigma }^{-1}\boldsymbol{e}_t \right] d\boldsymbol{p}\\
	&\quad -\frac{1}{\sqrt{2\pi \left| \mathbf{\Sigma } \right|}}\,\int_{(0,1)^N}{f_{\boldsymbol{r}}\left( \boldsymbol{p} \right) \cdot \left[ \frac{1}{2}\boldsymbol{e}_{s}^{\top}\mathbf{\Sigma }^{-1}\boldsymbol{e}_t+\frac{1}{2}\boldsymbol{e}_{t}^{\top}\mathbf{\Sigma }^{-1}\boldsymbol{e}_s \right] d\boldsymbol{p}}\\
	&=\frac{1}{\sqrt{2\pi \left| \mathbf{\Sigma } \right|}}\int_{(0,1)^N}{f_{\boldsymbol{r}}\left( \boldsymbol{p} \right) \cdot \left[ (\mathrm{logit}\boldsymbol{p}-\boldsymbol{\mu })^{\top}\mathbf{\Sigma }^{-1}\boldsymbol{e}_s\boldsymbol{e}_{t}^{\top}\mathbf{\Sigma }^{-1}(\mathrm{logit}\boldsymbol{p}-\boldsymbol{\mu }) \right] d\boldsymbol{p}}\\
	&\quad -\frac{1}{\sqrt{2\pi \left| \mathbf{\Sigma } \right|}}\,\int_{(0,1)^N}{f_{\boldsymbol{r}}\left( \boldsymbol{p} \right) \cdot \left[ \boldsymbol{e}_{t}^{\top}\mathbf{\Sigma }^{-1}\boldsymbol{e}_s \right] d\boldsymbol{p}}\\
	&=\frac{1}{\sqrt{2\pi \left| \mathbf{\Sigma } \right|}}\int_{(0,1)^N}{f_{\boldsymbol{r}}\left( \boldsymbol{p} \right) \cdot \left[ (\mathrm{logit}\boldsymbol{p}-\boldsymbol{\mu })^{\top}\mathbf{\Sigma }^{-1}\frac{\partial \mathbf{\Sigma }}{\partial \sigma _{st}}\mathbf{\Sigma }^{-1}(\mathrm{logit}\boldsymbol{p}-\boldsymbol{\mu }) \right] d\boldsymbol{p}}\\
	&\quad -\frac{1}{\sqrt{2\pi \left| \mathbf{\Sigma } \right|}}\,\int_{(0,1)^N}{f_{\boldsymbol{r}}\left( \boldsymbol{p} \right) \cdot tr\left( \mathbf{\Sigma }^{-1}\frac{\partial \mathbf{\Sigma }}{\partial \sigma _{st}} \right) d\boldsymbol{p}}\\
	&=\frac{1}{\sqrt{2\pi \left| \mathbf{\Sigma } \right|}}\int_{(0,1)^N}{f_{\boldsymbol{r}}\left( \boldsymbol{p} \right) \cdot \left[ (\mathrm{logit}\boldsymbol{p}-\boldsymbol{\mu })^{\top}\mathbf{\Sigma }^{-1}\frac{\partial \mathbf{\Sigma }}{\partial \sigma _{st}}\mathbf{\Sigma }^{-1}(\mathrm{logit}\boldsymbol{p}-\boldsymbol{\mu }) \right] d\boldsymbol{p}}\\
	&\quad -\frac{1}{\sqrt{2\pi \left| \mathbf{\Sigma } \right|}}\,tr\left( \mathbf{\Sigma }^{-1}\frac{\partial \mathbf{\Sigma }}{\partial \sigma _{st}} \right) \int_{(0,1)^N}{f_{\boldsymbol{r}}\left( \boldsymbol{p} \right) d\boldsymbol{p}} ,
\end{aligned}
$$
$$
\begin{aligned}
	\frac{\partial P_{\boldsymbol{r}}}{\partial \sigma _{st}}&=\frac{1}{\sqrt{2\pi \left| \mathbf{\Sigma } \right|}}\int_{(0,1)^N}{f_{\boldsymbol{r}}\left( \boldsymbol{p} \right) \cdot \left[ \frac{1}{2}(\mathrm{logit}\boldsymbol{p}-\boldsymbol{\mu })^{\top}\mathbf{\Sigma }^{-1}\frac{\partial \mathbf{\Sigma }}{\partial \sigma _{st}}\mathbf{\Sigma }^{-1}(\mathrm{logit}\boldsymbol{p}-\boldsymbol{\mu }) \right] d\boldsymbol{p}}\\
	&\quad -\frac{1}{2}\frac{1}{\sqrt{2\pi \left| \mathbf{\Sigma } \right|}}\,tr\left( \mathbf{\Sigma }^{-1}\frac{\partial \mathbf{\Sigma }}{\partial \sigma _{st}} \right) \int_{(0,1)^N}{f_{\boldsymbol{r}}\left( \boldsymbol{p} \right) d\boldsymbol{p}}\\
	&=\frac{1}{2}\frac{\partial ^2P_{\boldsymbol{r}}}{\partial \mu _s\partial \mu _t}  .
\end{aligned}
$$

Therefore, we can conclude the following.
$$
\frac{\partial P_{\boldsymbol{r}}}{\partial \sigma _{st}}=\frac{1}{2}\frac{\partial ^2P_{\boldsymbol{r}}}{\partial \mu _s\partial \mu _s}   . 
$$
Furthermore, it can be obtained that if $s \neq t$,
$$
\begin{aligned}
	\frac{\partial ^2P_{\boldsymbol{r}}}{\partial \mu _s\partial \mu _t}&=\left( r_s+1 \right) \frac{\partial P_{\boldsymbol{r}+\boldsymbol{e}_s}}{\partial \mu _t}-r_s\frac{\partial P_{\boldsymbol{r}}}{\partial \mu _t}\\
	&=\left( r_s+1 \right) \left[ \left( r_t+1 \right) P_{\boldsymbol{r}+\boldsymbol{e}_s+\boldsymbol{e}_t}-r_tP_{\boldsymbol{r}+\boldsymbol{e}_s} \right] -r_s\left[ \left( r_t+1 \right) P_{\boldsymbol{r}+\boldsymbol{e}_t}-r_tP_{\boldsymbol{r}} \right]\\
	&=r_sr_tP_{\boldsymbol{r}}-r_s\left( r_t+1 \right) P_{\boldsymbol{r}+\boldsymbol{e}_t}-r_t\left( r_s+1 \right) P_{\boldsymbol{r}+\boldsymbol{e}_s}+\left( r_s+1 \right) \left( r_t+1 \right) P_{\boldsymbol{r}+\boldsymbol{e}_s+\boldsymbol{e}_t};
\end{aligned}
$$
if $s=t$, 
$$
\begin{aligned}
	\frac{\partial ^2P_{\boldsymbol{r}}}{\partial \mu _s\partial \mu _s}&=\left( r_s+1 \right) \frac{\partial P_{\boldsymbol{r}+\boldsymbol{e}_s}}{\partial \mu _s}-r_s\frac{\partial P_{\boldsymbol{r}}}{\partial \mu _s}\\
	&=\left( r_s+1 \right) \left[ \left( r_s+2 \right) P_{\boldsymbol{r}+2\boldsymbol{e}_s}-\left( r_s+1 \right) P_{\boldsymbol{r}+\boldsymbol{e}_s} \right] -r_s\left[ \left( r_s+1 \right) P_{\boldsymbol{r}+\boldsymbol{e}_s}-r_sP_{\boldsymbol{r}} \right]\\
	&=r_{s}^{2}P_{\boldsymbol{r}}-\left( r_s+1 \right) \left( 2r_s+1 \right) P_{\boldsymbol{r}+\boldsymbol{e}_s}+\left( r_s+1 \right) \left( r_s+2 \right) P_{\boldsymbol{r}+2\boldsymbol{e}_s}  .
\end{aligned}
$$
So, 
$$
\frac{\partial P_{\boldsymbol{r}}}{\partial \sigma _{st}}=\begin{cases}
	0.5\left\{ r_sr_tP_{\boldsymbol{r}}-r_s\left( r_t+1 \right) P_{\boldsymbol{r}+\boldsymbol{e}_t}-r_t\left( r_s+1 \right) P_{\boldsymbol{r}+\boldsymbol{e}_s}+\left( r_s+1 \right) \left( r_t+1 \right) P_{\boldsymbol{r}+\boldsymbol{e}_s+\boldsymbol{e}_t} \right\}  \quad s\ne t\\
	0.5\left\{ r_{s}^{2}P_{\boldsymbol{r}}-\left( r_s+1 \right) \left( 2r_s+1 \right) P_{\boldsymbol{r}+\boldsymbol{e}_s}+\left( r_s+1 \right) \left( r_s+2 \right) P_{\boldsymbol{r}+\boldsymbol{e}_s+\boldsymbol{e}_t} \right\}     \,\,\,\, \quad \quad \quad \quad \quad s=t  .
\end{cases}
$$

\subsection*{A.8. Proof of Lemma \ref{gllemma2}}
$$
\begin{aligned}
	E\left( \frac{1}{p^k} \right) &=\int_0^1{\left[ \frac{1}{p} \right]}^k\cdot \frac{1}{\sqrt{2\pi \sigma ^2}}\cdot \frac{1}{p\left( 1-p \right)}\cdot \exp \left[ -\frac{1}{2}\left( \frac{\mathrm{logit}p-\mu}{\sigma ^2} \right) ^2 \right] dp\\
	&\overset{y=\mathrm{logit}p}{=}\int_{-\infty}^{+\infty}{\left[ 1+\exp \left( -y \right) \right] ^k}\frac{1}{\sqrt{2\pi \sigma ^2}}\cdot \exp \left[ -\frac{1}{2}\left( \frac{y-\mu}{\sigma ^2} \right) ^2 \right] dy\\
	&=\int_{-\infty}^{+\infty}{\sum_{i=0}^k{C_{k}^{i}\exp \left( -iy \right)}}\frac{1}{\sqrt{2\pi \sigma ^2}}\cdot \exp \left[ -\frac{1}{2}\left( \frac{y-\mu}{\sigma ^2} \right) ^2 \right] dy\\
	&=\sum_{i=0}^k{C_{k}^{i}}\int_{-\infty}^{+\infty}{\exp \left( -iy \right)}\frac{1}{\sqrt{2\pi \sigma ^2}}\cdot \exp \left[ -\frac{1}{2}\left( \frac{y-\mu}{\sigma ^2} \right) ^2 \right] dy\\
	&\overset{x=\frac{y-\mu}{\sigma}}{=}\sum_{i=0}^k{C_{k}^{i}}\int_{-\infty}^{+\infty}{\exp \left( -i\sigma x-i\mu \right)}\frac{1}{\sqrt{2\pi}}\exp \left[ -\frac{1}{2}x^2 \right] dx\\
	&=\sum_{i=0}^k{C_{k}^{i}}\cdot \exp\mathrm{(}-i\mu )\cdot \int_{-\infty}^{+\infty}{\frac{1}{\sqrt{2\pi}}}\cdot \exp \left[ -\frac{1}{2}\left( x^2+2i\sigma x+i^2\sigma ^2 \right) +\frac{1}{2}i^2\sigma ^2 \right] dx\\
	&=\sum_{i=0}^k{C_{k}^{i}}\exp\mathrm{(}-i\mu )\cdot \exp \left( \frac{1}{2}i^2\sigma ^2 \right)\\
	&=\sum_{i=0}^k{C_{k}^{i}\exp \left( -i\mu +\frac{1}{2}i^2\sigma ^2 \right)}.
\end{aligned}
$$

\subsection*{A.9. Proof of Lemma \ref{gllemma3}}
$$
\begin{aligned}
	&\quad  E\left( \left( \frac{1}{p_1} \right) ^{k_1}\left( \frac{1}{p_2} \right) ^{k_2}\cdots \left( \frac{1}{p_{\scrS N}} \right) ^{k_{\scrS N}} \right)
	\\
	 &=\int_{(0,1)^N}{\left( \frac{1}{p_1} \right) ^{k_1}\left( \frac{1}{p_2} \right) ^{k_2}\cdots \left( \frac{1}{p_{\scrS N}} \right) ^{k_{\scrS N}}}\cdot \frac{1}{\sqrt{2\pi \left| \mathbf{\Sigma } \right|}}\cdot \frac{1}{\prod_{s=1}^N{p_s\left( 1-p_s \right)}} 
	\\
	&\quad   \cdot \exp \left[ -\frac{1}{2}(\mathrm{logit}\boldsymbol{p}-\boldsymbol{\mu })^{\top}\mathbf{\Sigma }^{-1}(\mathrm{logit}\boldsymbol{p}-\boldsymbol{\mu }) \right] d\boldsymbol{p}
	\\
	&\text{Let}\,\boldsymbol{y}=\mathrm{logit}\boldsymbol{p}
	\\
	&=\int_{(-\infty ,+\infty )^N}{\left[ 1+\exp \left( -y_1 \right) \right] ^{k_1}\cdots \left[ 1+\exp \left( -y_{\scrS N} \right) \right] ^{k_{\scrS N}}}
	\\
	&\quad  \cdot \frac{1}{\sqrt{2\pi \left| \mathbf{\Sigma } \right|}}\cdot \exp \left[ -\frac{1}{2}(\boldsymbol{y}-\boldsymbol{\mu })^{\top}\mathbf{\Sigma }^{-1}(\boldsymbol{y}-\boldsymbol{\mu }) \right] d\boldsymbol{y}
	\\
	&=\sum_{i_1=0}^{k_1}{\sum_{i_2=0}^{k_2}{\cdots \sum_{i_{\scrS N}=0}^{k_{\scrS N}}{C_{k_1}^{i_1}C_{k_2}^{i_2}\cdots C_{k_{\scrS N}}^{i_{\scrS N}}}}}\int_{(-\infty ,+\infty )^N}{\exp \left( -\sum_{p=1}^N{i_py_p} \right)}\cdot \frac{1}{\sqrt{2\pi \left| \mathbf{\Sigma } \right|}}
	\\
	&\quad  \cdot \exp \left[ -\frac{1}{2}(\boldsymbol{y}-\boldsymbol{\mu })^{\top}\mathbf{\Sigma }^{-1}(\boldsymbol{y}-\boldsymbol{\mu }) \right] d\boldsymbol{y}
	\\
	&\text{Let}\,\boldsymbol{A}_{i_1,i_2,\dots ,i_{\scrS N}}=i_1\boldsymbol{e}_1+i_2\boldsymbol{e}_2+\cdots +i_{\scrS N}\boldsymbol{e}_{\scrS N}
	\\
	&=\sum_{i_1=0}^{k_1}{\sum_{i_2=0}^{k_2}{\cdots \sum_{i_{\scrS N}=0}^{k_{\scrS N}}{C_{k_1}^{i_1}C_{k_2}^{i_2}\cdots C_{k_{\scrS N}}^{i_{\scrS N}}}}}\int_{(-\infty ,+\infty )^N}{\exp \left( -\boldsymbol{A}_{i_1,i_2,\dots ,i_{\scrS N}}^{\prime}\boldsymbol{y} \right)}\cdot \frac{1}{\sqrt{2\pi \left| \mathbf{\Sigma } \right|}}
	\\
	&\quad  \cdot \exp \left[ -\frac{1}{2}(\boldsymbol{y}-\boldsymbol{\mu })^{\top}\mathbf{\Sigma }^{-1}(\boldsymbol{y}-\boldsymbol{\mu }) \right] d\boldsymbol{y}
	\\
	&\text{Let}\,\boldsymbol{y}=\boldsymbol{U\varLambda }^{\frac{1}{2}}\boldsymbol{x}+\boldsymbol{\mu }
	\\
	&=\sum_{i_1=0}^{k_1}{\sum_{i_2=0}^{k_2}{\cdots \sum_{i_{\scrS N}=0}^{k_{\scrS N}}{C_{k_1}^{i_1}C_{k_2}^{i_2}\cdots C_{k_{\scrS N}}^{i_{\scrS N}}}}}\int_{(-\infty ,+\infty )^N}{\exp \left( -\boldsymbol{A}_{i_1,i_2,\dots ,i_{\scrS N}}^{\prime}\boldsymbol{U\varLambda }^{\frac{1}{2}}\boldsymbol{x}-\boldsymbol{A}_{i_1,i_2,\dots ,i_{\scrS N}}^{\prime}\boldsymbol{\mu } \right)}
	\\
	&\quad  \cdot \frac{1}{\sqrt{2\pi}}\cdot \exp \left[ -\frac{1}{2}\boldsymbol{x}^{\prime}\boldsymbol{x} \right] d\boldsymbol{x}\\
	&=\sum_{i_1=0}^{k_1}{\sum_{i_2=0}^{k_2}{\cdots \sum_{i_{\scrS N}=0}^{k_{\scrS N}}{C_{k_1}^{i_1}C_{k_2}^{i_2}\cdots C_{k_{\scrS N}}^{i_{\scrS N}}}}}\exp \left( -\boldsymbol{A}_{i_1,i_2,\dots ,i_{\scrS N}}^{\prime}\boldsymbol{\mu } \right) \cdot \int_{-\infty}^{+\infty}{\frac{1}{\sqrt{2\pi}}}
	\\
	&\quad  \cdot \exp \left[ -\frac{1}{2}\boldsymbol{x}^{\top}\boldsymbol{x}-\boldsymbol{A}_{i_1,i_2,\dots ,i_{\scrS N}}^{\prime}\boldsymbol{U\varLambda }^{\frac{1}{2}}\boldsymbol{x} \right] d\boldsymbol{x}
	\\
	&=\sum_{i_1=0}^{k_1}{\sum_{i_2=0}^{k_2}{\cdots \sum_{i_{\scrS N}=0}^{k_{\scrS N}}{C_{k_1}^{i_1}C_{k_2}^{i_2}\cdots C_{k_{\scrS N}}^{i_{\scrS N}}}}}\exp \left( -\boldsymbol{A}_{i_1,i_2,\dots ,i_{\scrS N}}^{\prime}\boldsymbol{\mu } \right) \cdot \int_{-\infty}^{+\infty}{\frac{1}{\sqrt{2\pi}}}
	\\
	&\quad  \cdot \exp \left[ -\frac{1}{2}\boldsymbol{x}^{\top}\boldsymbol{x}-\frac{1}{2}\boldsymbol{A}_{i_1,i_2,\dots ,i_{\scrS N}}^{\prime}\boldsymbol{U\varLambda }^{\frac{1}{2}}\boldsymbol{x}-\frac{1}{2}\boldsymbol{x}^{\prime}\boldsymbol{\varLambda }^{\frac{1}{2}}\boldsymbol{UA}_{i_1,i_2,\dots ,i_{\scrS N}} \right] d\boldsymbol{x}
	\\
	&=\sum_{i_1=0}^{k_1}{\sum_{i_2=0}^{k_2}{\cdots \sum_{i_{\scrS N}=0}^{k_{\scrS N}}{C_{k_1}^{i_1}C_{k_2}^{i_2}\cdots C_{k_{\scrS N}}^{i_{\scrS N}}}}}\exp \left( -\boldsymbol{A}_{i_1,i_2,\dots ,i_{\scrS N}}^{\prime}\boldsymbol{\mu } \right) \cdot \int_{-\infty}^{+\infty}{\frac{1}{\sqrt{2\pi}}}
	\\
	&\quad  \cdot \exp \left[ -\frac{1}{2}\left( \boldsymbol{x}+\boldsymbol{A}_{i_1,i_2,\dots ,i_{\scrS N}}^{\prime}\boldsymbol{U\varLambda }^{\frac{1}{2}}\boldsymbol{x} \right) ^{\top}\left( \boldsymbol{x}+\boldsymbol{A}_{i_1,i_2,\dots ,i_{\scrS N}}^{\prime}\boldsymbol{U\varLambda }^{\frac{1}{2}}\boldsymbol{x} \right) \right]
	\\
	&\quad  \cdot \exp \left[ \frac{1}{2}\boldsymbol{A}_{i_1,i_2,\dots ,i_{\scrS N}}^{\prime}\boldsymbol{U\varLambda }^{\frac{1}{2}}\boldsymbol{\varLambda }^{\frac{1}{2}}\boldsymbol{UA}_{i_1,i_2,\dots ,i_{\scrS N}} \right] d\boldsymbol{x}
	\\
\end{aligned}
$$
$$
\begin{aligned}
	&=\sum_{i_1=0}^{k_1}{\sum_{i_2=0}^{k_2}{\cdots \sum_{i_{\scrS N}=0}^{k_{\scrS N}}{C_{k_1}^{i_1}C_{k_2}^{i_2}\cdots C_{k_{\scrS N}}^{i_{\scrS N}}}}}\exp \left( -\boldsymbol{A}_{i_1,i_2,\dots ,i_{\scrS N}}^{\prime}\boldsymbol{\mu }+\frac{1}{2}\boldsymbol{A}_{i_1,i_2,\dots ,i_{\scrS N}}^{\prime}\mathbf{\Sigma }\boldsymbol{A}_{i_1,i_2,\dots ,i_{\scrS N}} \right)
	\\
	&=\sum_{i_1=0}^{k_1}{\sum_{i_2=0}^{k_2}{\cdots \sum_{i_{\scrS N}=0}^{k_{\scrS N}}{C_{k_1}^{i_1}C_{k_2}^{i_2}\cdots C_{k_{\scrS N}}^{i_{\scrS N}}}}}\exp \left( -\sum_{s=1}^N{i_s\mu _s}+\frac{1}{2}\sum_{s=1}^N{\sum_{t=1}^N{i_s}}i_t\sigma _{st} \right) ,
\end{aligned}
$$
where $\boldsymbol{\Sigma}=\boldsymbol{U} \boldsymbol{\varLambda} \boldsymbol{U}^{\prime}$, $\boldsymbol{U}$ is an orthogonal matrix and $\boldsymbol{\varLambda}$ is a symmetric matrix.

\subsection*{A.10. Proof of Lemma \ref{gllemma4}}
For $\forall  k \in \left\{ 0 \right\}\cup  N$
$$
\begin{aligned}
	ER^{k+1}&=\sum_{\scrS   R=0}^{\infty}{r^{k+1}}\cdot p\left( 1-p \right) ^r\\
	&=\sum_{\scrS   R=1}^{\infty}{\left[ \left( r-1 \right) +1 \right] ^{k+1}}\cdot p\left( 1-p \right) ^r\\
	&=\sum_{\scrS   R=1}^{\infty}{\sum_{m=0}^{k+1}{C_{k+1}^{m}\left( r-1 \right) ^m}}\cdot p\left( 1-p \right) ^r\\
	&=\sum_{m=0}^{k+1}{C_{k+1}^{m}}\left( 1-p \right) \sum_{\scrS   R=1}^{\infty}{\left( r-1 \right) ^mp\left( 1-p \right) ^{r-1}}\\
	&=\left( 1-p \right) \sum_{m=0}^{k+1}{C_{k+1}^{m}ER^m}\\
	&=\left( 1-p \right) ER^{k+1} +  \left( 1-p \right) \sum_{m=0}^k{C_{k+1}^{m}ER^m}  .
\end{aligned}
$$
Furthermore, $ER^k=\sum_{\scrS   R=0}^{\infty}{r^k}\cdot \frac{\lambda ^r}{r!}e^{-\lambda}$, from which we can conclude
$$
\begin{aligned}
	ER^{k+1}&=\left( 1-p \right) ER^{k+1}+\left( 1-p \right) \sum_{m=0}^k{C_{k+1}^{m}ER^m}\\
	pER^{k+1}&=\left( 1-p \right) \sum_{m=0}^k{C_{k+1}^{m}ER^m}\\
	ER^{k+1}&=\frac{\left( 1-p \right)}{p}\sum_{m=0}^k{C_{k+1}^{m}ER^m}
\end{aligned}
$$

Finally, with some simple induction, we obtain
$$
E R^{k+1}= ER \cdot\sum_{m=0}^k{C_{k+1}^{m}ER^m} , \, \, k = 0,1,2,\dots.
$$

\subsection*{A.11. Proof of Theorem \ref{gltheo1}}
Without loss of generality, let $\forall \left( k_1,k_2,\dots ,k_{\scrS N} \right) \in \{ \left( k_1,k_2,\dots ,k_{\scrS N} \right) : k_1+k_2+\cdots +k_{\scrS N}=k,  k_s\in \left\{ 0,1,2,\dots ,k \right\} ,   s \in \left\{ 1,2,\dots ,N \right\} \} $, so we have
$$
\begin{aligned}
E\boldsymbol{R}_{k_1,k_2,\dots ,k_{\scrS N}}^{k}&=E\left( R^{k_1}R^{k_2}\cdots R^{k_{\scrS N}} \right) 
\\
&=E\left[ E\left( R^{k_1}R^{k_2}\cdots R^{k_{\scrS N}}|\boldsymbol{p } \right) \right] 
\end{aligned}
$$
and
$$
E\left( R^{k_1}R^{k_2}\cdots R^{k_{\scrS N}}|\boldsymbol{\lambda } \right) =E\left( R^{k_1}|\boldsymbol{\lambda } \right) E\left( R^{k_2}|\boldsymbol{\lambda } \right) \cdots E\left( R^{k_{\scrS N}}|\boldsymbol{\lambda } \right) .
$$
Therefore, by Lemma \ref{gllemma4}, we can obtain that $E\left( R^{k_1}R^{k_2}\cdots R^{k_{\scrS N}}|\boldsymbol{\lambda } \right)$ is a polynomial of degree at most $k$ composed of $\frac{1}{p_{2}}$, $\dots$, $\frac{1}{p_{\scrS N}}$.
Furthermore, for a given $\left( k_1,k_2,\dots ,k_{\scrS N} \right)$ as stated in Lemma \ref{gllemma3},
$$
E\left[ \left( \frac{1}{p_1} \right) ^{k_1}\left( \frac{1}{p_2} \right) ^{k_2}\cdots \left( \frac{1}{p_{\scrS N}} \right) ^{k_{\scrS N}} \right] =\sum_{i_1=0}^{k_1}{\sum_{i_2=0}^{k_2}{\cdots \sum_{i_{\scrS N}=0}^{k_{\scrS N}}{C_{k_1}^{i_1}C_{k_2}^{i_2}\cdots C_{k_{\scrS N}}^{i_{\scrS N}}}}}\cdot \exp \left[ -\sum_{s=1}^N{i_s\mu _s}+\frac{1}{2}\sum_{s=1}^N{\sum_{t=1}^N{i_s}}i_t\sigma _{st} \right] <\infty .
$$

\subsection*{A.12. Proof of Proposition \ref{glprop1}}
The proof can be obtained using Lemma \ref{gllemma2} and \ref{gllemma4}. Firstly by applying Lemma \ref{gllemma2} and the conditional expectation,
$$
\begin{aligned}
	E\left( R_i \right) &=E\left[ E\left( R_i|\boldsymbol{p} \right) \right]\\
	&=E\left[ \frac{1}{p_i}-1 \right]\\
	&=E\left[ \frac{1}{p_i} \right] -1\\
	&=1+\exp \left( -\mu _i+\frac{1}{2}\sigma _{ii} \right) -1\\
	&=\exp \left( -\mu _i+\frac{1}{2}\sigma _{ii} \right) .
\end{aligned}
$$
We obtain the proof of Equation (1). Furthermore by applying Lemma \ref{gllemma4},
$$
\begin{aligned}
	E\left( R_{i}^{2} \right) &=E\left[ E\left( R_{i}^{2}|\boldsymbol{p} \right) \right]\\
	&=E\left[ E\left( R_i|\boldsymbol{p} \right) +2\left( E\left( R_i|\boldsymbol{p} \right) \right) ^2 \right]\\
	&=E\left[ \frac{1-p_i}{p_i}+\frac{\left( 1-p_i \right) ^2}{p_{i}^{2}} \right]\\
	&=2E\left[ \left( \frac{1}{p_i} \right) ^2 \right] -3E\left[ \frac{1}{p_i} \right] +1\\
	&=2+4\exp \left( -\mu _i+\frac{1}{2}\sigma _{ii} \right) +2\exp \left( -2\mu _i+2\sigma _{ii} \right)\\
	&\quad  -3-3\exp \left( -\mu _i+\frac{1}{2}\sigma _{ii} \right) +1\\
	&=2\exp \left( -2\mu _i+2\sigma _{ii} \right) +\exp \left( -\mu _i+\frac{1}{2}\sigma _{ii} \right)  .
\end{aligned}
$$
We obtain the proof of Equation (2). By applying Lemma \ref{gllemma3}, we can infer that
$$
\begin{aligned}
	E\left( R_iR_j \right) &=E\left[ E\left( R_iR_j|\boldsymbol{p} \right) \right]\\
	&=E\left[ \left( \frac{1}{p_i}-1 \right) \left( \frac{1}{p_j}-1 \right) \right]\\
	&=E\left[ \frac{1}{p_i}\frac{1}{p_j} \right] -E\left[ \frac{1}{p_i} \right] -E\left[ \frac{1}{p_j} \right] +1\\
	&=\sum_{q_1=0}^1{\sum_{q_2=0}^1{C_{1}^{q_1}C_{2}^{q_2}}}\cdot \exp \left( -q_1\mu _i-q_2\mu _j+\frac{1}{2}q_1\cdot q_1\cdot \sigma _{ii}+\frac{1}{2}q_1\cdot q_2\cdot \sigma _{ij}+\frac{1}{2}q_2\cdot q_1\cdot \sigma _{ji}+\frac{1}{2}q_2\cdot q_2\cdot \sigma _{jj} \right)\\
	&=C_{1}^{0}C_{1}^{0}\exp \left( -0\cdot \mu _1-0\cdot \mu _1+\frac{1}{2}0\cdot 0\cdot \sigma _{ii}+\frac{1}{2}0\cdot 0\cdot \sigma _{ij}+\frac{1}{2}0\cdot 0\cdot \sigma _{ji}+\frac{1}{2}0\cdot 0\cdot \sigma _{jj} \right)\\
	&\quad +C_{1}^{0}C_{1}^{1}\exp \left( -0\cdot \mu _1-1\cdot \mu _2+\frac{1}{2}0\cdot 0\cdot \sigma _{ii}+\frac{1}{2}0\cdot 1\cdot \sigma _{ij}+\frac{1}{2}1\cdot 0\cdot \sigma _{ji}+\frac{1}{2}1\cdot 1\cdot \sigma _{jj} \right)\\
	&\quad +C_{1}^{1}C_{1}^{0}\exp \left( -1\cdot \mu _1-0\cdot \mu _2+\frac{1}{2}1\cdot 1\cdot \sigma _{ii}+\frac{1}{2}1\cdot 0\cdot \sigma _{ij}+\frac{1}{2}0\cdot 1\cdot \sigma _{ji}+\frac{1}{2}0\cdot 0\cdot \sigma _{jj} \right)\\
	&\quad +C_{1}^{1}C_{1}^{1}\exp \left( -1\cdot \mu _1-1\cdot \mu _2+\frac{1}{2}1\cdot 1\cdot \sigma _{ii}+\frac{1}{2}1\cdot 1\cdot \sigma _{ij}+\frac{1}{2}1\cdot 1\cdot \sigma _{ji}+\frac{1}{2}1\cdot 1\cdot \sigma _{jj} \right)\\
	&\quad -1-\exp \left( -\mu _i+\frac{1}{2}\sigma _{ii} \right)\\
	&\quad -1-\exp \left( -\mu _j+\frac{1}{2}\sigma _{jj} \right)\\
	&\quad +1\\
	&=\exp \left( -\mu _i-\mu _j+\frac{1}{2}\left( \sigma _{ii}+\sigma _{jj} \right) +\sigma _{ij} \right) .
\end{aligned}
$$
Equations (4) and (5) can be directly derived from Equations (1)-(3).

\subsection*{A.13. Proof of Proposition \ref{glprop3}}
The proof steps are similar to \citet{liu2021}. Assuming that Hypothesis \ref{glhypo1} holds, we can express the multivariate distribution of process \eqref{geneq1} using the distribution of the multivariate innovation process as follows:
$$
\boldsymbol{X}_t\overset{d}{=}\boldsymbol{A}^t\circ \boldsymbol{X}_0+\sum_{i=0}^{t-1}{\boldsymbol{A}^i}\circ \boldsymbol{R}_{t-i}. 
$$ 
Consequently, we have
$$
\boldsymbol{X}_t\overset{d}{=}\sum_{i=0}^{\infty}{\boldsymbol{A}^i}\circ \boldsymbol{R}_{t-i},
$$
which means that process \eqref{geneq1} possesses a unique stationary solution. Based on Hypothesis \ref{glhypo1} and Theorem \ref{pltheorem1}, it is known that $E\left( \boldsymbol{R}_{t}^{k} \right) <\infty $. Therefore, for any positive integer $k$ in any estimation, we have $E(\boldsymbol{X}_t^k )<\infty$.

\end{document}